\documentclass[11pt,onecolumn,lettersize,journal]{IEEEtran}
\usepackage[margin=1.05in]{geometry} 
\usepackage[utf8]{inputenc}                     

\usepackage{cite}                               
\usepackage{url}                                

\usepackage{amsmath,amssymb,amsfonts,mathtools} 
\allowdisplaybreaks[3]          
\usepackage{stmaryrd}                           
\usepackage{nicefrac}                           
\usepackage{mathrsfs}                           
\usepackage{dsfont}                             

\usepackage{amsthm}                             
\newtheorem{theorem}{Theorem}            
\newtheorem{lemma}{Lemma}                
\newtheorem{corollary}{Corollary}        

\theoremstyle{definition}
\newtheorem{definition}{Definition}      

\theoremstyle{remark}
\newtheorem{remark}{Remark}              

\usepackage[linesnumbered,ruled]{algorithm2e}   

\usepackage{graphicx}                           
\usepackage{epsfig,epstopdf}                    
\usepackage{stfloats}                           
\usepackage{wrapfig,floatflt,float}             

\usepackage{array}                              
\usepackage{booktabs}                           
\usepackage{multirow}                           
\usepackage{makecell}                           
\usepackage{adjustbox}                          

\usepackage{xcolor}                             
\usepackage{framed}                             
\usepackage{tcolorbox}                          

\usepackage{paralist}                           

\usepackage[hidelinks,unicode,pdfencoding=auto]{hyperref}
\usepackage{microtype}                          

\usepackage{rotating}                           
\usepackage{pdflscape}                          
\usepackage{placeins}                           
\usepackage[normalem]{ulem}                     
\usepackage{psfrag}                             

\makeatletter
\long\def\@makecaption#1#2{%
  \ifx\@captype\table
    \footnotesize\textbf{#1.}\ #2\par
  \else
    \footnotesize #1. #2\par
  \fi}
\makeatother
\usepackage[T1]{fontenc}
\usepackage{libertine}        
\usepackage{libertinust1math} 
\usepackage{inconsolata}      
\allowdisplaybreaks

\newcommand{\nodes}{\mathcal{N}}              
\newcommand{\comp}{^{\mathrm{c}}}             
\newcommand{\ii}{\mathds{1}}                  

\newcommand{\OO}{O}
\newcommand{\oo}{o}

\newcommand{\pr}{\mathbb{P}}
\newcommand{\bP}[1]{\mathbb{P}\!\bigl[#1\bigr]}

\newcommand{\N}{\mathbb{N}}
\newcommand{\R}{\mathbb{R}}

\newcommand{\limit}{\lim_{n\rightarrow\infty}}

\newcommand{\lnormalized}{\mathcal{L}}

\newcommand{\hh}{\mathbb{H}(n;K)}
\newcommand{\hhn}{\mathbb{H}(n;K_n)}

\newcommand{\hhdn}{\mathbb{H}(n;K_n,\gamma_n)}

\newcommand{\fsquare}{\vrule height6pt width7pt depth1pt}

\newcommand{\myendpf}{\hfill\fsquare\\[0.1in]}

\newcolumntype{L}{>{\centering\arraybackslash}m{1.7cm}}
\newcolumntype{M}{>{\centering\arraybackslash}m{1.45cm}}
\newcolumntype{N}{>{\centering\arraybackslash}m{3.45cm}}

\begin{document}

\title{\fontsize{24}{24}\selectfont On Balancing Sparsity with Reliable Connectivity  in Distributed Network Design\\ with Random K-out Graphs}

\author{
 \fontsize{11}{11}\selectfont Mansi~Sood$^*$, Eray~Can~Elumar$^+$, and Osman~Ya\u{g}an$^\dagger$\\[0.85em]
  $^*$Laboratory for Information \& Decision Systems, \\Massachusetts Institute of Technology, Cambridge, MA, USA\\ [-0.3em]
  $^+$Intel Corporation, Hillsboro, OR, USA\\ [-0.3em]
  $^\dagger$Department of Electrical \& Computer Engineering and Cylab Security \& Privacy Institute,\\ Carnegie Mellon University, Pittsburgh, PA, USA\\[0.85em]
  {\tt msood@mit.edu}, ~{\tt eray.can.elumar@intel.com}, ~{\tt oyagan@ece.cmu.edu}\\
 }

\maketitle

\begin{abstract}
In several applications in distributed systems, an important design criterion is ensuring that the network is \emph{sparse}, i.e., does not contain too many edges, while achieving \emph{reliable connectivity}. Sparsity ensures communication overhead remains low, while reliable connectivity is tied to reliable communication and inference on decentralized data reservoirs and computational resources. A class of network models called \emph{random K-out graphs} appear widely as a heuristic to balance connectivity and sparsity, especially in settings with limited trust, e.g., privacy-preserving aggregation of networked data in which networks are deployed. However, several questions remain regarding how to choose network parameters in response to different operational requirements, including the need to go beyond asymptotic results and the ability to model the stochastic and adversarial environments. To address this gap, we present  theorems to inform the choice of network parameters that guarantee reliable connectivity in regimes where nodes can be \emph{finite} or \emph{unreliable}. We first derive upper and lower bounds for probability of connectivity in random K-out graphs when the number of nodes is finite. Next, we analyze the property of \emph{$r$-robustness}, a stronger notion than connectivity that enables resilient consensus in the presence of malicious nodes. Finally, motivated by aggregation mechanisms based on \emph{pairwise masking}, we model and analyze the impact of a subset of adversarial nodes, modeled as \emph{deletions}, on connectivity and giant component size—metrics that are closely tied to privacy guarantees. Together, our results pave the way for end-to-end performance guarantees for a suite of algorithms for reliable inference on networks.\end{abstract}


\section{Introduction}

\subsection{Background}

As our world becomes increasingly interconnected, the informational landscape that drives decision making is characterized by ever‑expanding scale and complex interdependencies. Motivated by the communication and computational constraints of emerging computing paradigms, we study how leveraging network structure more effectively can push the frontiers of large‑scale inference. Of particular interest in this paper is the property of \emph{connectivity}, a fundamental driver of performance in distributed systems \cite{bondy1976graph, pirani2023graph, federated2019advances}. Connectivity requires the existence of a path between every pair of nodes, ensuring that all nodes participate in and can communicate across the network.  In addition to the simple interpretation of enabling communication (either through a direct edge or a sequence of edges) between node pairs, as we will see later in our discussion, a somewhat surprising implication of network connectivity is its connection with the privacy guarantees of several algorithms \cite{2022sabater-gopa,2020secureagg-polylog,10622678} for distributed inference (Figure~\ref{fig:partialsum}). Now, although the trivial way to guarantee connectivity is simply to add more edges, each edge introduces operational overhead. With \emph{sparsity} quantifying the number of edges or links in the network compared to the complete network with all possible edges, the sparsity-connectivity trade-offs manifest in different applications as the following question. How should the network designer choose parameters that ensure the desired connectivity properties without having to establish too many edges? 

 
\begin{figure}[t]
\centering
\includegraphics[scale=0.20]{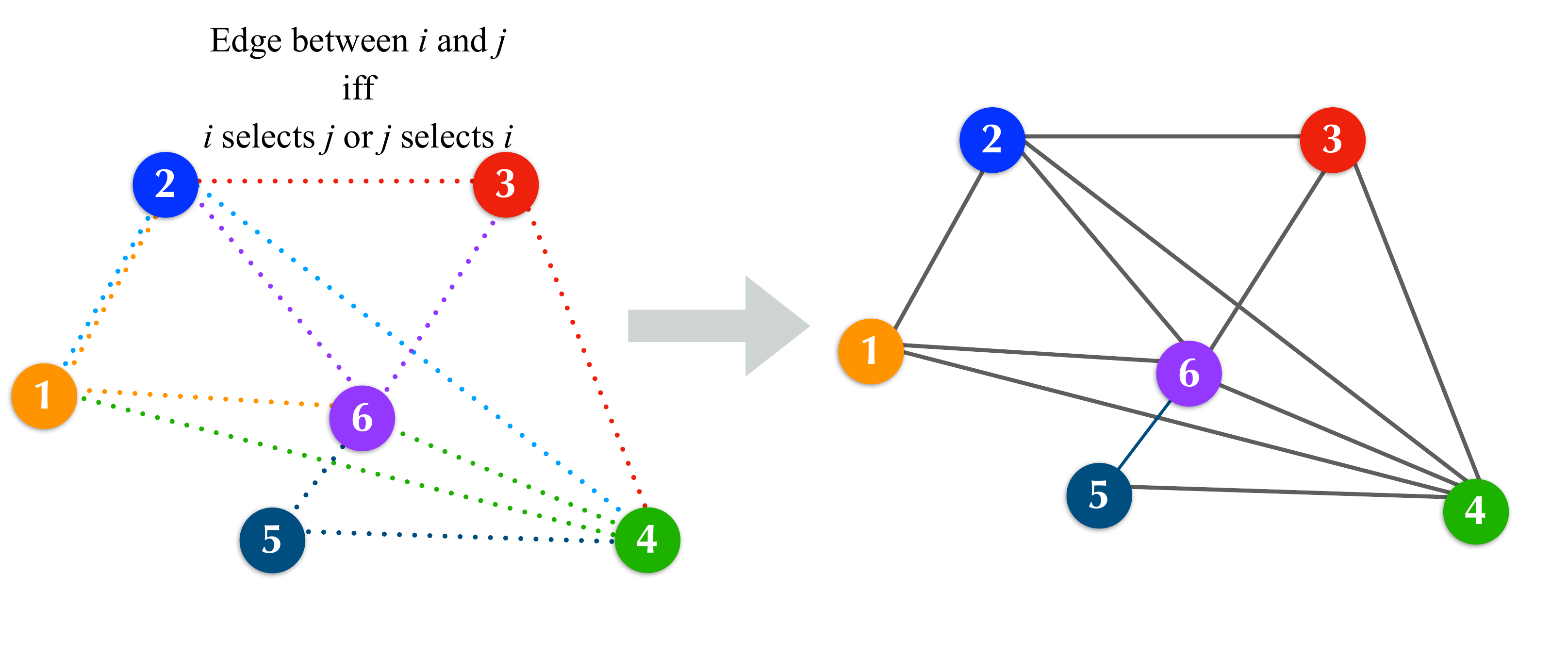}
\caption{A realization of the random K-out graph with $6$ nodes with parameter $K=2$. Each node selects $K=2$ nodes uniformly at random from all other nodes. An undirected edge is drawn between two nodes if at least one selects the other yielding the random K-out graph.} 
\label{fig:0}
\end{figure} 
\begin{table*}[t!]
  \centering
  \setlength{\tabcolsep}{0.2cm}
  \setlength{\extrarowheight}{6pt}
  \renewcommand{\arraystretch}{1.3}
  \begin{tabular}{|l|l|l|}
    \hline
    \makecell[c]{{Property of interest}}
      & \makecell[c]{{$K_n$ to guarantee property with high probability}}
      & {Results appear in} \\
    \hline \hline
            \makecell[c]{$1$‑connectivity (or connectivity) \\ $P(n;K_n)= \pr[{\mathbb{H}(n;K_n) ~\mbox{is connected}}] $}
      & \makecell[l]{ $K_n \ge 2$ \\ \\  $P(n;K_n)= 1 - \Theta\left(\frac{1}{n^{K^2-1}}\right)$ \\ \\  upper bound on $P(n;K_n)$ for finite $n$ \\ \\ lower bound on $P(n;K_n)$ for finite $n$ }
      & \makecell[l]{ \cite{FennerFrieze1982,Yagan2013Pairwise}  \\ \\\textbf{Corollary~\ref{corr:first}}\\ \\ \textbf{Theorem~\ref{thm:UpperBoundForConnectivity}} \\ \\\textbf{Theorem~\ref{thm:LowerBoundForConnectivity}}  \\ (improves \cite{FennerFrieze1982,Yagan2013Pairwise})} \\
      \hline
        \makecell[c]{Minimum node degree $\geq k$,\\ $k \geq 2$}
      & \makecell[l]{ $K_n \ge k$  }
      & \makecell[l]{ \cite{frieze2016introduction}}  \\
           \hline
        \makecell[c]{$k$‑connectivity,\\ $k \geq 2$ }
      & \makecell[l]{ $K_n \ge k$  }
      & \makecell[l]{ \cite{FennerFrieze1982}}  \\
    \hline
    \makecell[c]{$r$‑robustness,\\ $r \geq 2$ }
      & \makecell[l]{ \\$K_n \ge 2r$ \\ \\ $K_n \ge r \log r$\\}
      & \makecell[l]{ \textbf{Theorem~\ref{robustness-theorem:thmr_1}} \\ \\ \cite{can_cdc2021}}  \\
    \hline
\makecell[c]{1‑connectivity with \\$\gamma_n = \alpha n,\;\alpha\in(0,1)$ \\random node deletions\\
             }
  &
\makecell[l]{$K_n \ge (1+\varepsilon)\,\dfrac{\log n}{1 - \alpha - \log \alpha},\;\text{~for any~} \varepsilon>0$\\ \\ 
             $K_n \ge \dfrac{1}{1-\alpha} \cdot\,\dfrac{\log n}{1 - \alpha - \log \alpha}$}
  &
\makecell[l]{\textbf{Theorem~\ref{theorem:thmc_1}}\\ \\ \\
             \cite{YAGAN2013493}} \\
    \hline
    \makecell{1-connectivity with $\gamma_n = \oo(n) $ \\random node deletions\\  \\ $\gamma_n= \omega(\sqrt{n})$ \& $ \gamma_n=  \oo({n})$ \\ \\ $\gamma_n =\oo(\sqrt{n})$}
            &  \makecell[l]{ \\  \\ $K_n = \displaystyle \frac{\log(\gamma_n)}{\log 2 + \tfrac12} + \omega(1)$ \\ \\  $K_n \ge 2$} 
      &  \makecell[l]{\\ \\  \\ \\ \textbf{Theorem~\ref{theorem:thmc_2}} \\ \\ \textbf{Theorem~\ref{theorem:thmc_2}}} \\
        \hline
\makecell[c]{\\Largest connected component $C_{\rm max}$ s.t.\\ under $\gamma_n= \alpha n,\;\alpha\in(0,1)$ node deletions,  \\fewer than $\lambda_n~~\left(< \lfloor\tfrac{(1-\alpha)n}{3}\rfloor\right)$ nodes outside $C_{\rm max}$ \\ \\ 
             }
  &
\makecell[l]{%
  $\displaystyle
    K_n \ge 1
    + \frac{%
        \log\bigl(1 + \tfrac{\alpha n}{\lambda_n}\bigr)
        + \alpha + \log(1-\alpha)
      }{%
        \tfrac{1-\alpha}{2}
        - \log\bigl(\tfrac{1+\alpha}{2}\bigr)
      }$
}
  &
\makecell[l]{\textbf{Theorem~\ref{theorem:thmg_5}}} \\
    \hline
    \makecell{$C_{\rm max}$ with $\gamma_n = \oo(n) $ \\random node deletions, \\s.t. fwer than $\lambda_n$ nodes outside $C_{\rm max}$\\  \\ $\lambda_n=o(n)$ \\ \\ $\lambda_n=\beta n,~\beta \in (0, 1/3)$}
        & \makecell[l]{
  \\ \\ $\displaystyle K_n \ge 1
    + \frac{\log\bigl(1+\tfrac{\gamma_n}{\lambda_n}\bigr)}
           {\log 2 + \tfrac12}$\\[12pt]
  $K_n \ge 2$
}
      &  \makecell[l]{\\ \\  \\ \\ \textbf{Theorem~\ref{theorem:thmg_3}} \\ \\ \textbf{Theorem~\ref{theorem:thmg_3}}}\\ \hline
  \end{tabular}
  \vspace{2mm}
  \caption{A guide to distributed network design with random K-out graph $\hh$, we let $K$ scale with the number of nodes, denoted by $K_n$. Variants of random K-out graphs appear widely as a heuristic to simultaneously ensure reliable connectivity while not requiring too many edges \cite{2020secureagg-polylog,2022sabater-gopa, SCION2017book,SIAM2021paper,Haowen_2003} in distributed systems, especially in settings with limited trust. Note that the average node of a random K-out graph $\hhn$ scales as $2 K_n-{K_n^2}/{n-1}$, and consequently any finite choice of $K_n$ ensures the average node degree is bounded. We provide tight finite node upper and lower bounds for connectivity, with these bounds matching in order asymptotically, indicating no further improvement is possible. For the property of $r$-robustness defined in \ref{robustness-def:r-robust}, which is closely related to resilience of consensus algorithms, we show that $K_n \geq 2r$ (and thus a bounded average degree) suffices for imparting $r$-robustness whp.  Finally, motivated by adversarial models of capturing a random subset of nodes, such as in Figure~\ref{fig:partialsum}, where privacy guarantees are tied intimately to connectivity properties of the graph obtained by restricting to honest nodes and effectively removing adversarial nodes. We provide a suite of results on connectivity and size of the giant component of random K-out graphs in Theorems~\ref{theorem:thmc_1}-\ref{theorem:thmg_5}. }
  \label{table:compare}
\end{table*}
While the sparsity connectivity trade-off is a fundamental trade‑off appearing in many branches of mathematics to computing (such as the study of expanders \cite{hoory2006expander}), the increasingly decentralized nature of data in distributed systems, especially in settings with \emph{limited trust}, presents new challenges for navigating this trade‑off. In particular, we require simple, asynchronous constructions that are amenable to \emph{distributed construction} without any trusted centralized entity to organize the nodes into a graph topology. Another consideration is the \emph{ease of implementation} and analytical tractability to ensure that formal \emph{end-to-end} performance guarantees are met, especially in the presence of algorithmic \emph{randomness} (such as the use of additive noise) in \emph{stochastic} and \emph{adversarial} operational environments (such as nodes dropping out or being malicious). While several network models \cite{frieze2016introduction, hoory2006expander, law2003distributed} meet a subset of these requirements individually, variants of a class of stochastic network models known as \emph{random K-out graphs} have been proposed as a candidate to balance all of the above requirements in several distributed systems \cite{FennerFrieze1982,Bollobas,frieze2016introduction}. 


\begin{figure}[t]
\centering
\includegraphics[scale=0.22]{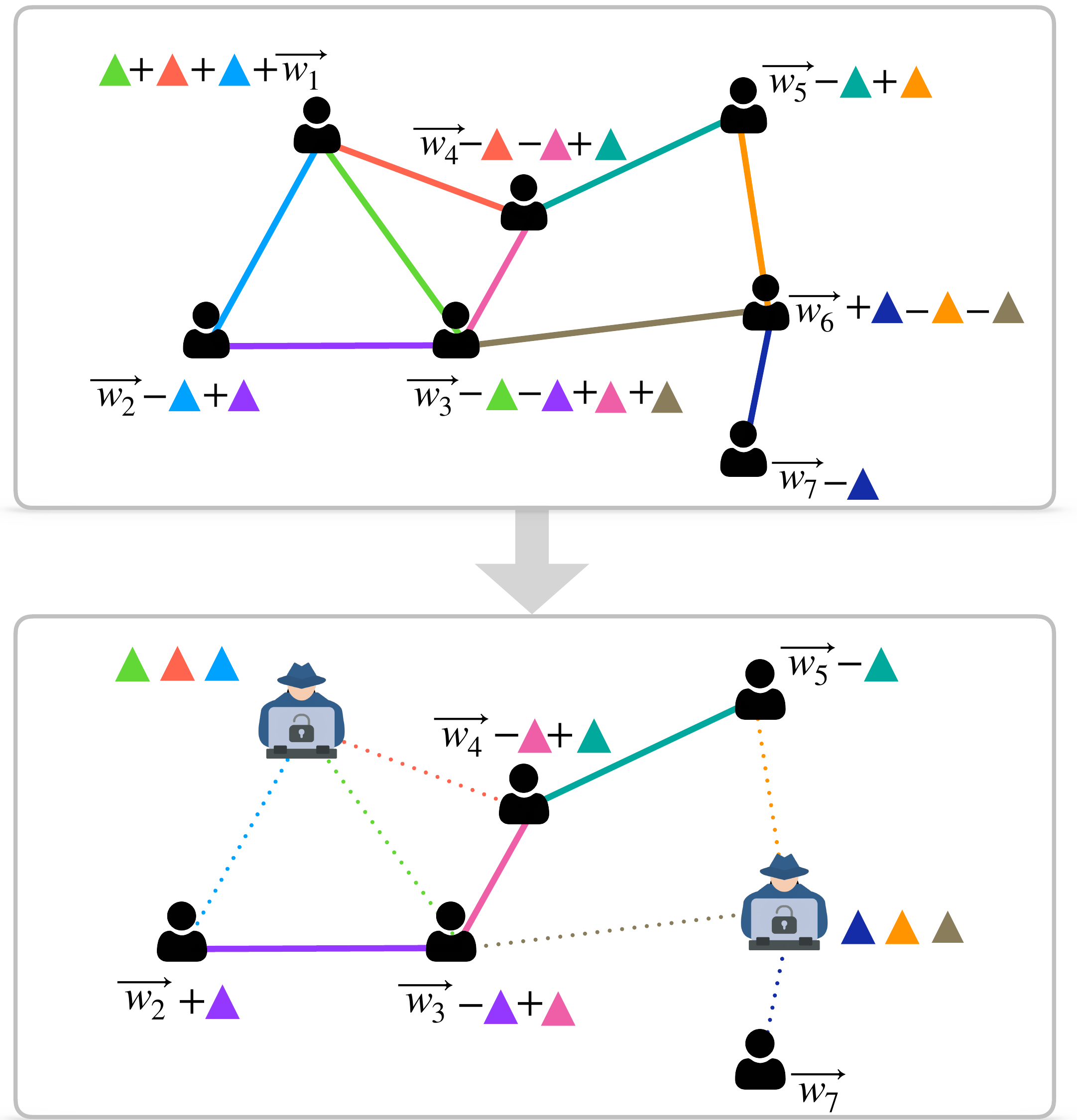}
\caption{Pairwise masks for fully-decentralized learning (e.g., \cite{2022sabater-gopa}): An illustration with seven nodes, each holding a private weight on an underlying communication graph. The goal is for the nodes to learn aggregate weights without revealing individual ones. Every directly connected node pair generates a pairwise noise term: one node adds it to its private weight before broadcasting, and the other subtracts it. In the second stage, two colluding nodes aim to recover the individual weights of the other users. Together, they can recover the masks from their neighbors and, in the process, learn the partial sum $w_2+w_3+w_4+w_5$ and the weight $w_7$, which correspond to the components of the honest-user subgraph after removing the adversarial nodes. This motivates the study of connectivity (ensuring no single user is cornered) and component sizes (in connection to partial sums exposed) under node deletions. Here, sparsity ensues bounded communication overhead while connectivity enables privacy guarantees.} 
\label{fig:partialsum}
\end{figure} 
\begin{figure}[t]
\centering
\includegraphics[scale=0.17]{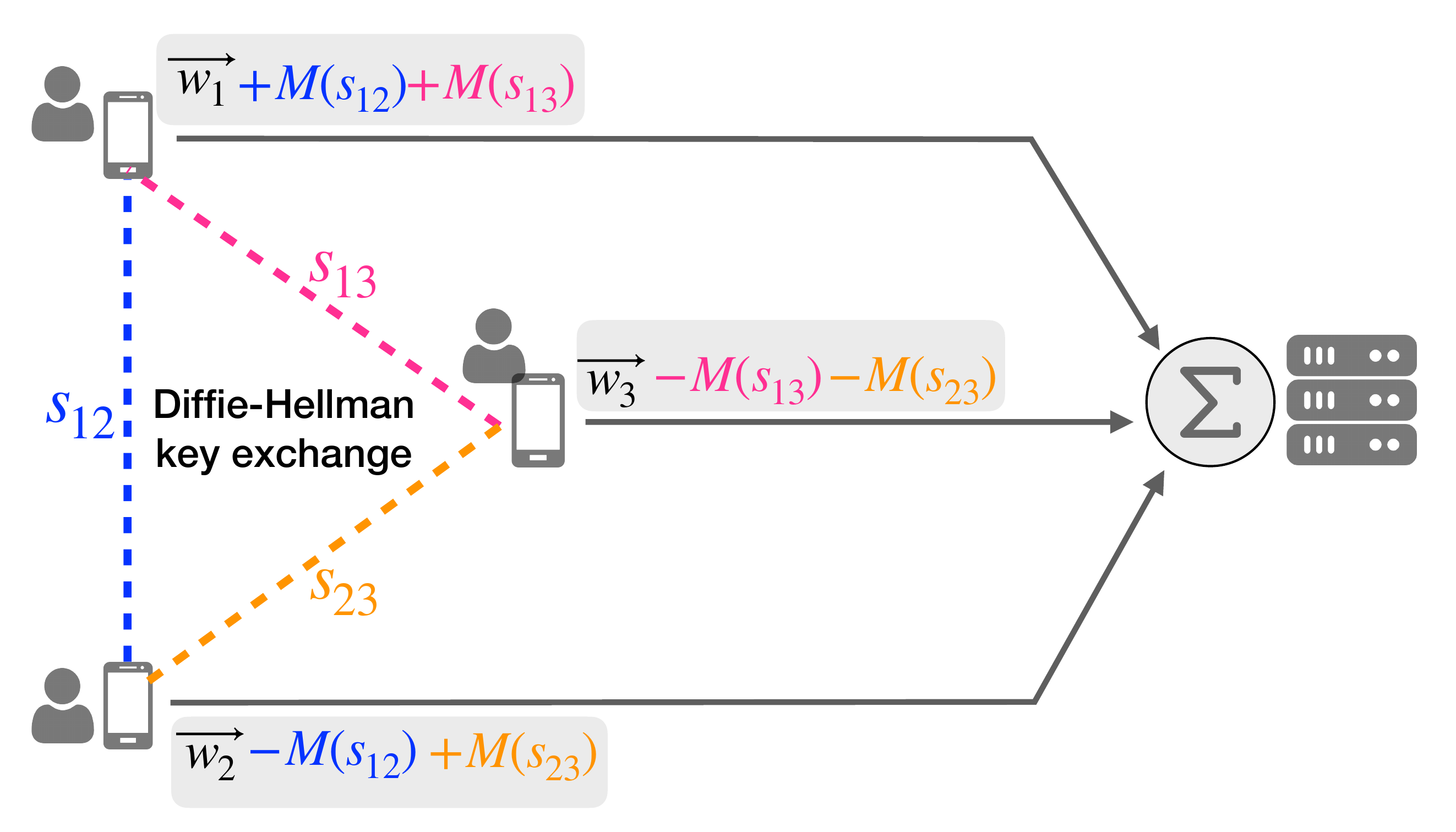}
\caption{Pairwise masks for federated learning: The basic idea behind \emph{pairwise} masking—obfuscating individual updates through the addition of pairwise masks with neighbors that cancel out in aggregate—extends to the federated learning setting through the secure-aggregation framework \cite{2020secureagg-polylog,yiping-ma-23}. Although the communication graph is a star graph and communication is facilitated by the server, clients create a virtual graph through a key exchange protocol such as the Diffie-Hellman protocol \cite{2017secureagg-linear,2020secureagg-polylog}, which enables the exchange of cryptographic primitives between node pairs. These masks are not revealed to a potentially semi-honest or malicious server. In this setting, the sparsity-connectivity trade-off manifests as designing networks that are \emph{sparse} (to minimize communication costs) and yet \emph{reliably connected} (to ensure each client has enough honest neighbors despite dropouts and corruptions).}
\label{fig:federated-motivation}
\end{figure}
\begin{figure}[t]
\begin{centering}
\includegraphics[scale=0.17]{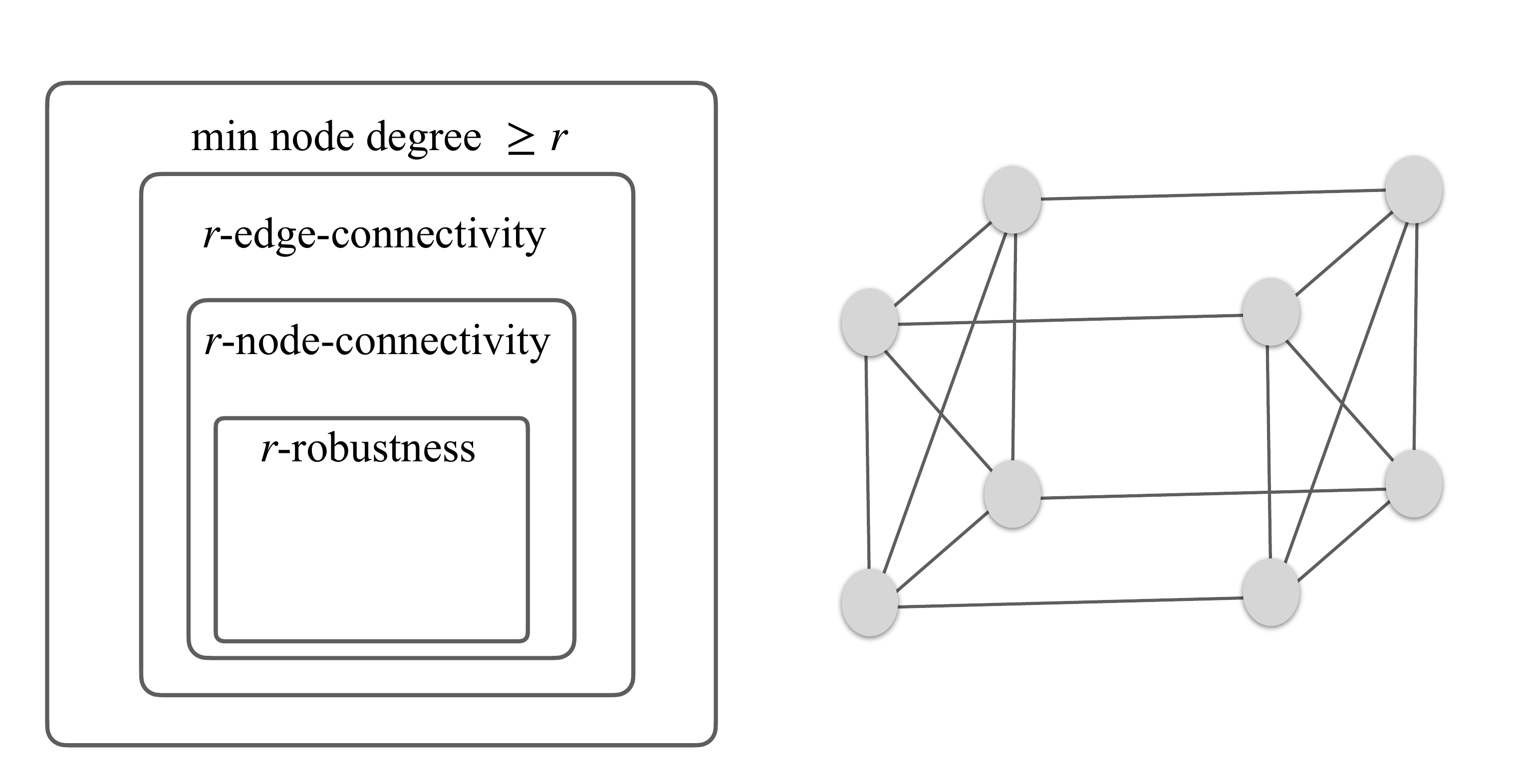}
\caption{Contextualizing different notions of connectivity. For any graph, $r$-robustness (Def.~\ref{robustness-def:r-robust}) implies $r$-connectivity (Def.~\ref{robustness-def:r_connectivity}) which in turn implies a minimum node degree of $r$ \cite{bondy1976graph,pirani2023graph}. Note that $r$-connectivity does not imply $r$-robustness. As an example, on the graph on the right, we have 8 nodes, with the network being $4$-connected; however, the network is only $1$-robust. Thus, $r$-robustness is the \emph{strongest} amongst these properties. One of the key implications of $r$-robustness is capturing redundancy of direct information exchange between subsets of nodes, which in turn imparts resilient consensus in a class of consensus algorithms \cite{robust_consensus,pirani2023graph,sundaram_robust}. More concretely, if at most $F$ adversarial nodes occupy the neighborhood of every honest node, consensus can be reached between the honest nodes if the network is $(2F + 1)$-robust.} 
\label{fig:r-robustness-motivation}
\end{centering}
\end{figure} 

Random K-out graphs are receiving increasing attention as a model to construct sparse yet well-connected topologies in a fully distributed fashion with applications such as the design of securely connected sensor networks \cite{Haowen_2003}, privacy-preserving distributed inference algorithms \cite{2022sabater-gopa}, and anonymity-preserving cryptocurrency networks \cite{FantiDandelion2018}. A { random K-out graph} over a set of $n$ nodes is constructed as follows (Figure~\ref{fig:0}). Each node draws an edge towards $K$ distinct nodes selected uniformly at random. The orientation of the edges is then ignored, yielding an {undirected} graph.  Random K-out graphs are known to achieve connectivity easily, i.e., with far fewer edges $(O(n))$ as compared to classical random graph models including Erd\H{o}s-R\'enyi (ER)  graphs \cite{erdos61conn,Bollobas}, random geometric graphs \cite{penrose2003random}, and random key graphs \cite{yagan2012zero}, which all require $O(n \log n)$ edges for connectivity.  In particular, it is known \cite{Yagan2013Pairwise, FennerFrieze1982} that random K-out graphs are connected with high probability ({\em whp}) when $K \geq 2$. 
Moreover, note that random K-out graphs are only suboptimal by a constant factor as compared to the sparsest network topologies that asymptotically achieve the desired properties. For example, line graphs for $1$-connectivity and Harary graphs \cite{harary1962maximum} for $k$-connectivity, i.e., the property of retaining connectivity in the presence of up to $k-1$ node or edge failures), but which may be more difficult to implement on a large scale in practice or require a more coordinated construction.

 \subsection{Applications of random K-out graphs in distributed network design and motivating network design questions}

 \subsubsection{Aggregation mechanisms based on pairwise additive masks} To more concretely illustrate why balancing sparsity and connectivity is important in distributed systems, we first consider two widely studied distributed inference algorithms: federated learning with a central server and fully decentralized learning without one \cite{federated2019advances}. In these settings, mechanisms based on \emph{{pairwise} additive masking} have been proposed to aggregate client updates in a privacy-preserving manner. Figure~\ref{fig:partialsum} depicts private gossip averaging for the fully decentralized setting, where the masks are pairwise statistical noise terms generated per edge of the communication graph \cite{2022sabater-gopa}. Figure~\ref{fig:federated-motivation} illustrates secure aggregation for the federated learning setting, where the additive masks are based on cryptographic primitives exchanged between a subset of client pairs through the Diffie-Hellman key exchange protocol \cite{maurer2000diffie,2020secureagg-polylog}. If the client dropouts are carefully accounted for, the masks get averaged out and do not affect the final aggregation result \cite{2017secureagg-linear, 2020secureagg-polylog, yiping-ma-23}. 
 
 An underlying phenomenon driving the trade-off between privacy guarantees and communication costs is the structure of the graph over which neighboring clients communicate and implement pairwise masks \cite{2020secureagg-polylog,2022sabater-gopa}. Intuitively, a sparser communication graph where each node has fewer connections entails lower communication overhead. At the same time, it is desirable that each node has enough neighbors that are {honest} to ensure their privacy. Across both these applications, for settings with malicious nodes, random K-out graphs have been used as a candidate to generate the graph over which masks are generated and shared pairwise between node pairs connected by an edge \cite{2022sabater-gopa,2020secureagg-polylog,yiping-ma-23}. As illustrated in Figure~\ref{fig:partialsum}, in the presence of colluding adversaries, the privacy guarantees are critically tied to the component sizes of the subgraph of honest users, after treating the colluding nodes as removed from the network. Ideally, we want to have a dense enough graph to ensure connectivity after removing the colluding nodes. However, if the colluding nodes occupy a large subset of nodes, ensuring connectivity of the subgraph of users may be impossible. In that case, we note that the colluding adversarial nodes are able to recover partial sums corresponding to connected components of subgraph of honest users. Moreover, the honest nodes outside the largest connected component, often referred to as a giant component (Defintion~\ref{robustness-def:giantcomp}), are most at risk of being compromised as their weights are obfuscated by only a small subset of nodes. The natural design question that emerges is the following:

%


\begin{quote}
\emph{Q1. Given that adversaries occupy a random subset of size $\gamma_n$, how to choose the parameter $K$ (as a function of $n$ and $\gamma_n$) to ensure $1$-connectivity of the subgraph of honest users?\\   Further, when establishing $1$-connectivity is infeasible, what is an upper bound on the number of nodes in the honest‐user subgraph that lie outside the largest connected component?}
\end{quote}
 \vspace{5pt}


\subsubsection{Resilient consensus algorithms}
Consensus dynamics involves the alignment of a subset of parameters of multiple agents through repeated local interactions \cite{pirani2023graph}. It is a canonical problem in multi-agent dynamics and has been studied in various communication-system applications. For example, in the Internet of Things (IoT), consensus algorithms have been applied to federated learning \cite{federated_learning_consensus2}, blockchain \cite{blockchain_consensus2}, and smart grid systems \cite{smart_energy_consensus}. In particular, \cite{smartgrid_consensus} employs a leader–follower consensus algorithm to synchronize heterogeneous energy storage devices and enable plug-and-play capability in smart grids. Consensus dynamics also underpins control-theoretic applications in multi-agent systems—such as flocking, swarming, synchronization of coupled oscillators, and load balancing in networks \cite{consensus_multi}.

Since consensus relies on interactions among nodes, resilience against adversarial agents is a key consideration. In \cite{consensus_security}, conditions on network connectivity required to tolerate misbehaving or faulty agents were derived for linear consensus networks. However, \cite{robust_consensus} shows that, in general,  when adversarial nodes are present, mere $r$-connectivity is \emph{insufficient} and we need a stronger property of $r$-robustness (Figure~\ref{fig:r-robustness-motivation}). Instead, if each correctly behaving node has at most $F$ adversarial neighbors, consensus can be achieved only if the network is $(2F+1)$-robust. Moreover, it is known that even when the network robustness condition is not satisfied at all times, resilient consensus can still be reached if the union of communication graphs over a bounded period of time is $(2f+1)$‑robust \cite{pirani2023graph}. Thus, \emph{$r$-robustness} (Definition~\ref{robustness-def:r-robust}) is especially relevant for consensus dynamics and has been explored in numerous applications. For instance, \cite{rrobust_robot_control} derives control laws to arrange robots in $r$-robust graphs, enabling consensus despite malicious robots. Consequently, analysis of $r$-robustness of different graph models is being actively studied \cite{can_cdc2021,sundaram_robustness,lee2024construction,lee2024maintaining} to identify those that satisfy this property as efficiently (i.e., with few edges) as possible, however tight thresholds for random K-out graphs remains an open problem. \\

\begin{quote}
    \emph{Q2. Given the design requirement of $r$-robustness, how do we choose $K$ as a function of $r$ to ensure the random K-out graph is $r$-robust with high probability?}
\end{quote}
 \vspace{10pt}

\subsubsection{More applications} The underlying principle of balancing sparsity and connectivity translates to various application domains. In addition to applications in privacy-preserving learning, network connectivity is a common structural assumption in decentralized learning. Here, the trade-off is between ensuring fast convergence (by ensuring high connectivity) while keeping per-round communication costs low (which requires sparsity)\cite{song2022communication}. We direct the reader to the following salient applications in distributed systems where variants of random K-out graphs, (including directed variants and $K$-out subgraphs of arbitrary graphs \cite{frieze2016introduction}) have been used: {secure connectivity in wireless sensor networks} \cite{Haowen_2003}, 
{secure inter-domain routing protocols for next generation Internet architectures} \cite{SIAM2021paper, SCION2017book}, and {anonymity preserving routing in cryptocurrency networks} \cite{FantiDandelion2018}. Across the above applications, going beyond asymptotic results typical in the literature \cite{bollobas2007phase}, we consider the following fundamental question: \\

\begin{quote}
\emph{Q3. Given a finite and fixed number of nodes $n$ and the mission requirement of guaranteeing a probability of connectivity of at least~$\Delta$, how should a network designer choose $K$ as a function of $n$ and $\Delta$?}
\end{quote}
 \subsection{Main contributions and related results}
 We present a summary of related results and contextualize our contributions in Table~\ref{table:compare}. A preliminary version of these results was presented at \cite{sood2021tight,can_isit21,canicc2023}
 \subsubsection{1-connectivity}
 In this work, we derive {\em matching} upper and lower bounds prove that the probability of connectivity is $1-\Theta({1}/{n^{K^2-1}})$ for all $K \geq 2$. To the best of our knowledge, our work is the first to provide an upper bound on the probability of connectivity. Our corresponding lower bound significantly improves the existing ones; see Section~\ref{sec:main-results-tight-bounds} for a detailed comparison of the bounds. These matching bounds are made possible by invoking a variant of the Stirling formula \cite{stirlingremark} and the {Bonferroni inequality} \cite{bonferroni}. Our bounds give a precise characterization of connectivity for the finite node setting and help answer design {Q3.}, e.g., through \eqref{eq:koutbounds:ourlb2} show that with $n=16$ nodes and $\Delta=99.9\%$, setting $K=2$ ensures 1‑connectivity with probability at least $\Delta$ for random K‑out graphs.

\subsubsection{$r$-robustness}
 In this work, using a novel proof technique, we show that $r$-robustness can be achieved with a much smaller threshold than previously known \cite{can_cdc2021}, and find that $K \geq 2r$ is sufficient to ensure that the random K-out graph $\hh$ is $r$-robust {\emph whp}. This improves the previous known condition of $K=\OO(r \log(r))$ for $r$-robustness {\em with high probability} (whp) and provides a tighter answer to design Q2. Distinct from prior work, which relies on commonly used upper bound for the binomial coefficients
$\binom{n}{k} \le \Bigl({e n}/{k}\Bigr)^k$
and the union bound\cite{sundaram_robustness,can_cdc2021} to bound the probability of a subset of given size being not \(r\)-reachable, our proof uses the Beta function \(B(a,b)\)\cite{NIST:DLMF} and leverages its properties to achieve tighter bounds. An investigation of $r$-robustness forms the focus of Section~\ref{sec:main-results-r-robustness}.

\subsubsection{1-connectivity under node deletions}
We provide a comprehensive set of results on the connectivity and giant component size of \(\mathbb{H}(n;K_n,\gamma_n)\) when \(\gamma_n\) of its nodes, selected uniformly at random, are deleted. First, we derive conditions for \(K_n\) and \(n\) that ensure, with high probability (whp), the connectivity of the remaining graph when the number of deleted nodes is \(\gamma_n=\Omega(n)\) and \(\gamma_n=o(n)\), respectively. For the regime $\gamma_n = \alpha n,~ \alpha \in (0,1)$, our results correspond to a sharp threshold corresponding to a \emph{sharp} zero-one law for connectivity, filling in the gap of a factor of $1/(1-\alpha)$ in existing results \cite{YAGAN2013493}. This improvement emanates from bounding the respective binomial coefficient with a tighter bound, a simpler Stirling‐style bound used in previous works. Also note that the factor $\frac{1}{1-\alpha}$ for large values of $\alpha$ corresponds to large attack sizes, and makes a very loose estimate for the required $K$. Moreover, for $\gamma_n=o(n)$, we require a $K_n = \Omega (\gamma_n)$; more details for other regimes are presented in Section~\ref{sec:main-results-deletions-connectivity}.

\subsubsection{Giant component under node deletions}
Next, we derive conditions for \(\mathbb{H}(n;K_n,\gamma_n)\) to admit a {\em giant component}, i.e., a connected subgraph with \(\Omega(n)\) nodes, whp. This is also done for different scalings of \(\gamma_n\) and  {\em upper} bounds are provided for the number of nodes {\em outside} the giant component.  For the finite node regime, we present an empirical investigation demonstrating the usefulness of the results; see Sections~\ref{sec:main-results-deletions-connectivity} and \ref{sec:main-results-deletions-gc}.  While we follow the framework introduced in \cite{IT23sood}, we note that the case of inhomogeneous random K-out graphs \cite{IT23sood, eletreby2019connectivity}, does not readily yield results for our setting of homogeneous random K-out graphs because in the former there is a strict probability with which nodes make one selection which is strictly tied to the analysis presented in \cite{IT23sood}. Together, our results for 1-connectivity and giant component size directly presented an answer to design Q1. for different sizes of subsets of nodes targeted by the adversary. 

\subsection{Organization}
In Section~\ref{sec:setup} we give the formal setup including the notation,network models with and without deletions, and different notions of connectivity. In Section~\ref{sec:main-results-tight-bounds} we present our main results on tight bounds for probability of $1$-connectivity and in Section~\ref{sec:main-results-r-robustness}, we characterize $r$-robustness. We consider the case of random node deletions in Section~\ref{sec:main-results-deletions-gc} and \ref{sec:main-results-deletions-connectivity}, we present a suite of results on connectivity and giant component sizes under different sizes of subsets of node removals. We present partial proofs in Section~\ref{sec:1-con-partial} and \ref{sec:r-robustness-proof} and provide additional details in the Supplementary Appendix. Conclusions and future directions are presented in Section~\ref{sec:conc}.

\section{Problem Setup}
\label{sec:setup}
\subsection{Notation} We adopt the following notational convention. While comparing asymptotic behavior of a pair of sequences $\{a_n\},\{b_n\}$, we use $a_n = \oo(b_n)$, $a_n=\omega(b_n)$,  $a_n = \OO(b_n)$, $a_n=\Theta(b_n)$, {\color{black}and $a_n = \Omega(b_n)$} with their meaning in the standard Landau notation. All limits are understood with the number of nodes $n$ going to infinity. All random variables are defined on the same probability triple $(\Omega, {\mathcal{F}}, \mathbb{P})$. Probabilistic statements are made with respect to this probability measure $\mathbb{P}$, and we denote the corresponding expectation operator by $\mathbb{E}$. For an event $A$, its complement is denoted by $A\comp$. We let $\ii[\mathcal{E}]$ denote the indicator random variable which takes the value 1 if event $\mathcal{E}$ occurs and 0 otherwise. The cardinality of a discrete set $A$ is denoted by $|A|$ and the set of all positive integers by $\N_0$. We say that an event occurs with high probability (whp) if it holds with \emph{probability tending to one} as \(n \to \infty\).

\subsection{Random K-out graph construction}
\label{sec:simple-graph-construction}
 Consider a network consisting of $n$ nodes. Let $\nodes:=\{v_1,v_2,\dots,v_n\}$ denote the set of nodes. For $i \in \{1,2,\dots n\}$, let $\nodes_{-i}:=\{v_1,v_2, \dots, v_n\}\setminus v_i$. In its simplest form, the random K-out graph is constructed on the vertex set $\{v_1, \ldots, v_n\}$  as follows.
For each $v_i \in \nodes$, let $\Gamma_{n,i} \subseteq \nodes_{-i}$ denote the subset of nodes selected by node $v_i$ uniformly at random from the set $\nodes_{-i}$. Since by construction all nodes select $K$ other nodes, for any subset $A \subseteq {\cal N}_{-i}$, we have
\begin{equation}
\pr{[\Gamma_{n,i}  = A ]} = \left \{
\begin{array}{cl}
{{n-1}\choose{K}}^{-1} & \mbox{if $|A|=K$} \\
0             & \mbox{otherwise.}
\end{array}
\right .
\label{eq:main_eqn_for_gamma}
\end{equation}
Thus, the selection of $\Gamma_{n,i}$ is done {\em uniformly}
amongst all subsets of ${\cal N}_{-i}$ which are of
size exactly $K$. Let $v_i \sim v_j$ indicate the presence of an edge between nodes $v_i$ and $v_j$. With $n=2,3, \ldots $ and
positive integer $K < n$, we have
\begin{align}
v_i \sim v_j ~ \mbox{if} ~ v_j \in \Gamma_{n,i} ~\vee~ v_i \in \Gamma_{n,j}. 
\label{eq:Adjacency}
\end{align}
The adjacency condition (\ref{eq:Adjacency}) gives a precise construction of edges on the node set $\{v_1, \ldots, v_n\}$.
Let $\mathbb{H}(n;K)$ denote the undirected random graph on the
vertex set $\{ v_1, \ldots , v_n \}$ induced by the adjacency notion
(\ref{eq:Adjacency}).
In the literature on random graphs,  $\mathbb{H}(n;K)$
is often referred to as a random $K$-out graph \cite{Bollobas,FennerFrieze1982,philips1990diameter,Yagan2013Pairwise,yavuz2017k,yagan2013scalability}. In our subsequent results, if we allow $K$ to scale with $n$ to achieve the desired property, we indicate that explicitly through the notation $K_n$.


\subsection{Notions of connectivity}
Throughout let $\mathcal{N}$ and $\mathcal{E}$ respectively denote the node and edge set of an undirected graph $\mathbb{G}$, and let $n= |\mathcal{N}|$ be the number of nodes. Following, \cite{bondy1976graph, pirani2023graph}, we define

\begin{definition}[Connected Components]
A {connected component} of \(\mathbb{G}\equiv(\mathcal{N},\mathcal{E})\) is a maximal subset \(S \subseteq \mathcal{N}\) such that for every pair \(u,v \in S\), there exists a path in \(\mathbb{G}\) from \(u\) to \(v\).
\label{def:connected-component}
\end{definition}

\begin{definition}[Giant Component]\label{robustness-def:giantcomp}
For a graph \(\mathbb{G}\equiv(\mathcal{N},\mathcal{E})\) with $|\mathcal{N}|=n$ nodes, a {giant} component exists if the {largest} connected component, denoted by $C_{\rm max}$, has size $\Omega(n)$. In that case, the largest connected component is referred to as a giant component of the graph.
\end{definition}

\begin{definition}[Connectivity]
A graph $\mathbb{G}\equiv(\mathcal{N},\mathcal{E})$ is {connected} or $1$-connected if for every pair of distinct nodes $u,v \in \mathcal{N}$ there exists a path in $\mathbb{G}$ joining $u$ and $v$. Equivalently, the giant component of $\mathbb{G}$ exists uniquely and has size precisely $|C_{\rm max}|=n$. \footnote{Note that we focus on combinatorial notions of connectivity motivated by the applications considered.  Algebraic connectivity, given by the Laplacian’s second‑smallest eigenvalue, is discussed in the Appendix, where we empirically demonstrate the usefulness of random \(K\)-out graphs in balancing sparsity and algebraic connectivity.}

\end{definition}

\begin{definition}[$k$-(node)-connectivity]\label{robustness-def:r_connectivity}
A graph $\mathbb{G}\equiv(\mathcal{N},\mathcal{E})$ is $k$-connected for a fixed and finite $k$, if it remains connected after the removal of any set of $k-1$ (or fewer) nodes.
\end{definition}

\begin{definition}[$r$-reachable Set]\label{robustness-def:r-reachable}
Given a graph $\mathbb{G}\equiv(\mathcal{N},\mathcal{E})$ and a subset of nodes $S \subseteq \nodes$, we say $S$ is $r$-reachable if there exists an $i \in S$ such that 
\[
  \bigl|\mathcal{V}_i \setminus S\bigr| \;\ge\; r,
\]
where $\mathcal{V}_i$ denotes the neighbor set of node $i$. In other words, $S$ contains a node with at least $r$ neighbors outside of $S$.
\end{definition}

\begin{definition}[$r$-robustness]\label{robustness-def:r-robust}
A graph \(\mathbb{G}=(\mathcal{N},\mathcal{E})\) is called \emph{\(r\)-robust} if for every pair of nonempty, disjoint subsets of nodes \(S_1,S_2 \subseteq \mathcal{N}\), at least one of \(S_1\) or \(S_2\) is \(r\)-reachable.
\end{definition}

\subsection{Random K-out graph with node deletions}
Motivated by settings in which a subset of nodes in the network is adversarial, we study the connectivity and giant component size by removing compromised nodes from the initial network. ~Given 
\[
  \nodes = \{v_1,\dots,v_n\},
\]
we first construct the random $K_n$‑out graph $\hhn$ as outlined in Section~\ref{sec:simple-graph-construction}.  
Then we simulate an attack that can be abstracted as a removal of $\gamma_n$ nodes: choose 
\[
 \mathcal{N}_{D} \subseteq \nodes,\quad | \mathcal{N}_{D} | = \gamma_n,
\]
uniformly at random.  Let $\hhdn$ be the subgraph of $\hhn$ induced by $\nodes\setminus  \mathcal{N}_{D} $.  Two nodes $u_1,u_2\in \nodes\setminus  \mathcal{N}_{D} $ are adjacent in $\hhdn$ if and only if they were adjacent in $\hhn$. Thus, $\hhdn$ represents the sub-network of reliable nodes that remains after the removal of adversarial nodes. Note that the property of $k$-connectivity only investigates the case when $k$ is finite and it can be thought of as the worst-case finite-sized attack by an adversary targeting a subset of nodes. Note that modeling adversaries as an unbounded set of colluding nodes spread over the network is a canonical benchmark for privacy guarantees across many domains, including privacy‑preserving distributed computations \cite{2022sabater-gopa} and anonymity‑preserving cryptocurrency transactions \cite{FantiDandelion2018}. Therefore, modeling the cardinality of the subset of deleted nodes as scaling with $\gamma_n$ provides a flexible framework to capture the size of the network attacked by an adversary.

Another closely related notion is that of {bond or site percolation}, which primarily characterizes {giant component sizes} as a function of the probability with which each node (or edge) is removed from the network \cite{ne:newman2018networks}. However, there remains a limited understanding of such analyses for 
random K-out graphs in different attack sizes \cite{frieze2016introduction}. As we will see in Section~\ref{sec:main-results-deletions-gc}, the above model goes beyond the mere existence of a giant component. It helps the network designer choose a value of $K_n$ needed to ensure that, even when a random subset of $\gamma_n$ nodes is deleted, the largest connected component  has size at least $n - \lambda_n$, for different values of $\lambda_n$.



\section{Results on $1$-connectivity: Finite $n$ and asymptotically order-wise optimal bounds}
In this section, we present our main results-- upper and lower bounds for the probability of connectivity of $\hh$, and compare them with existing results. Throughout, we write
\[
P(n;K)
:=
\pr[{~ \mathbb{H}(n;K) ~\mbox{is connected}~}] .
\]
\label{sec:main-results-tight-bounds}
\vspace{-20pt}
\subsection{Main Theorems}
We provide our first technical result-- an upper bound for the probability of connectivity $P(n;K)$.
\begin{theorem}[Upper Bound on Connectivity]
\label{thm:UpperBoundForConnectivity}
For any fixed positive integer $K\geq 2$,  we have
\begin{align}
  P(n;K)
\leq  1- \frac{({K!})^K e^{-K(K+1)}}{{K+1}}\cdot \frac{1}{n^{K^2-1}}  (1+\oo(1)) .
\label{eqn:UpperBoundForConnectivity}
\end{align}{}
\end{theorem}{}
We present the above asymptotic version of the upper bound in Theorem~\ref{thm:UpperBoundForConnectivity} to make it easier to interpret; see the Appendix for the more detailed bound with an explicit function of $n$ replacing the $(1+o(1))$ term in (\ref{eqn:UpperBoundForConnectivity}). The dependence of the upper bound on the order of $n$ in terms of the parameter $K$ is succinctly captured in the remark below.
\begin{remark}
For a fixed positive integer $K\geq 2$,
\begin{align}
    P(n;K)=1 - \Omega\left(\frac{1}{n^{K^2-1}}\right).
    \label{eq:koutbounds:upperbound-asymptotic}
\end{align}{}
\end{remark}{}
Given a fixed value of the parameter $K$ ($K \geq 2$), we derive the upper bound on the probability of connectivity by computing a lower bound on the probability of existence of isolated components comprising $K+1$ nodes using the \emph{Bonferroni inequality} \cite{bonferroni}. Note that in random K-out graphs, each node makes $K$ selections and therefore there exist no isolated components of size strictly less than $K$. 
We outline the proof for the case of $K=2$ in Section~\ref{sec:upperbound-bounds} and present the full proof ($K\geq2$) in the Appendix. To the best of our knowledge, our work is the \emph{first} to derive an upper bound on $P(n;K)$. 

In our second main result, we derive an order-wise {\em matching} lower bound and show that $P(n;K)=1-O\left(\frac{1}{n^{K^2-1}}\right)$, which indicates that further improvement on the order of $n$ is not possible for the bounds on connectivity.

\begin{theorem}[Lower Bound on Connectivity]
{\sl For any fixed positive integer $K\geq 2$, for all $ n \geq 4(K+2)$, we have
\begin{align}
P(n;K)
\geq 1 - c(n;K) Q(n;K)
\label{eq:koutbounds:LowerBoundForConnectivity}
\end{align}
where,
}
\label{thm:LowerBoundForConnectivity}
\begin{align}
c(n;K)&= \frac{e^{- (K^2-1)(1- \frac{K+1}{n})}}{\sqrt{2 \pi (K+1)}}\sqrt{\frac{n}{(n-K-1)}},
\label{eq:koutbounds:c(K)}\\
 Q(n;K) 
&=
\left ( \frac{K+1}{n} \right )^{K^2-1}
+ \frac{n}{2}
\left ( \frac{K+2}{n} \right )^{(K+2)(K-1)}.
\label{eq:koutbounds:q(K)}
\end{align}
\end{theorem}
We discuss one of the key steps which distinguishes our proof from existing bounds \cite{Yagan2013Pairwise,FennerFrieze1982} and provide a brief outline here; full details are presented in the Appendix. In contrast to the standard bound ${n \choose r} \leq \left(\frac{ne}{r} \right)^r$ used in the prior works, we upper bound ${n \choose r}$ using a \emph{variant \cite{stirlingremark} of Stirling formula}. For all $x=1,2,\dots$, we have
\begin{align}
    \sqrt{2 \pi} x^{x+0.5} e^{-x} e^{\frac{1}{12x+1}} < x! <    \sqrt{2 \pi} x^{x+0.5} e^{-x} e^{\frac{1}{12x}}, \label{eq:koutbounds:stirling}
\end{align}{}
which gives
\begin{align}
{n \choose{ r}}& \leq \frac{1}{\sqrt{2 \pi}} \left(\frac{n}{n-r}\right)^{n-r} \left( \frac{n}{r}\right)^r \frac{\sqrt{n}}{\sqrt{n-r}\sqrt{r}} \nonumber\\
&\qquad \cdot \exp\left\{ \frac{1}{12n}-\frac{1}{12(n-r)+1}-\frac{1}{12r+1}\right\}\nonumber\\
& \leq \frac{1}{\sqrt{2 \pi}} \left(\frac{n}{n-r}\right)^{n-r} \left( \frac{n}{r}\right)^r \frac{\sqrt{n}}{\sqrt{n-r}\sqrt{r}},
\label{eq:koutbounds:stirlingtakeaway}
\end{align}{}
since $$\frac{1}{12n}-\frac{1}{12(n-r)+1}-\frac{1}{12r+1} < 0.$$
The proof for the lower bound for connectivity follows the framework in \cite{FennerFrieze1982,Yagan2013Pairwise}, is based on the observation that if $\mathbb{H} (n;K)$ is {\em not} connected, then there exists a non-empty subset $S$ of nodes that is isolated. Further, since each node is paired with at least $K$ neighbors, $|S|\geq K+1$. 
Using the upper bound for ${n \choose r}$ as presented in (\ref{eq:koutbounds:stirlingtakeaway}) eventually leads to the factor 
$e^{- (K^2-1)(1- \frac{K+1}{n})}$ improvement in the lower bound on probability of connectivity in Theorem~\ref{thm:LowerBoundForConnectivity}.  We present the  sequence of steps leading to the final bound in Theorem~\ref{thm:LowerBoundForConnectivity} in the appendix. 
\begin{remark}
For any fixed positive integer $K\geq 2$ we have
\begin{align}
    P(n;K)
= 1 - O\left(\frac{1}{n^{K^2-1}}\right).\label{eq:koutbounds:lowerbound-asymptotic}
\end{align}{}
\end{remark}{}
This shows that our lower bound (\ref{eq:koutbounds:LowerBoundForConnectivity})  for connectivity {\em matches} our upper bound (\ref{eqn:UpperBoundForConnectivity}), and is therefore order-wise optimal. Combining (\ref{eq:koutbounds:upperbound-asymptotic}) and (\ref{eq:koutbounds:lowerbound-asymptotic}), we obtain the following result.
\begin{corollary}
{\sl For any positive integer $K\geq 2$, for all $ n \geq 4(K+2)$, we have
\begin{align}
P(n;K)
=  1 - \Theta\left(\frac{1}{n^{K^2-1}}\right)\label{eq:koutbounds:corollary-asymptotic}.
\end{align}
}\label{corr:first}
\end{corollary}{}
The above expression 
indicates how rapidly $P(n;K)$ converges to one as $n$ grows large. As expected, a larger value of $K$ results in a faster convergence of $P(n;K)$ to 1. 

 \subsection{Discussion}
We present a summary of the related lower bounds \cite{Yagan2013Pairwise, FennerFrieze1982} on the probability of connectivity. \\
\emph{Earlier results by Ya\u{g}an and Makowski \cite{Yagan2013Pairwise}}
: It was established \cite[Theorem 1]{Yagan2013Pairwise} that for $K \geq 2$,
{
\begin{equation}
P(n;K)
\geq 1 - a(K) Q(n;K)
\label{eq:koutbounds:oyam_lb}
\end{equation}
holds for all $n \geq n(K)$ with $n(K) = 4(K+2)$, where
}
\begin{equation}
a(K)= e^{-\frac{1}{2} (K+1)(K-2)}.
\label{eq:koutbounds:a(K)}
\end{equation}

\emph{Earlier results by Fenner and Frieze \cite{FennerFrieze1982}}: A lower bound for probability of connectivity can be inferred from the proof of \cite[Theorem 2.1, p. 348]{FennerFrieze1982}.
Upon inspecting Eqn. 2.2 in
\cite[p. 349]{FennerFrieze1982} with $p=0$; it can be inferred that
\begin{equation}
P(n;K) \geq 1 - b(n;K) Q(n;K)
\label{eq:koutbounds:BoundByFF}
\end{equation}
holds for all $n$ and $K$ such that $K < n$, where
\begin{align}
    b(n;K) = \frac{12 n}{12 n -1}\sqrt{ \frac{1}{ 2\pi (K+1) } }
\sqrt{\frac{n}{n-K-1}}.
\label{eq:koutbounds:b(K)}
\end{align}
Observe from
(\ref{eq:koutbounds:LowerBoundForConnectivity}), (\ref{eq:koutbounds:oyam_lb}) and (\ref{eq:koutbounds:BoundByFF}) that the smaller the value of $c(n;K), a(K)$ and $b(n;K)$, the better is the corresponding lower bound. As discussed in \cite{Yagan2013Pairwise}, the bound (\ref{eq:koutbounds:BoundByFF}) by Fenner and Frieze is tighter than (\ref{eq:koutbounds:oyam_lb}) when $K=2$, while (\ref{eq:koutbounds:oyam_lb}) is tighter than (\ref{eq:koutbounds:BoundByFF}) for all $K \geq 3$.
We illustrate the performance of these bounds in the succeeding discussion.

\begin{figure}[htbp]
\centering
\includegraphics[width = 0.55 \textwidth]{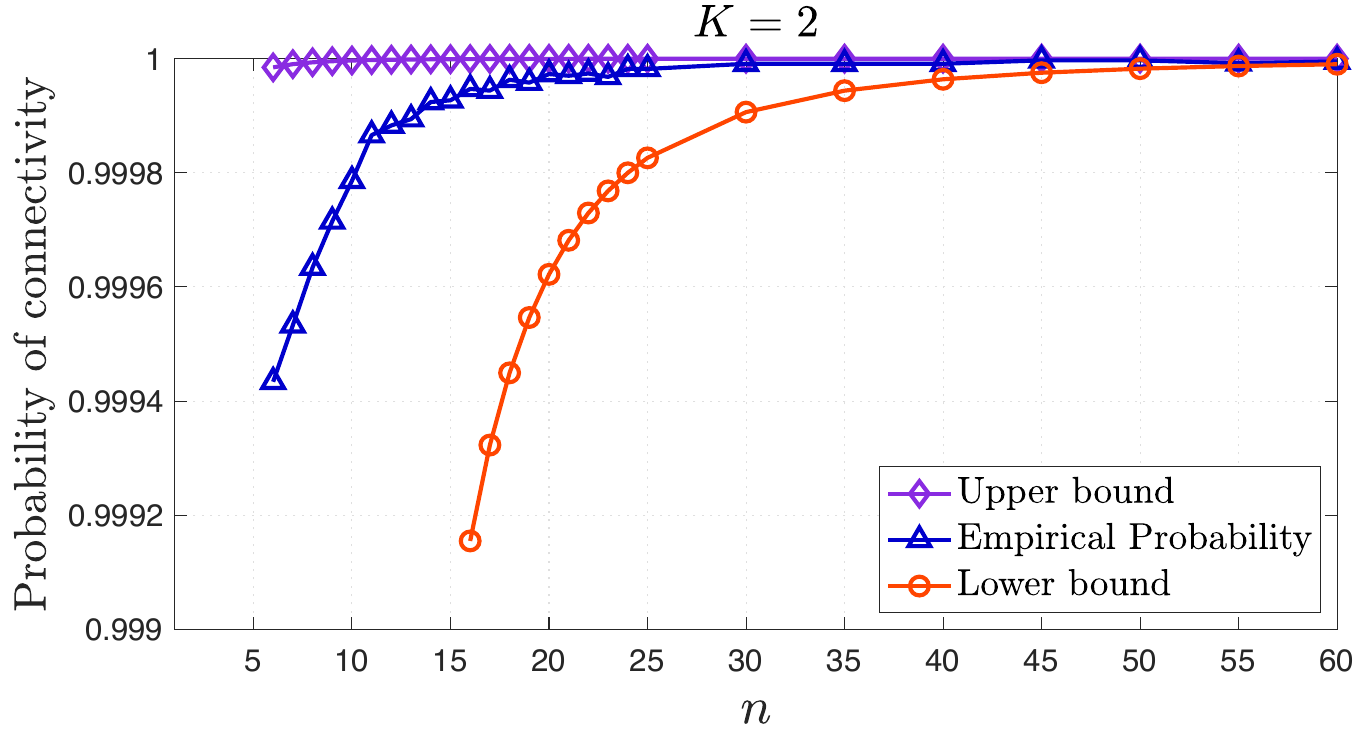}
\caption{A zoomed-in view of our  results and empirical probability of connectivity (computed by averaging $10^6$ independent experiments for each data point) for $K=2$ as a function of $n$ for $n\geq 16$. The lower bound corresponds to Theorem \ref{thm:LowerBoundForConnectivity} and the upper bound corresponds to Theorem \ref{thm:UpperBoundForConnectivity}.
}\label{fig:bounds2}
\end{figure}

Through simulations, first we study how our upper and lower bounds compare with the empirically observed probability of connectivity. We consider a random K-out graph with parameter $K=2$ and compute the empirical probability of connectivity as we vary the number of nodes $n$. For each parameter pair $(n,K)$, we generate $10^6$ independent realizations of $\hh$. To obtain the empirical probability of connectivity, we divide the number of instances for which the generated graph is connected by the total number ($10^6$) of instances generated; see Figure~\ref{fig:bounds2}. Next, we compare the lower bound for $P(n;K)$ presented in Theorem \ref{thm:LowerBoundForConnectivity} 
with the corresponding bounds in \cite{Yagan2013Pairwise,FennerFrieze1982}. Recall that $K=2$ is the critical threshold for connectivity of $\hh$ in the limit of large network size; thus, we focus on the case $K=2$ for the discussion below.
\begin{align}
 & \textrm{YM \cite{Yagan2013Pairwise}}:  P(n;2)  \geq 1-  \frac{155}{n^3} \label{eq:koutbounds:YM2}\\ 
& \textrm{FF \cite{FennerFrieze1982}}:    P(n;2) \geq 1-  \frac{155}{n^3}\cdot 
\frac{12 n/(12 n -1)}{\sqrt{ 6 \pi }}
\sqrt{\frac{n}{n-3}}
\label{eq:koutbounds:ff2}
\\
& \textrm{This work}  : P(n;2) \geq 1-  \frac{155}{n^3}\cdot  \frac{e^{-(3-\frac{9}{n})}}{\sqrt{6 \pi }}
    \sqrt{\frac{n}{n-3}}
    \label{eq:koutbounds:ourlb2}
\end{align}{}

\begin{figure}[htbp]
\centering
\includegraphics[width = 0.55 \textwidth]{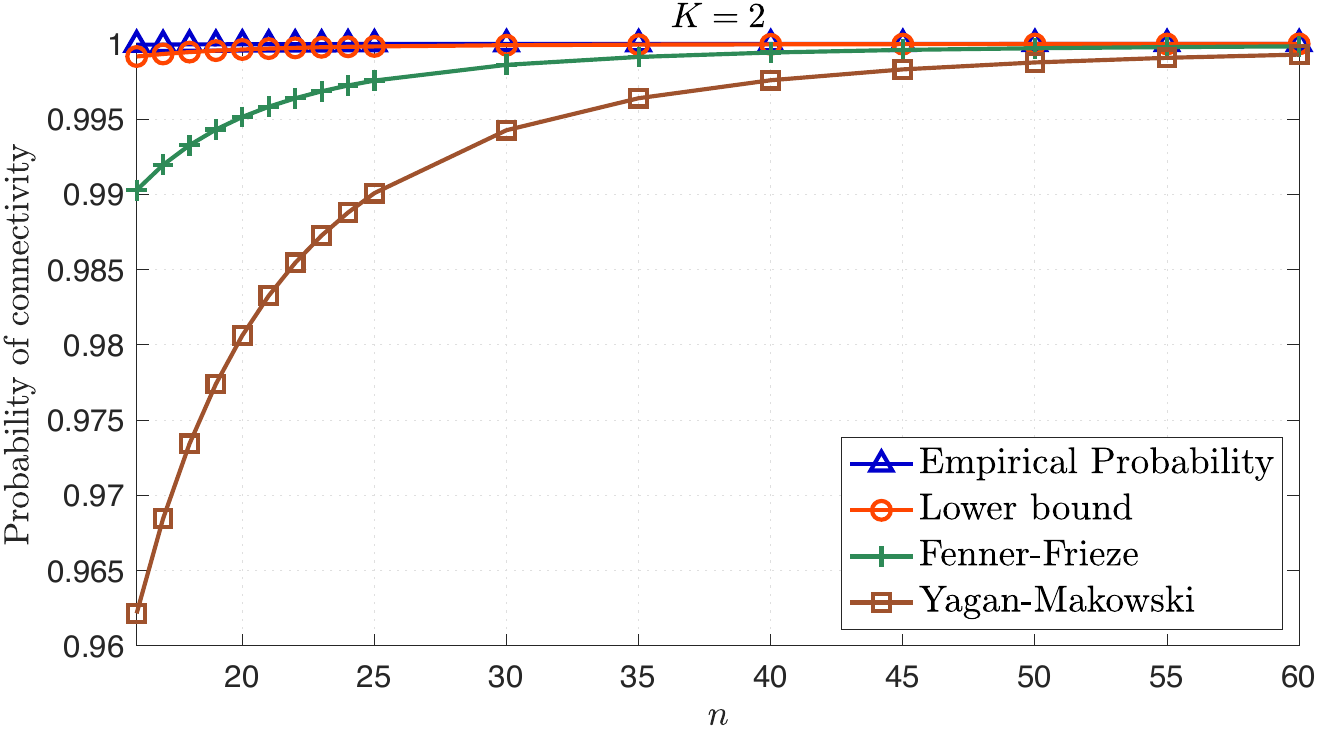}
\caption{ Lower bounds and empirical probability
         of connectivity (computed by averaging $10^6$ independent experiments for each data point) for $K=2$ as a function of $n$ for $n\geq 16$. Our lower bound  given in Theorem \ref{thm:LowerBoundForConnectivity} 
         significantly improves the existing lower bounds by Ya\u{g}an and Makowski \cite{Yagan2013Pairwise}, and Fenner and Frieze \cite{FennerFrieze1982}. }
         \vspace{-1mm}
         \label{fig:bounds3}
\end{figure}


With $K=2$, we plot the lower bounds (\ref{eq:koutbounds:LowerBoundForConnectivity}), (\ref{eq:koutbounds:oyam_lb}) and (\ref{eq:koutbounds:BoundByFF}) for comparison in Figure~\ref{fig:bounds3}. Corresponding to the lower bounds (\ref{eq:koutbounds:YM2}), (\ref{eq:koutbounds:ff2}) and (\ref{eq:koutbounds:ourlb2}) for $K=2$, we compare the mean number of realizations of $\hh$ that can be generated until one disconnected realization is observed in Table~\ref{tab:tab1}.  Our results complement the existing asymptotic zero-one laws for random K-out graphs \cite{FennerFrieze1982,Yagan2013Pairwise} in providing a design guide for network designers working with a finite set of nodes.

\begin{table}[!h]
\begin{center}
     \label{tab:tab1}
\begin{tabular}{|c|c|c|c|}
   \hline
   & \multicolumn{3}{c|}{Mean number of disconnected realizations}
   \\
   \hline
$n$& Theorem~\ref{thm:LowerBoundForConnectivity}& YM \cite{Yagan2013Pairwise} & FF \cite{FennerFrieze1982}\\
   \hline
     16&1 in {1183} &1 in {26} & 1 in {102} \\
     \hline
     20&1 in {2645} &  1 in {51} & 1 in { 205}\\
     \hline
          25&1 in { 5753} & 1 in {100}&1 in {409} \\
           \hline
     35&1 in { 17834 } & 1 in { 276} & 1 in { 1145 }\\
     \hline
\end{tabular}{}
\end{center}
     \caption{
          Comparison of the lower bound (\ref{eq:koutbounds:LowerBoundForConnectivity}) with existing lower bounds (\ref{eq:koutbounds:oyam_lb}) and (\ref{eq:koutbounds:BoundByFF}) from \cite{Yagan2013Pairwise} and \cite{FennerFrieze1982}, respectively for $K=2$. The entries in the table corresponds to the mean number of realizations of $\hh$ generated until one disconnected realization is observed, as predicted by the respective lower bounds.}
\vspace{-4mm}
     \end{table}
\renewcommand{\arraystretch}{1.0}

\section{Results on $r$-robustness}
\label{sec:main-results-r-robustness}

\subsection{Main Theorem}
Below we establish are main result on how to choose the parameter $K$ to ensure $\hh$ is $r$-robust for a given value of $r$.
{
\begin{theorem}[Ensuring $r$-robustness]
{\sl 
Define 
$$r^{\star}(K) = \max_{r =1, 2, 3\ldots} r \text{~~s.t.~~}\{  \lim_{n \to \infty} P(\hh \text{ is } r \text{-robust}) = \hspace{-0.5mm} 1 \}$$
Then, we have 
$$r^{\star}(K) \geq \lfloor K/2 \rfloor.$$
}  \label{robustness-theorem:thmr_1}
\end{theorem}
\vspace{-2mm}
In other words, we find that {\em with high probability}, a random K-out graph is $r$-robust when $K \geq 2r, r \geq 2, r \in \mathbb{Z}^+$, and $n \to \infty$. This threshold is much smaller than the previously established threshold \cite{can_cdc2021} of $ K > \frac{2r\left(\log(r) + \log(\log(r) + 1\right)}{\log(2) + 1/2 - \log\left(1+ \frac{\log(2)+1/2}{2\log(r)+5/2+\log(2)}\right) } $ which scales with $r\log r$. Hence, Theorem \ref{robustness-theorem:thmr_1} constitutes a sharper one-law for $r$-robustness. 
This tighter result  was made possible through several novel steps introduced here. While the proofs in prior work \cite{sundaram_robustness,can_cdc2021} also rely on finding upper bounds on the probability of having at least one subset that is not $r$-reachable, they tend to utilize standard upper bounds for the binomial coefficients ${n \choose k}\leq \left(\frac{en}{k}\right)^k$ and a union bound to establish them. Instead, our proof uses extensively  the Beta function $B(a,b)$ and its properties to obtain tighter upper bounds on such probabilities, which then enables us to establish a much sharper one-law for $r$-robustness of random K-out graphs. 

}

\subsection{Discussion}
It is of interest to compare the threshold of $r$-robustness with the closely related property of $r$-connectivity (Definition~\ref{robustness-def:r_connectivity}) in random K-out graphs. 
For Erd\H{o}s-R\'enyi graphs, the threshold for $r$-connectivity and $r$-robustness have been shown  \cite{sundaram_robustness} to coincide with each other. For random K-out graphs, we know from \cite{FennerFrieze1982} that $\hhn$ is $r$-connected {\em whp} whenever $K_n \geq r$, and it is {\em not} $r$-connected {\em whp} if $K_n < r$.  This leaves a factor of 2 difference between the condition $K_n \geq 2r$ we established for $r$-robustness here and the threshold of $r$-connectivity. Put differently, we know from \cite{FennerFrieze1982} and Theorem \ref{robustness-theorem:thmr_1} that for any $r=2, 3, \ldots$
\begin{align}
& \lim_{n \to \infty} \pr \left[ \hh \text{ is $r$-robust}\right] =  \begin{cases}
    1, & \mathrm{if}\quad K_n \geq 2r\\
    0, & \mathrm{if}\quad K_n< r.
  \end{cases}
\label{eq:osy_new}
\end{align}
For $r=1$, 1-robustness is equivalent to 1-connectivity \cite{sundaram_robustness} and therefore $\lim_{n \to \infty} \pr \left[ \hhn \text{ is $1$-robust}\right]=1$ if only if $K_n \geq 2r=2$. 
Since the currently established conditions for the zero-law and one-law of $r$-robustness are not the same for random K-out graphs (unlike ER graphs where the two thresholds coincide),  there is a question as to whether our threshold of $2r$ is the tightest possible for $r$-robustness. To the best of our knowledge, this is currently an open problem.

  Next, we contextualize our results by comparing them to the Erd\H{o}s-R\'enyi graph $G(n,p)$. As shown in~\cite{pirani2023graph}, $G(n,p)$ is $r$-robust \emph{whp} if
$
p_n = \frac{\log(n) + (r-1) \log(\log(n)) + \omega(1)}{n},
$
which corresponds to an average degree $\langle k \rangle \sim \log(n) + (r-1) \log(\log(n))$. Since the random K-out graph is $r$-robust \emph{whp} when $K \geq 2r$, which means an average node degree of $<k> = 4r$, we can conclude that the average node degree required for a random K-out graph to be $r$-robust is significantly smaller than the average node degree required for an Erd\H{o}s-R\'enyi graph, demonstrating that the random K-out graph is more edge efficient for $r$-robustness. 

Determining the $r$-robustness of a graph involves checking all subsets of a graph, and it was shown in \cite{sundaram_robustness} that this is a co-NP-complete problem. Hence, for our empirical study, we focus on the small $n$ regime. In the simulations, we generate instantiations of the random graph  $\hh$,
 with $n=20$, and $K$ in the range $[1, 12]$. For each $K$ value, we generate 500 independent realizations of $\hh$. In each experiment, i.e., for each realized instance of $\hh$, we record the empirical $r^*(K)$ as  $ \max \{r =1, 2, : \text{generated graph } \hh \text{ is } r \text{-robust}) \}$. The minimum, and maximum empirical $r^*(K)$ observed in these 500 experiments for each $K$ value, $r_{min}(K)$ and $r_{max}(K)$, are plotted in Fig.~\ref{robustness-fig:fig_sim1} along with the corresponding theoretical plot obtained from Theorem \ref{robustness-theorem:thmr_1}.  Also, the theoretical upper bound for $r$-robustness is plotted. This plot is obtained from the upper bound of $r \leq K$ for $r$-connectivity \cite{ FennerFrieze1982}. Since $r$-robustness is a stronger property than $r$-connectivity, it can also be used as an upper bound on $r$-robustness. Hence, combining this with the lower bound of Theorem \ref{robustness-theorem:thmr_1}, we can write  $ \lfloor K/2 \rfloor \leq r^{\star}(K) \leq  K$.  As can be seen from Fig.~\ref{robustness-fig:fig_sim1}, both empirical plots, $r_{min}(K)$ and $r_{max}(K)$, are between the plots of theoretical lower and upper bounds for all tested $K$ values, validating both the lower bound asserted by Theorem \ref{robustness-theorem:thmr_1} and the upper bound obtained from \cite{ FennerFrieze1982} when the number of nodes is small. Moreover, in Figure \ref{robustness-fig:fig_sim1}, the empirical plots, $r_{min}(K)$ always lie below the theoretical upper bound for $r$-robustness. And thus, the simulation suggests that having a threshold of $r^*(K) \geq K$ may not be feasible; however, it remains to be examined if this is an artifact of the small node regime. Still, determining whether a threshold tighter than $\lfloor K/2 \rfloor \leq r^*(K) \leq K$ exists for $r$-robustness is currently an open problem and a direction for future work.

{\color{blue}
\begin{figure}[h]
\centering
\includegraphics[scale=0.62]{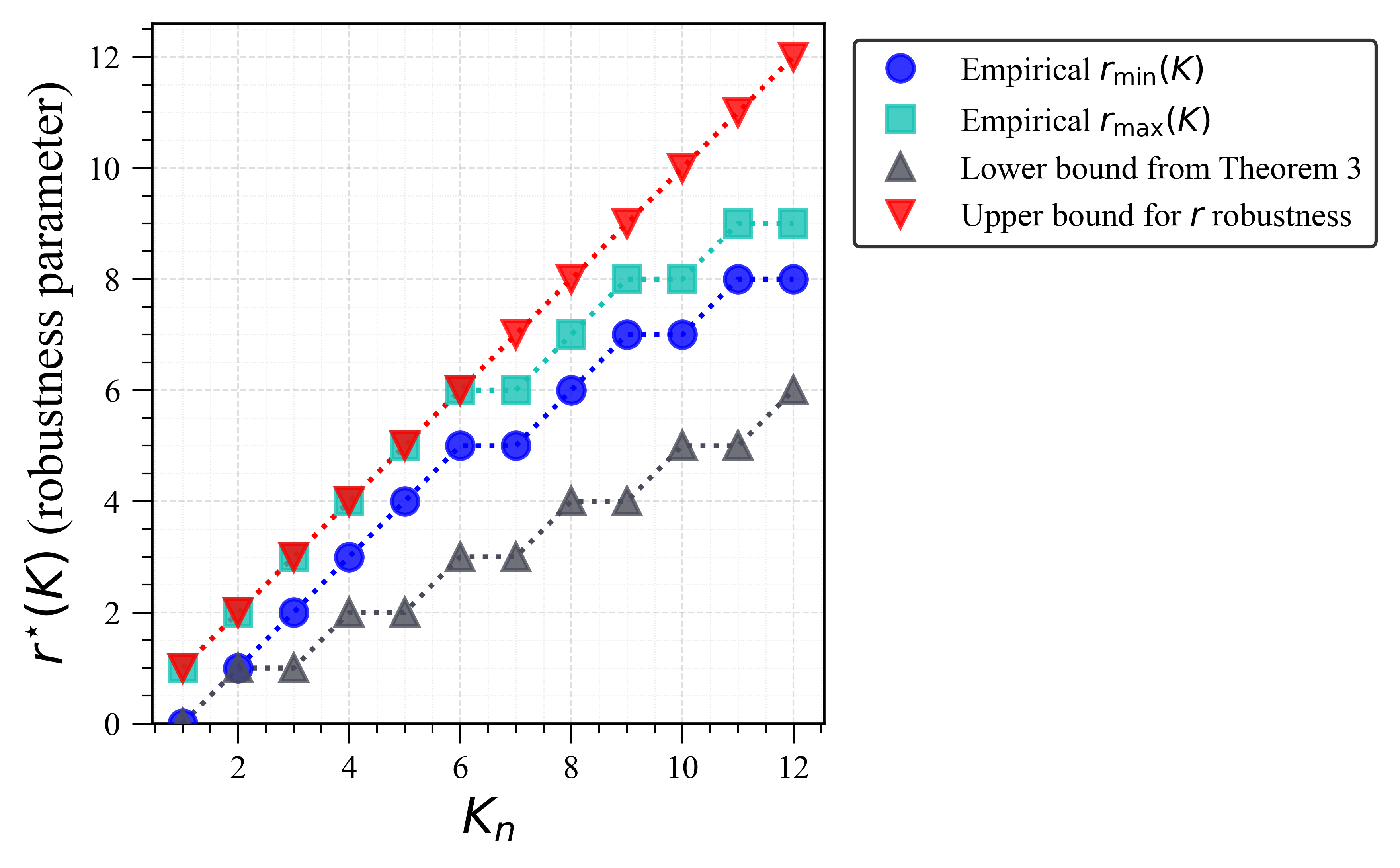}\label{robustness-fig:sim1} \vspace{-2mm}
\caption{ Empirically observed minimum and maximum $r^*(K)$ value for each $K$ value across 500 experiments for $n=20$. We also plot  the theoretical minimum robustness value asserted by Theorem \ref{robustness-theorem:thmr_1} and the theoretical threshold for $r$-connectivity as an upper bound on $r^*(K)$.} 
\label{robustness-fig:fig_sim1}
\vspace{-3mm}
\end{figure}


}

\section{Results on $1$-connectivity under random node deletions}
\label{sec:main-results-deletions-connectivity}
Recall that the random K-out graph will be connected whp if $K \geq 2$ and disconnected whp for $K=1$. In this section, we analyze how this connectivity threshold changes when a subset of nodes is removed from the graph after construction, leading to the graph $\hhdn$. \\
Let $P(n,K_n,\gamma_n) = \pr \left[ ~\hhdn \text{ is connected}~\right]$.

\subsection{Main Theorems}
We first present our main results, which elucidate how to choose $K_n$ to ensure that $\hhdn$ remains connected under various node removal scenarios, as characterized by the size of $\gamma_n$. For the regime $\gamma_n = \alpha n$ with $\alpha \in (0,1)$, our results result lead to a zero-one law for 1-connectivity. For additional regimes of $\gamma_n = o(\sqrt{n})$ and $\gamma_n = o(n)$, we present sufficient conditions on the scaling of $K_n$ to guarantee the desired level of connectivity.


\begin{theorem}[Connectivity with $\gamma_n = \alpha n$ node deletions]
{\sl 
Let $\gamma_n = \alpha n$ with $\alpha$ in $(0,1)$, and consider a scaling $K_n:\mathbb{N}_0 \to \mathbb{N}_0$ such that with $c>0$ we
have 
\begin{align}
K_n \sim c \cdot r_1(\alpha, n), \ \  \textrm{where} \quad r_1(\alpha, n) = \frac{\log n}{1 - \alpha - \log \alpha}
\label{eq:threshold_1}
\end{align}
is the threshold function. Then, we have
\begin{align}
& \lim_{n \to \infty} P(n,K_n,\gamma_n) =  \begin{cases}
    1, & \mathrm{if}\quad c > 1\\
    0, & \mathrm{if}\quad 0< c < 1.
  \end{cases}
\label{eq:c_1}
\end{align}
}  \label{theorem:thmc_1}
\end{theorem}
The proof of the {\em one-law} in (\ref{eq:c_1}), i.e., that 
$\lim_{n \to \infty} P(n,K_n,\gamma_n)=1$
if $c>1$, is given in the Appendix. 

\begin{theorem}[Connectivity with $\gamma_n = \oo(n), \oo(\sqrt{n})$ node deletions]
{\sl  Consider a scaling $K_n:\mathbb{N}_0 \to \mathbb{N}_0$. \\
a) If $\gamma_n = o(\sqrt{n})$, then we have
\begin{align}
\lim_{n \to \infty} P(n,K_n,\gamma_n) = 
    1, \quad \mathrm{if}\quad K_n \geq 2 \ \ \forall n
\label{eq:c_2a}
\end{align}
b) If $\gamma_n = \Omega(\sqrt{n})$ and $\gamma_n = o(n)$, and if for some sequence $w_n$, it holds that
$$
K_n = r_2(\gamma_n) + \omega_n, \ \ \textrm{where} \quad 
r_2(\gamma_n) = \frac{\log (\gamma_n)}{\log 2 + 1/2}$$
is the threshold function, then we have
\begin{align}
\lim_{n \to \infty} P(n,K_n,\gamma_n) = 
    1, \quad \mathrm{if}\quad \lim_{n \to \infty}\omega_n = \infty
\label{eq:c_2b}
\end{align}
}  \label{theorem:thmc_2}
\end{theorem}

The {\em zero-law} of (\ref{eq:c_1}),
i.e., that 
$\lim_{n \to \infty} P(n,K_n,\gamma_n)=0$
if $c<1$,
was  established previously in \cite[Corollary 3.3]{YAGAN2013493}. There,
a one-law was also provided: 
under
(\ref{eq:threshold_1}), it was shown that
$\lim_{n \to \infty} P(n,K_n,\gamma_n)=1$
if $c>\frac{1}{1-\alpha}$, leaving a gap between the thresholds of the zero-law and the one-law. Also note that the factor $\frac{1}{1-\alpha}$ for large values of $\alpha$ corresponding to large attack sizes, and making a very loose estimate for the required $K$.Theorem \ref{theorem:thmc_1} presented here fills this gap by establishing a tighter one-law, and constitutes a {\em sharp} zero-one law; e.g., when $\alpha=0.5$, 
 the one-law in \cite{YAGAN2013493} is given with $c>2$, while we show that it suffices to 
have $c>1$.  The key technical innovation of our proof as compared to \cite{YAGAN2013493} is the use of tighter upper bounds on the probability of not being connected emanating from using the bound 
$
\binom{n}{m}
\le
\bigl(\tfrac{n}{m}\bigr)^{m}
\,
\bigl(\tfrac{n}{\,n-m\,}\bigr)^{\,n-m}
$ \cite{IT23sood}
instead of the simpler “Stirling‐style” bound
$
\binom{\lfloor\gamma n\rfloor}{r}
\le
\bigl(\tfrac{\lfloor\gamma n\rfloor\,e}{r}\bigr)^{r}
\quad r=1,\dots,\lfloor\gamma n\rfloor
$; see Appendix for a detailed proof.

With this, we close the gap between the zero law and the one law, and hence establish a sharp zero-one law for connectivity when $\gamma_n= \Omega(n)$ nodes are deleted from $\hhdn$. 

\subsection{Discussion}

 In Theorem \ref{theorem:thmc_2}, we establish that the graph $\hhdn$ with $\gamma_n = o(n)$ is connected whp when $K_n \sim \log(\gamma_n)$; and when $\gamma_n = o(\sqrt{n})$, $K_n \geq 2$ is sufficient for connectivity whp. The latter result is especially important, since $K_n \geq 2$ is the previously established threshold for connectivity \cite{FennerFrieze1982}, we improve this result by showing that the graph is still connected with $K_n \geq 2$ even after $o(\sqrt{n})$ nodes (selected randomly) are deleted.  To put these results in perspective, we compare them with an Erd\H{o}s-R\'enyi graph $G(n,p)$, which is connected whp if $p > \log n / n$. This translates to having an average node degree of $<k> \sim \log n$ \cite{erdHos1960evolution}. The $<k>$ required for the random K-out graph to be connected whp is much lower, with $<k> = O(1)$ when $o(\sqrt{n})$ nodes are removed, and $<k> \sim \log(\gamma_n)$ when $\gamma_n=\Omega(\sqrt{n})$ nodes are removed.  
 
\begin{figure}[!t]
\centering
\includegraphics[scale=0.50]{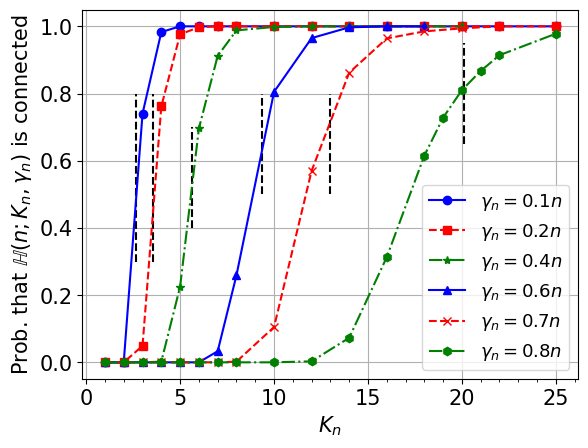}\label{fig:conn1}
\caption{ Empirical probability that $\hhdn$ is connected for a network of size $n = 5000$, with $K_n$ varying in $\{0,1,\dots,25\}$, averaged over 1000 independent realizations. Here, $\gamma_n = \alpha n $ denotes the size of the subset of nodes removed from the network, where $\alpha \in \{0.1,\dots, 0.8\}$. The vertical lines represent the theoretical thresholds given by Theorem~\ref{theorem:thmc_1}.
} 
\label{fig:figcondelete}
\end{figure}

 Further, to assess the usefulness of our results when the number $n$ of nodes is finite, we examine the probability of connectivity under node failure empirically. We set $n=5000$, $\gamma_n = \alpha n$, with $\alpha$ in $(0,1)$. We generate  instantiations of the random graph $\hhdn$ with $n=5000$, varying $K_n$  in the interval $[1, 25]$ and several $\alpha$ values in the interval $[0.1,0.8]$. Then, we record the empirical probability of connectivity from  1000 independent experiments for each $(K_n,\alpha)$ pair. The results of this experiment are shown in Fig.~\ref{fig:figcondelete}. In each curve, $P(n,K_n,\gamma_n)$ exhibits a threshold behaviour as $K_n$ increases, and the transition from $P(n,K_n,\gamma_n)=0$ to $P(n,K_n,\gamma_n)=1$ takes place around $K_n = \frac{\log n}{1 - \alpha - \log \alpha}$, aligning with the thresholds predicted by Theorem~\ref{theorem:thmc_1}.

\section{Results on largest connected component sizes under node deletions}
\label{sec:main-results-deletions-gc}


Let $C_{\rm max}(n,K_n,\gamma_n)$ denote the set of nodes in the {\em largest}  connected component  of  $\hhdn$. Define $$P_G(n,K_n,\gamma_n,\lambda_n) :=\mathbb{P}[|C_{\rm max}(n,K_n,\gamma_n)| > n - \gamma_n- \lambda_n].$$ Namely, $P_G(n,K_n,\gamma_n,\lambda_n)$ is the probability that less than $\lambda_n$ nodes are {\em outside} the largest  connected component of $\hhdn$.  We have

\begin{theorem}[$|C_{\rm max}|$ with $\gamma_n = \oo( n)$ node deletions]
{\sl 
Let $\gamma_n = o(n)$, $\lambda_n = \Omega(\sqrt{n})$ and $\lambda_n \leq \lfloor (n-\gamma_n)/3 \rfloor$. Consider a scaling $K_n:\mathbb{N}_0 \to \mathbb{N}_0$ and let
$$r_3(\gamma_n,\lambda_n) = 1 +  \frac{\log(1+\gamma_n / \lambda_n)}{\log 2 + 1/2}$$
be the threshold function. Then, we have
\begin{align*}
& \lim_{n \to \infty}  P_G(n,K_n,\gamma_n,\lambda_n) =  1, \quad \mathrm{if}\quad K_n > r_3(\gamma_n, \lambda_n),  \ \ \forall n.
\end{align*}

}  \label{theorem:thmg_3}
\end{theorem}

\begin{theorem}[$|C_{\rm max}|$ with $\gamma_n = \alpha n$ node deletions]
{\sl 
Let $\gamma_n = \alpha n$ with $\alpha$ in  $(0,1)$, and $\lambda_n \leq \lfloor \frac{(1-\alpha)n}{3} \rfloor$. Consider a scaling $K_n:\mathbb{N}_0 \to \mathbb{N}_0$ and let
$$r_4(\alpha, \lambda_n) = 1 +  \frac{\log(1 + \frac{n \alpha}{\lambda_n}) +\alpha + \log(1-\alpha)}{\frac{1-\alpha}{2} - \log\left(\frac{1+\alpha}{2}\right)}$$
be the threshold function. Then, we have
\begin{align*}
& \lim_{n \to \infty}  P_G(n,K_n,\alpha,\lambda) =  1, \quad \mathrm{if}\quad K_n > r_4(\alpha, x_n), \ \ \forall n.
\end{align*}
}  \label{theorem:thmg_5}
\end{theorem}
In Table~\ref{table:compare}, we provide a comparative summary of implications of Theorems~\ref{theorem:thmc_1}, ~\ref{theorem:thmc_2}, ~\ref{theorem:thmg_3}, and  ~\ref{theorem:thmg_5}. 
We also note that together Theorems \ref{theorem:thmg_3} and \ref{theorem:thmg_5} constitute the first investigation concerning the size of the largest connected components of random K-out graphs under randomly deleted nodes. Our proof builds upon the framework for giant component sizes from \cite{IT23sood} and adapts it to the homogeneous random K-out graph setting; full proof details are presented in the Appendix. 
 
\begin{figure}[!t]
\centering
\includegraphics[scale=0.48]{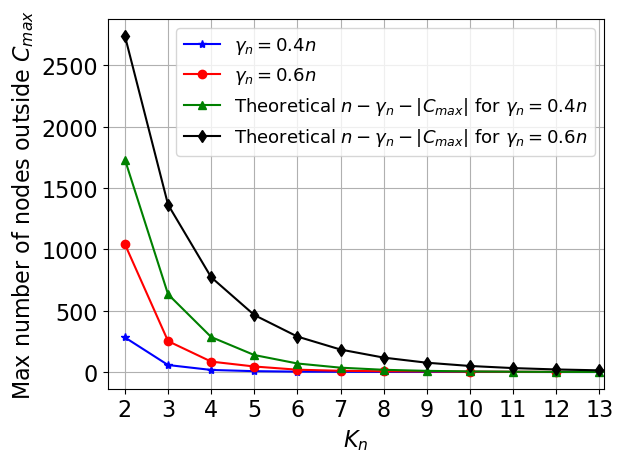}\label{fig:gc2a}  
\caption{{Maximum number of nodes outside the giant component of $\hhdn$ for $n = 5000$, $\gamma_n = 0.4n$ and $\gamma_n = 0.6n$, observed through 1000 independent realizations. We also plot heuristic upper bounds using Theorem~\ref{theorem:thmg_3}.}} 
\label{fig:fig3}
\end{figure}

\subsection{Discussion}
Invoking Theorem \ref{theorem:thmg_3}, we note that when $\gamma_n=o(n)$ nodes are removed,   setting $\lambda_n=\epsilon n$ with $\epsilon>0$, then  $r_3(\gamma_n,\lambda_n) = 1+o(1)$. This shows that when $\gamma_n=o(n)$, by choosing $K_n \geq 2$ and making $\epsilon$ arbitrarily small, we ensure that even the rest of the network contains a connected component whose {\em fractional} size is arbitrarily close to 1 whp. 
If we let $\lambda_n = 1$ in this result, we get the same scaling factor of $K_n =\Omega (\log n)$ as in Theorem \ref{theorem:thmc_1}. 
We also remark that the threshold $r_4(\alpha,\lambda_n)$ is finite when $\lambda_n=\Omega(n)$. This shows that even when a positive fraction of the nodes of the random K-out graph are removed, a finite $K_n$ is still sufficient to have a giant component of size $\Omega(n)$ in the graph.  

Also note that the largest connected component emanating from Theorem~\ref{theorem:thmg_3} \textbf{ }is infact the unique giant component, since under the regimes considered $(\lambda_n \leq \lfloor (n-\gamma_n)/3 \rfloor)$, $C_{\rm max}(n,K_n,\gamma_n) \geq 2 (n - \gamma_n)/3)$. Note that Theorems \ref{theorem:thmg_3} and \ref{theorem:thmg_5} present sufficient conditions on $K_n$ as a function of $n, \gamma_n$, and $\lambda_n$, which ensures that after a subset of $\gamma_n$ nodes is deleted, the largest connected component in the remaining subgraph is at least of size $n- \gamma_n-\lambda_n$ whp.  Further analysis into making these sufficient conditions less conservative is an open future direction.

Next we investigate the applicability of Theorems~\ref{theorem:thmg_3} and \ref{theorem:thmg_5} for finite node regimes. In Figure~\ref{fig:fig3}, we set $n=5000$ throughout and vary $K_n \in \{2,3,\dots,8\}$ and $\alpha \in \{0.4,0.6\}$. For each value of $\alpha$, for each $K_n$, we generate $1000$ independent realizations of $\hhdn$ with $\gamma_n= \alpha_n$. We first plot the maximum number of nodes observed outside the largest connected component observed in 1000 experiments for each parameter pair $(K_n, \alpha)$. Next, invoking Theorem \ref{theorem:thmg_5}, we compute a \emph{heuristic} upper bound bounding the number of nodes outside the largest connected component whp. We do this by identifying the largest value of $\lambda_n$ such that the chosen $K_n$ exceeds the threshold $r_4(\alpha, \lambda_n)$ and plot this as well.  

\begin{figure}[!t]
\centering
\includegraphics[scale=0.40]{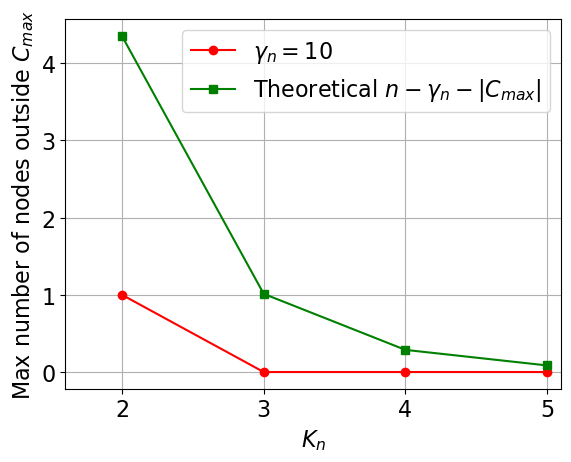}\label{fig:gc1} 
\hspace{0.4mm}
\includegraphics[scale=0.40]{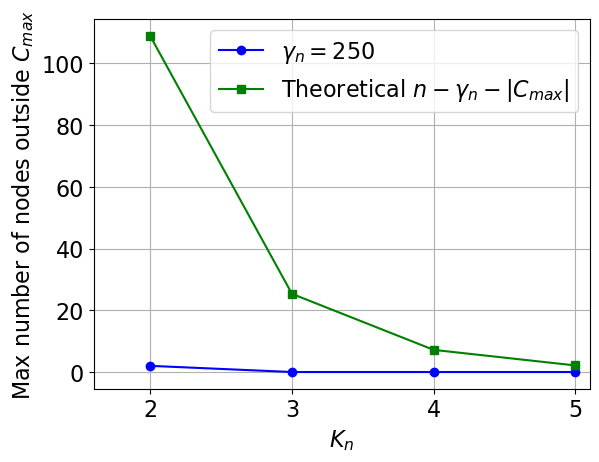}\label{fig:gc3} 
\caption{{ Maximum number of nodes observed outside the giant component over 1000 independent experiments with $n = 50,000$ and $\gamma_n = 10$ and  $250$ along with heuristic upper bounds through Theorem~\ref{theorem:thmg_3}.} \vspace{-2mm}} 
\label{fig:fig2}
\end{figure}

In Figure~\ref{fig:fig2}, we run a second set of experiments to simulate the case where $\gamma_n =  o(n)$. As before, we generate  instantiations of the random graph  $\hhdn$,
 but with a larger $n=50,000$, varying $K_n$ in  $\{2,\dots, 5\}$ and varying $\lambda_n$ in $\{10,2000\}$. For each $(K_n,\gamma_n)$ pair, we generate 1000 experiments and record the maximum number of nodes seen outside the giant component. We note that in some cases, no nodes are seen outside the giant component, indicating that the graph is connected. We also plot a heuristic upper bound on the number of nodes outside the largest connected by identifying the largest value of $\lambda_n$ such that the specified $K_n$ exceeds the threshold $r_3(\alpha, \lambda_n)$ in Theorem~\ref{theorem:thmg_3}. Across Figure~\ref{fig:fig3} and \ref{fig:fig2}, the experimentally observed maximum number of nodes outside the largest connected component is smaller than the heuristic upper bounds, reinforcing the usefulness of our results in finite node settings.


\subsection{Proof of 1-connectivity upper bound (Theorem~\ref{thm:UpperBoundForConnectivity})}
\label{sec:1-con-partial}
\label{sec:upperbound-bounds}
For easier exposition, we give a proof 
of Theorem~\ref{thm:UpperBoundForConnectivity} here
for $K=2$. 
The general version of our proof for $K\geq 2$ is given in the Appendix. For $K=2$, each node selects at least two other nodes and there can be no isolated nodes or node pairs in $\hh$. Thus, for $K=2$, the smallest possible isolated component is a {\em triangle},  i.e., a complete sub-networks over {\em three} nodes such that each node selects the other two nodes.  To derive the upper bound on connectivity, we first derive a lower bound on the probability of existence of isolated triangles in $\hh$. In the proof for the general case $(K \geq 2)$ presented in the Appendix, we investigate the existence of isolated components of size $K+1$.

Let $\Delta_{ijk}$ denote the event that nodes $v_i,v_j$ and $v_k$ form an isolated triangle in $\hh$. The number of isolated triangles in $\hh$, denoted by $Z_n$ is given by
\begin{align}
    Z_n = \sum_{1\leq i<j<k \leq n} \ii\{ \Delta_{ijk}\}
\end{align}{}Note that the existence of one or more isolated triangles  ($Z_n \geq 1$), implies that $\hh$ is \emph{not} connected. Thus, we can upper bound the probability of connectivity of $\hh$ as
\begin{align}
    &\pr [~\hh \text{~is connected}~]\nonumber \\
    &=1- \pr [~\hh \text{~is \emph{not} connected}~]\nonumber \\
   &=1-\pr[~\exists \text{~at least one isolated \emph{sub-network} in~} \hh ~]\nonumber \\
    &\leq 1-\pr[~ \exists \text{~at least one {isolated} \emph{triangle} in~} \hh ~] \nonumber \\
     &= 1-\pr[~ Z_n \geq 1 ~].
    \label{eq:koutbounds:p_con_upper_bound}
\end{align}{}
where,
\begin{align}
[Z_n \geq 1] = \cup_{1\leq i<j<k \leq n} \{\Delta_{ijk}\}.
\end{align}{}

In the succeeding discussion, we assume $K=2$ and
 use the Bonferroni inequality \cite{bonferroni} to lower bound the union of the events $\ii\{\Delta_{ijk}\}$, where $1\leq i<j<k \leq n$.
\begin{align}
&\pr[~ Z_n \geq 1 ~] \nonumber\\
& \geq \sum_{ i<j<k } \pr [~\Delta_{ijk}~]-\sum_{ i<j<k }\sum_{ x<y<z } \pr [~\Delta_{ijk}\cap\Delta_{xyz}~]
\label{eq:koutbounds:2out-union-lb}
\end{align}{}For all $1\leq i<j<k\leq n$ and $1 \leq x<y<z \leq n$, we have
\begin{align}
    \pr [~\Delta_{ijk}~]= \left(\frac{{1}}{{n-1 \choose 2}}\right)^3\left(\frac{{n-4 \choose 2}}{{n-1 \choose 2}}\right)^{n-3}
    \label{eq:koutbounds:2out-singlesum}
\end{align}{}
Moreover, note that if the sets $\{i,j,k\}$ and $\{x,y,z\}$ have one or more nodes in common, then these sets cannot simultaneously constitute isolated triangles; i.e., the events $\Delta_{ijk}$, $\Delta_{xyz}$ are mutually exclusive if $\{i, j, k\} \cap \{x, y, z\} \neq \emptyset $.
Thus, 
\begin{align}
\pr [~\Delta_{ijk} \cap\Delta_{x y z}~]=
\begin{cases}
& 0  \text{~~~if~} \{i,j,k\}\cap\{x,y,z\} \neq \phi, \\
& \left(\frac{{1}}{{n-1 \choose 2}}\right)^6\left(\frac{{n-7 \choose 2}}{{n-1 \choose 2}}\right)^{n-6} \text{otherwise}.
\label{eq:koutbounds:2out-doublesum}
\end{cases}
\end{align}{}We now calculate the summations term appearing in (\ref{eq:koutbounds:2out-union-lb}).
\begin{align}
    \sum_{ i<j<k } \pr [~\Delta_{ijk}~]
    &= {n \choose 3} \pr [~\Delta_{ijk}~]\nonumber\\
    &= {n \choose 3}\left(\frac{{1}}{{n-1 \choose 2}}\right)^3\left(\frac{{n-4 \choose 2}}{{n-1 \choose 2}}\right)^{n-3}\nonumber\\
   &= \frac{{4n}}{{3(n-1)^2 (n-2)^2}}
   \prod_{\ell=1}^2\left( 1-\frac{3}{n-\ell}\right)^{n-3} 
   \nonumber\\
   &\geq \frac{{4}}{{3n^3}}\left( 1-\frac{3}{n-2}\right)^{2n-6}
   \label{eq:koutbounds:2out-bound-singlesum},
\end{align}
and
\begin{align}
    &\sum_{ i<j<k }\sum_{ x<y<z } \pr [~\Delta_{ijk}\cap\Delta_{xyz}~] \nonumber\\
    &= {n \choose 3}{n-3 \choose 3}\left(\frac{{1}}{{n-1 \choose 2}}\right)^6\left(\frac{{n-7 \choose 2}}{{n-1 \choose 2}}\right)^{n-6}\nonumber\\
     &= \frac{{16n(n-3)(n-4)(n-5)}}{{9 (n-1)^5(n-2)^5}}
     \prod_{\ell=1}^2\left( 1-\frac{6}{n-\ell}\right)^{n-6} 
    \nonumber\\
     &\leq \frac{{16n^4}}{{9 (n-2)^{10}}}\left( 1-\frac{6}{n-1}\right)^{2n-12}
      \label{eq:koutbounds:2out-bound-doublesum}.
\end{align}
Substituting (\ref{eq:koutbounds:2out-bound-singlesum}) and (\ref{eq:koutbounds:2out-bound-doublesum}) in (\ref{eq:koutbounds:2out-union-lb}), we obtain
\begin{align}
&\pr[~ Z_n \geq 1 ~] \nonumber\\
& \geq \sum_{ i<j<k } \pr [~\Delta_{ijk}~]-\sum_{ i<j<k }\sum_{ x<y<z } \pr [~\Delta_{ijk}\cap\Delta_{xyz}~]\nonumber\\
& \geq \frac{{4}}{3{n^3}}\hspace{-.5mm}\left( \hspace{-.5mm}1\hspace{-.5mm}-\hspace{-.5mm}\frac{3}{n-2}\hspace{-.5mm}\right)^{2n\hspace{-.5mm}-6} \hspace{-1.5mm}-\frac{{16n^4}}{{9 (n-2)^{10}}}\left(\hspace{-.5mm} 1\hspace{-.5mm}-\hspace{-.5mm}\frac{6}{n-1}\hspace{-.5mm}\right)^{2n-12}\nonumber\\
& =\frac{{4 e^{-6}}}{3{n^3}}(1+\oo(1)) \label{eq:koutbounds:upperrbound}
\end{align}{}
Substituting (\ref{eq:koutbounds:upperrbound}) into (\ref{eq:koutbounds:p_con_upper_bound}) establishes Theorem \ref{thm:UpperBoundForConnectivity}
for $K=2$. \myendpf
More compactly, this result can be stated
as $$P(n;2)=1-\Omega\left(\frac{1}{n^3}\right).$$
We prove the more general result for $K \geq 2$ 
in the Appendix. 

\section{Proof for $r$-robustness (Theorem~\ref{robustness-theorem:thmr_1})}
\label{sec:r-robustness-proof}

\subsection{Preliminaries for Proof of Theorem \ref{robustness-theorem:thmr_1} }

We start with a few definitions and properties that will be useful throughout the rest of the proof. First, let $B(a,b)$ denote the beta function, $B_x(a,b)$ denote the incomplete beta function, and $I_x(a,b)$ denote the regularized incomplete beta function, where $a$ and $b$ are non-negative integers. These functions are defined as follows \cite{NIST:DLMF}:
\begin{align}
    B(a,b)&= \int_{0}^{1} t^{a-1}(1-t)^{b-1} dt = \frac{(a-1)!(b-1)!}{(a+b-1)!} \nonumber \\
    B_x(a,b)&= \int_{0}^{x} t^{a-1}(1-t)^{b-1} dt, \quad 0\leq x \leq 1 \nonumber \\
    I_x(a,b)&=\frac{B_x(a,b)}{B(a,b)}, \quad 0\leq x \leq 1
\end{align}

Using these definitions, it can easily be shown that 
 \begin{align}
     I_{1/2}(r,r) = 1/2, \quad r > 0
 \end{align}
Proof: $
    B(r,r) =  \int_{0}^{1} t^{r-1}(1-t)^{r-1}dt  = 2 \int_{0}^{\frac{1}{2}} t^{r-1}(1-t)^{r-1}dt 
     = 2 B_{1/2}(r,r)  $
  where we divided the integral into two parts since the function $(t-t^2)^{r-1}$ is symmetric around 1/2.
 Using the fact that $B_{1/2}(r,r) = I_{1/2}(r,r) B(r,r)$, we can conclude that  $I_{1/2}(r,r) = 1/2$.

The cumulative distribution function $F(a;n,p)$ of a Binomial random variable $X \sim B(n,p)$ can be expressed using the regularized incomplete beta function as:
\begin{align}
    F(a;n,p) &= \pr[X \leq a] = I_{1-p}(n-a,a+1) \nonumber \\
    &=(n-a){n \choose a} \int_{0}^{1-p} t^{n-a-1}(1-t)^a dt
    \label{eq:binom_prob}
\end{align}

 \begin{lemma}
\cite[Eq.~8.17.4]{NIST:DLMF}: For $a,b > 0, \ 0 \leq x \leq 1$,
\begin{align}
    I_x(a,b) = 1 - I_{1-x}(b,a)  \label{eq:I_oneminusx}
\end{align}
 \end{lemma}
 
 \begin{lemma}
\cite[Eq.~8.17.20]{NIST:DLMF}: For $a,b > 0, \ 0 \leq x \leq 1$,
\begin{align}
    I_x(a+1,b) = I_x(a,b) - \frac{x^a(1-x)^b}{a B(a,b)}  \label{eq:I_KandR}
\end{align}
 \end{lemma}


 \begin{lemma}
 The equation $I_{\alpha}(r,r) = c\alpha$ has only one solution when $r>0$, $r \in \mathbb{Z}^+$, $0<\alpha \leq 1/2$ and $0<c\leq 1$. 
 \label{corr:one_soln}
 \end{lemma}

\subsection{Proof of Theorem \ref{robustness-theorem:thmr_1}}

To prove Theorem \ref{robustness-theorem:thmr_1}, we  need to show that   for any $r \in \mathbb{Z}^+$, the random K-out graph $\hhn$ is $r$-robust {\em whp} if $K_n \geq 2r$. 
To do this, similar to the proof given in \cite{sundaram_robustness} for Erd\H{o}s-R\'enyi graphs, we will first find an upper bound on the probability of a subset of given size being not $r$-reachable, and then use this result to show that the probability of not being $r$-robust goes to zero when $n \to \infty$ and $K_n \geq 2r$. 

First, let $\mathcal{E}_n (K_n, r; S)$ denote the event that  $S \subset V$ is an $r$-reachable set as per Definition~\ref{robustness-def:r-reachable}. The event $\mathcal{E}_n (K_n, r; S)$ occurs if there exists at least one node in $S$ that is adjacent to at least $r$ nodes in $S^c$, the subset comprised of nodes outside the subset $S$. Thus, we have
\begin{align}
\mathcal{E}_n (K_n, r; S) = 
\bigcup_{i \in \nodes_{S}}  \left \{ \left( \sum_{j \in \nodes_{S^c}} \mathds{1} \left \{ v_i \sim v_j  \right \} \right) \geq r  \right \}  \nonumber
\end{align}
with $\nodes_{S}$,  $\nodes_{S^c}$ denoting the set of labels of the vertices in  $S$ and $S^c$, respectively, and $\mathds{1} \{ \}$ denoting the indicator function.
We are also interested in the complement of this event, denoted as $\left (\mathcal{E}_n (K_n, r; S) \right)\comp$, which occurs if all nodes in $S$ are adjacent to less than r nodes in $S^c$. This can be written as
\begin{align}
\left( \mathcal{E}_n\comp (K_n, r; S) \right)  = 
\bigcap_{i \in \nodes_{S}} \left \{ \left( \sum_{j \in \nodes_{S^c}} \mathds{1} \left \{ v_i \sim v_j  \right \} \right) < r  \right \}.  \nonumber
\end{align}
Note that at least one subset in every disjoint subset pairs needs to be $r$-reachable per the definition of $r$-robustness, hence one of the events $\mathcal{E}_n (K_n, r; S)$ or $\mathcal{E}_n (K_n, r; S')$ need to hold {\em with high probability} for every disjoint subset pairs $S, S'$ of $V$. Now, let $\mathcal{Z}(K_n, r)$ denote the event that both subsets in at least one of the disjoint subset pairs $S, S' \subset V$ are $r$-reachable. Thus, we have  

\begin{align}
\mathcal{Z}(K_n, r) & = \hspace{-4mm} \bigcup_{S, S' \in \mathcal{P}_n: ~  |S| \leq \lfloor \frac{n}{2} \rfloor} \hspace{-3mm} \left[ (\mathcal{E}_n({K}_n,r; S))\comp \wedge (\mathcal{E}_n({K}_n,r; S\comp))\comp \right], \nonumber
\end{align}
where $S \cap S' = \emptyset$, $\mathcal{P}_n$ is the collection of all non-empty subsets of $V$ and since for each $S$ we check the $r$-reachability of both $S$ and $S\comp$, the condition $| S | \leq \lfloor \frac{n}{2} \rfloor$ is used to prevent counting each subset twice. From this defintion, it can be seen that the graph $\hhn$ is $r$-robust if the event $\mathcal{Z}(K_n, r)$ does not occur. Using union bound, we get
\begin{align}
P_Z &\leq \hspace{-3mm}  \sum_{ |S| \leq \lfloor \frac{n}{2} \rfloor } \pr[ \left( \mathcal{E}_n (K_n, r; S)\right) \comp \wedge \left( \mathcal{E}_n (K_n, r; S')\right) \comp ] \nonumber \\
&\leq \hspace{-3mm}  \sum_{ |S| \leq \lfloor \frac{n}{2} \rfloor } \pr[ \left( \mathcal{E}_n (K_n, r; S)\right) \comp   ] \nonumber \\
&=\hspace{-1mm} \sum_{m=1}^{ \left\lfloor \frac{n}{2} \right\rfloor }
 \sum_{S_m \in \mathcal{P}_{n,m} } \pr[\left( \mathcal{E}_n (K_n, r; S_m)\right) \comp  ]  \label{eq:UnionBound},
\end{align}
where $\mathcal{P}_{n,m}$ denotes the collection of all subsets of $V$ with exactly $m$ elements, and let $S_m \in \mathcal{P}_{n,m}$ be a subset of the vertex set $V$ with size $m$, i.e. $S_m \subset V$ and $|S_m| =m$. Further,   $\pr\left[\mathcal{Z}(K_n,r) \right]$ is abbreviated as $P_Z := \pr\left[\mathcal{Z}(K_n,r) \right]$. 
 From the exchangeability of the node labels and associated random variables, we have

\begin{align}
\sum_{S_m \in \mathcal{P}_{n,m} }   \pr[\left( \mathcal{E}_n (K_n, r; S_m)\right) \comp  ]   = {n\choose m} \pr[\left( \mathcal{E}_n (K_n, r; S_m)\right) \comp  ]. 
\label{eq:ForEach=r}
\end{align}

since $|\mathcal{P}_{n,m} | = {n \choose m}$, as there are ${n \choose m}$ subsets of $V$ with $m$ elements. Substituting this into (\ref{eq:UnionBound}), we obtain 
\begin{align}
P_Z \leq \sum_{m=1}^{ \left\lfloor \frac{ n}{2} \right\rfloor } {n\choose m} \pr[\left( \mathcal{E}_n (K_n, r; S_m)\right) \comp  ] \nonumber
\label{eq:Zbound}
\end{align}
\vspace{-2mm}

Before evaluating this expression, we will start with evaluating the probability that the set $S_m$ is not $r$-reachable, abbreviated as $\pr[\left( \mathcal{E}_n (K_n, r; S_m)\right) \comp ] :=  P_{S_m}$. Since a node $v \in S_m$ can have neighbors in $S_m^c$ if it forms an edge with nodes in $S_m^c$ or if nodes in $S_m^c$ forms edges with node $v$, let $P_{S_{m,1}}$ denote the probability that all nodes $v \in S_m$ form an edge with less than $r$ nodes in $S_m^c$, and let $P_{S_{m,2}}$ denote the probability that for each node $v \in S_m$, nodes in $S_m^c$ form less than $r$ edges with them. Evidently, $P_{S_m} \leq P_{S_{m,1}} \cdot P_{S_{m,2}}$. Further, let $P_{v_{m,1}}$ denote the probability that a node $v \in S_m$ forms an edge with less than $r$ nodes in $S_m^c$, and let $P_{v_{m,2}}$ denote the probability that nodes in $S_m^c$ form less than $r$ edges with the node $v \in S_m$.

\begin{lemma}
The probability that the node $v \in S_m$ chooses less than $r$ nodes in the set $S_m^c$, denoted as $P_{v_{m,1}}$, can be upper bounded by the cumulative distribution function $F(r-1;K_n,p)$ of a binomial random variable with $K_n$ trials and success probability $p=\frac{n-m-r+1}{n-r}$. 
\label{corrol:prob1}
\end{lemma}

A proof is presented in the Appendix

Next, using this upper bound, we plug in $n=K_n$ and $p=\frac{n-m-r+1}{n-r}$ to \eqref{eq:binom_prob}, then we have
\begin{align}
    P_{v_{m,1}} & \leq F\left(r-1;m,1- \frac{m-1}{n-r}\right) = I_{\frac{m-1}{n-r}}(K_n-r+1,r) \nonumber \\ & = (K_n -r+1){K_n \choose r-1} \int_{0}^{\frac{m-1}{n-r}} t^{K_n-r}(1-t)^{r-1}dt 
\end{align}

The selections of each node in $S_m$ are independent, hence we can use $(P_{S_{m,1}}) =(P_{v_{m,1}})^m$.

In order to find $P_{v_{m,2}}$, a node in $S_m^c$ forming an edge with the node $v$ can be modeled as a Bernoulli trial with probability $p=\frac{K_n}{n-1}$ so the event that nodes in $S_m^c$ forming less than $r$ edges with the node $v$ can be represented by a Binomial model with $n-m$ trials and $p=\frac{K_n}{n-1}$. Hence,
\begin{align}
    P_{v_{m,2}} & = F\left(r-1;n-m, \frac{K_n}{n-1}\right)\nonumber\\
    &= I_{\frac{n-K_n-1}{n-1}}(n-m-r+1,r)
\end{align}


Since the nodes in $S_m^c$ forming edges with nodes in $S_m$ are not independent of the other nodes in $S_m$, we cannot write $(P_{S_{m,2}}) \leq (P_{v_{m,2}})^m$. To find $(P_{S_{m,2}})$, we will decompose it into the following conditional probabilities.
\begin{align}
    P_{S_{m,2}} = & \pr[d_{v_1}<r] \cdot \pr[d_{v_2}<r | d_{v_1}<r] \cdot \nonumber \\ & \ldots \pr[d_{v_m}<r | d_{v_1}<r, d_{v_2}<r, \ldots, d_{v_{m-1}}<r]
\end{align}

where $v_1, v_2, \ldots, v_m  \in S_m$ represent all the nodes in $S_m$, and $d_{v_i}$ is used to denote the number of nodes in $S_m^c$ that form an edge with the node $v_i$. To find an upper bound on $P_{S_{m,2}}$, we consider the worst case. In the worst case, all the preceding nodes are selected by nodes in $S_m^c$ exactly $r-1$ times. This reduces the number of available edges in $S_m^c$ that can make connections with the remaining nodes in $S_m$, hence increases the probability of nodes in $S_m^c$  forming less than $r$ edges with the remaining nodes in $S_m$. Hence, we can write:

\begin{align}
    P_{S_{m,2}} & \leq  \pr[d_{v_1}<r] \cdot \pr[ d_{v_2}<r | d_{v_1}=r-1] \cdot \nonumber \\ & \ldots \pr[d_{v_m}<r | d_{v_1}=r-1,  \ldots, d_{v_{m-1}}=r-1]
\end{align}

Consider the general case for $\pr[d_{v_{a+1}}<r | d_{v_1}=r-1,  \ldots, d_{v_{a}}=r-1]$ where $1\leq a \leq m-1$. Assume that $q_1$ nodes in $S_m^c$ formed an edge with only one node among the nodes $v_1,\ldots,v_a$. Similarly, assume $q_2$ nodes in $S_m^c$ formed an edge with only two nodes, and so on ($q_{K_n}$ nodes in $S_m^c$ formed edges with $K_n$ nodes in $v_1,\ldots,v_a$). Also define $q=q_1 + q_2 + \ldots + q_{K_n}$. It can be seen that $a = \frac{q_1 + 2q_2 + \ldots + K_n q_{K_n}}{r-1}$. Here, we have $n_0 = n-m-q$ nodes that did not use any of its selections yet, so their probability of choosing the node $v_{a}$ is $p_0 = \frac{K_n}{n-a-1}$. Similarly, we have $n_1 = q_1$ nodes that used one of its selections, so their probability is $p_1 = \frac{K_n -1 }{n-a-1}$ (for $n_{K_n}=q_{K_n}$ nodes $p_{K_n}=\frac{K_n - K_n }{n-a-1}=0$).

Considering the selection of each node in $S_m^c$ as a Bernoulli trial with different probabilities, the collection of all such trials defines a Poisson Binomial distribution $X \sim PB([p_0]^{n_0},\ldots,[p_{K_n}]^{n_{K_n}})$. The mean of this distribution is:
\begin{align}
    \mu_X &= (n-m-q)*\frac{K_n}{n-a-1} + \sum_{i=1}^{K_n} q_i\frac{K_n-i}{n-a-1} \nonumber \\
    &= \frac{(n-m)K_n-a(r-1)}{n-a-1}
\end{align}

Assuming $K_n > 2r-2$ and using $m \leq \lfloor n/2 \rfloor$, we have:
\begin{align}
    \mu_X = \frac{(2n-2m-a)}{n-a-1} *(r-1) +   \frac{K_n - 2r + 2}{n-a-1}
\end{align}

Since the mode of the Poisson Binomial distribution satisfies $\psi_X \leq \mu_X + 1$ \cite{poissonbinomial_mode}, we have $\psi_X > r-1$ for any $1\leq a \leq m-1$ if $K_n > 2r-2$. Since for any distribution, the cumulative distribution function is equal to 1/2 at the mode of distribution, we have that $\pr[d_{v_{a+1}} < r | d_{v_1}=r-1,  \ldots, d_{v_{a}}=r-1] < 1/2$. Hence,
\begin{align}
    P_{S_{m,2}} < \left(\frac{1}{2}\right)^m
\end{align}


Now, using the fact that $P_{S_m} \leq P_{S_{m,1}} \cdot P_{S_{m,2}}$, we have:

\begin{align*}
    P_{S_m} & \leq P_{S_{m,1}} \cdot P_{S_{m,2}} \leq (P_{v_{m,1}})^m \cdot \left(\frac{1}{2}\right)^m \nonumber \\ & \leq \hspace{-1mm} \left( \frac{1}{2} I_{\frac{m-1}{n-r}}(K_n \hspace{-0.3mm} - \hspace{-0.3mm} r \hspace{-0.3mm} + \hspace{-0.3mm} 1,r) \hspace{-0.3mm}  \hspace{-0.3mm}\right)^{m} 
\end{align*}



Let $P_m := {n\choose m}  P_{S_m} $. Then, we have:
\begin{align}
    P_Z \leq \sum_{m=1}^{ \left\lfloor \frac{ n}{2} \right\rfloor } {n\choose m} P_{S_m} = \sum_{m=1}^{ \left\lfloor \frac{ n}{2} \right\rfloor } P_m
\end{align}

We will divide the summation into three parts as follows:
\begin{align}
    P_Z = \sum_{m=1}^{\lfloor n/2 \rfloor} P_m = \sum_{m=1}^{\lfloor \log(n) \rfloor} P_m + \sum_{m=\lceil \log(n) \rceil}^{\lfloor \alpha^*(n - r) \rfloor} P_m \nonumber \\ + \sum_{m=\lceil \alpha^*(n - r)  \rceil}^{\lfloor n/2 \rfloor} P_m = P_1 + P_2 + P_3
\end{align}
where $\alpha^*$ is the solution to the equation $I_{\alpha}(r,r) = \frac{n-r}{ne}\cdot \alpha$ in the range $0<\alpha<\frac{1}{2}$. (The purpose of defining $\alpha^*$ this way will be given later in the proof.) 
Start with the first summation $P_1$ and use  \eqref{eq:binomial_property} along with ${n \choose m} \leq \left( \frac{en}{m} \right)^m$, then we have:

\begin{align}
    P_m & = {n \choose m} P_{S_m}  \leq  {n \choose m}  P_{S_{m,1}} \cdot P_{S_{m,2}} \leq {n \choose m}  P_{S_{m,1}}  \nonumber \\
   & \leq  \left(\frac{en}{m} I_{\frac{m-1}{n-r}}(K_n-r+1,r) \right)^{m} 
\nonumber  \\ &
 \leq  \left(\frac{en}{m B(K_n-r+1,r)} \int_{0}^{\frac{m-1}{n-r}} t^{K_n-r}(1-t)^{r-1}dt \right)^{m} 
   \nonumber  \\ &
     \leq  \left(\frac{en}{m B(K_n-r+1,r)} \int_{0}^{\frac{m-1}{n-r}} t^{K_n-r}dt \right)^{m} \nonumber  \\ &
   \leq  \left(\frac{en \left( \frac{m-1}{n-r}\right)^{K_n-r+1}}{m B(K_n-r+1,r) (K_n-r+1)}  \right)^{m}  \nonumber  \\ &  \leq  \left( \frac{e (1+r/m)}{B(K_n-r+1,r)(K_n-r+1)} \left( \frac{m-1}{n-r}\right)^{K_n-r} \right)^{m} \nonumber  \\ &  \leq  \left( \frac{e (1+r) \left( \frac{\log(n)-1}{n-r}\right)^{K_n-r}}{B(K_n-r+1,r)(K_n-r+1)}  \right)^{m} := (a_n)^m \nonumber
\end{align}

For $K_n > r$, since $B(K_n-r+1,r)$ and $r$ are finite values, we have $\limit a_n = 0$ by virtue of $\limit \left(\frac{\log(n)-1}{n-r}\right)^{K_n-r} = 0$.  Using this, we can express the summation as:
\begin{align}
    P_1 = \sum_{m=1}^{\lfloor \log(n) \rfloor} (a_n)^m \leq a_n \cdot \frac{1 - (a_n)^{\log(n)}}{1 - a_n}
\end{align}
where the geometric sum converges by virtue of $\lim _{n \to \infty} a_n = 0$, leading to $P_1$ converging to zero for large $n$. 

Now, similarly consider the second summation $P_2$. Using ${n \choose m} \leq \left( \frac{en}{m} \right)^m$, we have
\begin{align}
    P_m & \leq  {n \choose m} P_{S_m} \leq {n \choose m} (P_{S_{m,1}})^m  \nonumber \\
    &\leq  \left(\frac{en}{m} I_{\frac{m-1}{n-r}}(K_n-r+1,r) \right)^{m} := (a_m)^m
\end{align}
where $a_m := \frac{en}{m} \cdot  I_{\frac{m-1}{n-r}}(K_n-r+1,r)$. Assume that $K_n > 2r-1$. It can be shown that $I_{\frac{m+r}{n}}(K_n -r+1) < I_{\frac{m+r}{n}}(r,r)$ as a consequence of the property (\ref{eq:I_KandR}). Hence, we have
\begin{align}
    a_m & <  \frac{en}{m}\cdot I_{\frac{m-1}{n-r}}(r,r) \leq \frac{en}{m}\cdot I_{\frac{m}{n-r}}(r,r)
\end{align}

Assume that $\alpha=\alpha^*$ is the solution of the equation $I_{\alpha}(r,r) = \frac{n-r}{ne}\cdot \alpha$ in the range $0<\alpha<\frac{1}{2}$. From Lemma \ref{corr:one_soln}, we know that this equation has only one solution in this range, and $I_{\alpha}(r,r) \leq \frac{\alpha^* (n-r)}{ne}$ for $0<\alpha \leq \alpha^*$. Plugging in $\alpha = \frac{m}{n-r}$, we have $I_{\frac{m}{n-r}}(r,r) \leq \frac{\alpha^* m}{ne}$, which leads to $a_m < 1$ when $m \leq \lfloor \alpha^*(n-r) \rfloor$. Denoting $a:= \max_m (a_m)$, we have 
\begin{align}
    P_2 =   \sum_{m=\lceil \log(n) \rceil}^{\lfloor \alpha^*(n - r) \rfloor} (a_m)^m \leq \sum_{m=\lceil \log(n) \rceil}^{\infty} (a)^m \leq \frac{a^{\log(n)}}{1-a}
\end{align}
where the geometric sum converges by virtue of $\lim _{n \to \infty} a < 1$, leading to $P_2$ converging to zero as $n$ gets large.

Now, similarly consider the third summation $P_3$. Assuming $K_n \geq 2r-1$, we have

\begin{align}
    P_m & \leq  {n \choose m} (P_{S_{m,1}})^m \cdot (P_{S_{m,2}})^m  \nonumber \\
    & <  \left(\frac{1}{2} {n \choose m}^{\frac{1}{m}} I_{\frac{m-1}{n-r}}(K_n-r+1,r) \right)^{m} \nonumber \\
    & <  \left(\frac{1}{2} {n \choose m}^{\frac{1}{m}} I_{\frac{m}{n-r}}(r,r) \right)^{m} := (a_m)^m   
\end{align}

From the previous summation $P_2$ and the proof of Lemma \ref{corr:one_soln}, we know that in the range $\lceil \alpha^*n - r  \rceil \geq m \geq \lfloor n/2 \rfloor$; $\frac{n}{m}$ is decreasing, $I_{\frac{m}{n-r}}(r,r)$ is increasing, and the overall expression $\frac{n}{m} \cdot I_{\frac{m}{n-r}}(r,r)$ is also increasing. Hence, $\frac{n}{m}\cdot I_{\frac{m}{n-r}}(r,r)$ takes its maximum value when $m=\lfloor n/2 \rfloor$. Since $\frac{en}{m}$ is an upper bound of ${n \choose m}^{\frac{1}{m}}$, the expression $a_m$ also takes its maximum value when $m=\lfloor n/2 \rfloor$. Denoting $a:= \max_m (a_m) = a_{\lfloor n/2 \rfloor}$, and using Stirling's approximation $\limit {2n \choose n} \sim \frac{4^n}{\sqrt{\pi n}} < \frac{4^n}{\sqrt{\pi}}$ as a finer upper bound, we have 

\begin{align}
    \limit a = \limit a_{\lfloor n/2 \rfloor} <  \frac{1}{2} \cdot \frac{4}{\sqrt{\pi} } \cdot \frac{1}{2} < 1 
\end{align}

From this, we have:

\begin{align}
     P_3 & \leq  \sum_{m=\lceil \alpha^*(n - r)  \rceil}^{\lfloor \frac{n}{2} \rfloor} (a_m)^n \leq \frac{n}{2} (a)^n
\end{align}

where the sum converges by virtue of $\limit a = 1$ since $\limit n(a)^n = 0 $ in this case, leading to $P_3$ converging to zero as $n$ gets large. Since $P_1$, $P_2$, and $P_3$ all converge to zero as $n$ gets large, $P_Z = P_1 + P_2 + P_3$ also converges to zero as $n$ gets large. This concludes the proof of Theorem \ref{robustness-theorem:thmr_1}. \myendpf

\section{Conclusions}\label{sec:conc}
In this work, we provide formal performance guarantees for random K‑out graphs by stating precise conditions on network parameters to achieve the desired notion of connectivity in different operational contexts. First, we derive {matching} upper and lower bounds for connectivity for random K-out graphs when the number of nodes is finite. We prove that the probability of connectivity is $1-\Theta({1}/{n^{K^2-1}})$ for all $K \geq 2$. To the best of our knowledge, our work is the first to provide an upper bound on the probability of connectivity, which shows that further improvement on the order of $n$ in the lower bound is not possible. Next, motivated by applications in resilient consensus, we establish that for large $n$, $r^{\star}(K) \geq \lfloor K/2 \rfloor$, in other words, $K \geq 2r$ ensures, {\em with high probability}, the $r$-robustness of the random K-out graph $\hh$, improving the previously known condition of $K=\OO(r \log(r))$. Further investigating if the threshold of $2r$ is the tightest possible for $r$-robustness remains an open future direction. Motivated by settings in which nodes fail or are unreliable, we analyze the random K‑out graph $\mathbb{H}(n;K_n,\gamma_n)$ with $\gamma_n$ uniformly random node deletions and derive a sharp one-law for $K_n$ (as a function of $n$) for connectivity when $\gamma_n=\Omega(n)$. Additionally we give a suite of results to guide how to set the parameter $K_n$ as a function of $\gamma_n$ to ensure the desired level of connectivity or size of the largest connected component. Through numerical simulations, we demonstrate the efficacy of our theoretical results as practical guidelines for selecting $K_n$ across a wide range of settings. While our analysis of random K‑out graphs bridges several gaps in aligning the model with operational requirements in distributed systems, its implications can be explored further. For instance, in pairwise masking–based privacy mechanisms—whether based on cryptographic primitives \cite{2020secureagg-polylog} or statistical noise \cite{2022sabater-gopa}—an important future direction is to understand how the component‐size distribution governs an adversarial subset’s ability to recover partial sums when occupying a fraction of the network. It would also be of interest to extend our connectivity analysis to other random K‑out variants—such as subgraphs of given graphs where nodes are constrained to select only from a subset of peers \cite{frieze2016introduction}.

\section*{Acknowledgements}
 The authors gratefully acknowledge support from the Carnegie Mellon University Cylab Seed Grant; the Cylab Presidential Fellowship; the Philip and Marsha Dowd Fellowship; the Knight Foundation and IDeaS Knight Fellowship; the Lee–Stanziale Ohana Endowed Fellowship; the David H. Barakat and LaVerne Owen–Barakat Carnegie Institute of Technology (CIT) Dean’s Fellowship; and the CIT Dean’s Fellowship.  This work was supported in part by the National Science Foundation under grants CCF~\#2225513, \#2026985, \#1813637, and \#1617934; the Office of Naval Research under grant N00014‑23‑1‑2275; the Army Research Office under grants W911NF‑22‑1‑0181, W911NF‑20‑1‑0204, and W911NF‑17‑1‑0587; and the Air Force Office of Scientific Research under grant FA9550‑22‑1‑0233. We acknowledge Dr. Giulia Fanti, Dr. Carlee Joe-Wong, and Dr. Chai Wah Wu for helpful discussions regarding applications in distributed inference.


\bibliographystyle{IEEEtran}
\bibliography{ref}


\section*{Appendix}

\subsection{Upper bound on Probability of Connectivity $K \geq 2$}
\label{sec:genupperbound-bound}
In Section~\ref{sec:upperbound-bounds} we proved Theorem~\ref{thm:UpperBoundForConnectivity} for the case $K=2$. In this section, we prove the upperbound for the general case $K \geq 2$. Let $K$ be a fixed positive integer such that $ K \geq 2$. Let $\Delta_{i_1 \dots i_{K+1}}$ denote the event that nodes $v_i,v_j,\dots,v_{K+1}$ form an isolated component in $\hh$. The number of such isolated components of size $K+1$ in $\hh$, denoted by $Z_n$ is given by
\begin{align}
    Z_n = \sum_{1\leq i_1<i_2 \dots < i_{K+1} \leq n} \ii\{ \Delta_{i_1 \dots i_{K+1}}\}
\end{align}{}
Note that the existence of one or more isolated components of size $K+1$  ($Z_n \geq 1$), implies that $\hh$ is \emph{not} connected. We can upper bound the probability of connectivity of $\hh$ as
\begin{align}
    &\pr [~\hh \text{~is connected}~]\nonumber \\
    &=1- \pr [~\hh \text{~is \emph{not} connected}~]\nonumber \\
  &=1-\pr[~\exists \text{~at least one isolated \emph{sub-network}}  ~]\nonumber \\
    &\leq 1-\pr[~ \exists \text{~at least one {isolated} component of size $K+1$}  ~] \nonumber \\
     &= 1-\pr[~ Z_n \geq 1 ~].
    \label{eq:koutbounds:gen_p_con_upper_bound}
\end{align}{}
where,
\begin{align}
\{Z_n \geq 1\} = \bigcup_{1\leq i_1<i_2 \dots < i_{K+1} \leq n} \ii\{\Delta_{i_1 \dots i_{K+1}}\}.
\end{align}{}

In the succeeding discussion, we use the Bonferroni inequality \cite{bonferroni} to lower bound the union of the events given in (\ref{eq:koutbounds:gen_p_con_upper_bound}).
\begin{align}
&\pr[~ Z_n \geq 1 ~] \nonumber\\
& \geq \sum_{ i_1<i_2 \dots < i_{K+1} } \pr [~\Delta_{i_1 \dots i_{K+1}}~]\\
&\qquad -\sum_{i_1<i_2 \dots < i_{K+1} }\sum_{ j_1<j_2 \dots < j_{K+1}} \pr [~\Delta_{i_1 \dots i_{K+1}}\cap\Delta_{j_1 \dots j_{K+1}}~]
\label{eq:koutbounds:gen_2out-union-lb}
\end{align}{}
For all $1\leq i_1<i_2 \dots < i_{K+1}\leq n$ and $1 \leq j_1<j_2 \dots < j_{K+1}\leq n$, we have
\begin{align}
    \pr [~\Delta_{i_1 \dots i_{K+1}}~]= \left(\frac{{1}}{{n-1 \choose K}}\right)^{K+1}\left(\frac{{n-K-2 \choose K}}{{n-1 \choose K}}\right)^{n-K-1}
    \label{eq:koutbounds:gen_2out-singlesum}
\end{align}{}
Moreover, note that if the sets $\{i_1,\dots,i_{K+1}\}$ and $\{j_1,\dots,j_{K+1}\}$ have one or more nodes in common, then these sets cannot simultaneously constitute isolated components.
Thus, $\pr [~\Delta_{i_1 \dots i_{K+1}} \cap\Delta_{j_1 \dots j_{K+1}}~]=$
\begin{align}
\begin{cases}
& 0  \text{~~~if~~~} \{i_1,\dots,i_{K+1}\}\cap\{j_1,\dots,j_{K+1}\} \neq \phi, \\
& \left(\frac{{1}}{{n-1 \choose K}}\right)^{2(K+1)}\left(\frac{{n-2K-3 \choose K}}{{n-1 \choose K}}\right)^{n-2(K+1)} \text{otherwise}. \label{eq:koutbounds:gen_2out-doublesum}
\end{cases}
\end{align}
We now calculate the term appearing in (\ref{eq:koutbounds:gen_2out-union-lb}) in turn. We have
\begin{align}
    &\sum_{ i_1<i_2 \dots < i_{K+1}} \pr [~\Delta_{i_1\dots i_{K+1}}~]\nonumber\\
    & = {n \choose {K+1}} \pr [~\Delta_{i_1 \dots i_{K+1}}~]\nonumber\\
    & = {n \choose {K+1}}\left(\frac{{1}}{{n-1 \choose K}}\right)^{K+1}\left(\frac{{n-K-2 \choose K}}{{n-1 \choose K}}\right)^{n-K-1}\nonumber\\
      & = \frac{({K!})^K n}{{K+1}}\cdot \left(\frac{(n-K-1)!}{(n-1)!}\right)^{K} \prod_{\ell=1}^{K}\cdot\left(1-\frac{K+1}{n-\ell}\right)^{n-K-1} \nonumber\\
  & \geq  \frac{({K!})^K}{{K+1}}\cdot \frac{1}{n^{(K^2-1)}} \prod_{\ell=1}^{K}\cdot\left(1-\frac{K+1}{n-\ell}\right)^{n-K-1} \nonumber\\
      & \geq  \frac{({K!})^K}{{K+1}}\cdot \frac{1}{n^{(K^2-1)}} \cdot\left(1-\frac{K+1}{n-K}\right)^{K(n-K-1)} \label{eq:koutbounds:gen_2out-bound-singlesum}\\
      & = \frac{({K!})^K e^{-K(K+1)}}{{K+1}}\cdot \frac{1}{n^{(K^2-1)}}  (1+\oo(1)) \nonumber
\end{align}{}
where (\ref{eq:koutbounds:gen_2out-bound-singlesum}) is plain from the observation that for all $\ell$ in $1, \dots, K$, 
\begin{align*}
    1-\frac{K+1}{n-\ell} \geq 1-\frac{K+1}{n-K}.
\end{align*}
Next,
\begin{align}
    &\sum_{ i_1< \dots < i_{K+1}}\sum_{  j_1< \dots < j_{K+1} } \pr [~\Delta_{ i_1, \dots, i_{K+1}}\cap\Delta_{j_1, \dots , j_{K+1}}~] \nonumber\\
    &= {n \choose {K+1}}{n-K-1 \choose {K+1}}\left(\frac{{1}}{{n-1 \choose K}}\right)^{2(K+1)}\nonumber\\
    &\qquad \cdot \left(\frac{{n-2K-3 \choose K}}{{n-1 \choose K}}\right)^{n-2(K+1)}\nonumber\\
     &\leq 
      {n \choose {K+1}}{n-K-1 \choose {K+1}}\left(\frac{{1}}{{n-1 \choose K}}\right)^{2(K+1)}\nonumber\\
    &\qquad \cdot \left(\frac{n-2K-3}{n-1}\right)^{K(n-2(K+1))}\label{eq:koutbounds:gen_2out-bound-doublesum-2}\\
    &=
      \frac{n!}{(n-2(k+1))!((K+1)!)^2}\left(\frac{{K!}}{{n-1 \choose K}}\right)^{2(K+1)}\nonumber\\
    &\qquad \cdot \left(\frac{n-2K-3}{n-1}\right)^{K(n-2(K+1))}\nonumber\\
     &= \frac{(K!)^{2(K+1)}}{((K+1)!)^2}
      \frac{n(n-1)\dots (n-2K-3)}{(n (n-1) \dots (n-K))^{2(K+1)}}\nonumber\\
    &\qquad \cdot \left(\frac{n-2K-3}{n-1}\right)^{K(n-2(K+1))}\nonumber\\
    &\leq \frac{(K!)^{2(K+1)}}{((K+1)!)^2} \cdot \frac{n^{2(K+1)}}{(n-K)^{2K(K+1)}}\nonumber\\
    &\qquad \cdot \left(1-\frac{2(K+1)}{n-1}\right)^{K(n-2(K+1))}
      \label{eq:koutbounds:gen_2out-bound-doublesum},
\end{align}{}
where (\ref{eq:koutbounds:gen_2out-bound-doublesum-2}) follows from (\ref{eq:koutbounds:ratio}).Substituting (\ref{eq:koutbounds:gen_2out-bound-singlesum}) and (\ref{eq:koutbounds:gen_2out-bound-doublesum}) in (\ref{eq:koutbounds:gen_2out-union-lb}), we obtain
\begin{align}
&\pr[~ Z_n \geq 1 ~] \nonumber\\
& \geq \sum_{  i_1<i_2\dots<i_{K+1}} \pr [~\Delta_{i_1\dotsi_{K+1}}~]\nonumber\\
&-\sum_{ i_1<i_2\dots<i_{K+1}}\sum_{ j_1<j_2\dots<j_{K+1}} \pr [~\Delta_{i_1\dotsi_{K+1}}\cap\Delta_{j_1,\dots,j_{K+1}}~]\nonumber\\
& \geq \frac{({K!})^K}{{K+1}}\cdot \frac{1}{n^{(K^2-1)}} \left(1-\frac{K+1}{n-K}\right)^{K(n-K-1)}\nonumber\\
&\qquad -\frac{(K!)^{2(K+1)}}{((K+1)!)^2} \cdot \frac{n^{2(K+1)}}{(n-K)^{2K(K+1)}}\nonumber\\
    &\qquad \cdot \left(1-\frac{2(K+1)}{n-1}\right)^{K(n-2(K+1))}\nonumber\\
& = \frac{({K!})^K e^{-K(K+1)}}{{K+1}}\cdot \frac{1}{n^{(K^2-1)}}  (1+\oo(1))  \label{eq:koutbounds:gen_upperrbound}\\
&= \Omega\left(\frac{1}{n^{K^2-1}} \right).\nonumber
\end{align}{}
\myendpf
In view of (\ref{eq:koutbounds:gen_p_con_upper_bound}), we then obtain for $K\geq2$ that
\[
\pr [~\hh \text{~is connected}~]  = 1-\Omega\left(\frac{1}{n^{K^2-1}} \right).
\]

 \subsection{Lower bound on probability of connectivity (Theorem~\ref{thm:LowerBoundForConnectivity})}
The lower bound proof follows \cite{FennerFrieze1982,Yagan2013Pairwise}, with a key difference: we upper bound ${n \choose r}$ using a variant of Stirling’s formula \eqref{eq:koutbounds:stirling}, instead of the standard bound ${n \choose r} \leq \left(\frac{ne}{r} \right)^r$ typically used in the literature.

Fix $n=2,3, \ldots $ and consider a fixed positive integer $K$.
Throughout, we assume the following:
\begin{equation}
2 \leq K  \quad \mbox{and} \quad e(K+2) < n.
\label{eq:koutbounds:OneLawConditions}
\end{equation}
Note that the condition $e(K+2) < n$ automatically implies $K<n$.

\label{sec:lowerbound-bounds}
\subsubsection{Preliminaries}
Next, we note that for $0\leq K \leq x \leq y$, 
\begin{align}
 \frac{ {x \choose K} }{ {y \choose K} }
= \prod_{\ell=0}^{K-1}
\left (
\frac{x-\ell}{y-\ell} \right ) \leq \left ( \frac{x}{y}
\right )^K \label{eq:koutbounds:ratio}   
\end{align}{}
since $\frac{x-\ell}{y-\ell}$ decreases as $\ell$ increases from
$\ell = 0$ to $\ell=K-1$.
Lastly, for all $x \in \R$, we have
\begin{align}
    1  \pm x  &\leq e^{\pm x}. \label{eq:koutbounds:kcon_1pmx}
\end{align}

 \subsubsection{Proof of Theorem~\ref{thm:LowerBoundForConnectivity} (Lower bound on Connectivity)}

If $\mathbb{H} (n;K)$ is {\em not} connected, then there exists a non-empty subset $S$ of nodes that is isolated. Further, since each node is paired with at least $K$ neighbors, $|S|\geq K+1$. Let $C_n (K)$ denote the event that $\mathbb{H}(n;K)$ is
connected. We have 
\begin{equation}
C_n(K)^c
\subseteq
\bigcup_{S \in \mathcal{P}_n: ~ |S| \geq K+1} ~ B_n (K ; S)
\label{eq:koutbounds:BasicIdea}
\end{equation}
where $\mathcal{P}_n$ stands for the collection of all non-empty
subsets of ${\cal N}$. Let $\mathcal{P}_{n,r} $ denotes the collection of all subsets
of ${\cal N}$ with exactly $r$ elements.
A standard union bound argument yields
\begin{eqnarray}
\bP{ C_n(K)^c }
&\leq &
\sum_{ S \in \mathcal{P}_n: K+1 \leq |S| \leq \lfloor \frac{n}{2} \rfloor }
\bP{ B_n (K ; S) }
\nonumber \\
&=&
\sum_{r=K+1}^{ \lfloor \frac{n}{2} \rfloor }
\left ( \sum_{S \in \mathcal{P}_{n,r} } \bP{ B_n (K; S) } \right ).
\label{eq:koutbounds:BasicIdea+UnionBound}
\end{eqnarray}
For each $r=1, \ldots , n$, let
$B_{n,r} (K) = B_n (K ; \{ 1, \ldots , r \} )$. Exchangeability implies
\[
\bP{ B_n (K ; S) } = \bP{ B_{n,r} (K ) }, \quad S \in
\mathcal{P}_{n,r}
\]
and since $|\mathcal{P}_{n,r} | = {n \choose r}$, we have
\begin{equation}
\sum_{S \in \mathcal{P}_{n,r} } \bP{ B_n (K ; S) } 
= {n \choose r} ~ \bP{ B_{n,r}(K ) } 
\label{eq:koutbounds:ForEach=r}
\end{equation}
Substituting (\ref{eq:koutbounds:ForEach=r})
into (\ref{eq:koutbounds:BasicIdea+UnionBound}) we obtain 
\begin{align}
\bP{ C_n(K)^c }& \leq \sum_{r=K+1}^{ \lfloor \frac{n}{2} \rfloor }{n \choose r} ~ \bP{ B_{n,r}(K ) } .
\nonumber\\
&\leq 
\sum_{r=K+1}^{ \lfloor \frac{n}{2} \rfloor } {n \choose r}\hspace{-1mm}
\left({ {r-1}\choose K }\over { {n-1} \choose K } \right)^{\hspace{-.5mm}r}\hspace{-1mm}
\left(\hspace{-1mm}{ {n-r-1}\choose K }\over { {n-1} \choose K } \hspace{-1mm}\right)^{n-r}.\label{eq:koutbounds:BasicIdea+UnionBound2}
\end{align}



 Using (\ref{eq:koutbounds:ratio}) in
(\ref{eq:koutbounds:BasicIdea+UnionBound2}) together with  (\ref{eq:koutbounds:stirlingtakeaway}), 
we conclude that
\begin{align}
& \bP{ C_n(K)^c }
\nonumber \\
& \leq \sum_{r=K+1}^{ \lfloor \frac{n}{2} \rfloor } {n \choose r} \left(\frac{r-1}{n-1}\right)^{rK}
\left(1-\frac{r}{n-1}\right)^{(n-r)K}\\
& \leq \sum_{r=K+1}^{ \lfloor \frac{n}{2} \rfloor } {n \choose r} \left(\frac{r}{n}\right)^{rK}
\left(1-\frac{r}{n}\right)^{(n-r)K}\nonumber\\
& \leq \sum_{r=K+1}^{ \lfloor \frac{n}{2} \rfloor }
\frac{1}{\sqrt{2 \pi}} \left(\frac{n}{n-r}\right)^{n-r} \left( \frac{n}{r}\right)^r \frac{\sqrt{n}}{\sqrt{n-r}\sqrt{r}}\nonumber\\ &\qquad \cdot \left(\frac{r-1}{n-1}\right)^{rK}
\left(1-\frac{r}{n-1}\right)^{(n-r)K}
\nonumber \\
& =
\sum_{r=K+1}^{ \lfloor \frac{n}{2} \rfloor }\frac{\sqrt{n}}{\sqrt{2 \pi}\sqrt{n-r}\sqrt{r}}
\left(\frac{r}{n}\right)^{r(K-1)}
\left(1-\frac{r}{n}\right)^{(n-r)(K-1)}
\nonumber \\
&\leq
\hspace{-1mm}\sum_{\hspace{-.5mm}r=K+1}^{ \lfloor \frac{n}{2} \rfloor }\hspace{-.5mm}\frac{\sqrt{n}}{\sqrt{2 \pi}\sqrt{n-r}\sqrt{r}}
\left(\hspace{-.5mm}\frac{r}{n}\hspace{-.5mm}\right)^{r(K-1)}
\hspace{-1.5mm}e^{-\left(\frac{r}{n}\right){(n-r)(K-1)}}
\label{eq:koutbounds:eapply},
\end{align}
where (\ref{eq:koutbounds:eapply}) follows from (\ref{eq:koutbounds:kcon_1pmx}). For $K+1 \leq r \leq  \lfloor \frac{n}{2} \rfloor,$ we have
\begin{align}
r(n-r) \geq (K+1)(n-K-1) 
\end{align}
Substituting in (\ref{eq:koutbounds:eapply}),
\begin{align}
&{\bP{ C_n(K)^c }}\nonumber \\
 & \leq \sum_{r=K+1}^{ \lfloor \frac{n}{2} \rfloor }\frac{\sqrt{n}}{\sqrt{2 \pi}\sqrt{n-K-1}\sqrt{K+1}}
\left(\frac{r}{n}\right)^{r(K-1)}\nonumber\\ &\qquad \cdot
e^{-\left(\frac{K+1}{n}\right){(n-K-1)(K-1)}}
\nonumber \\
& =
\sum_{r=K+1}^{ \lfloor \frac{n}{2} \rfloor }
\left(\frac{r}{n}\right)^{r(K-1)} 
\frac{e^{- (K^2-1)(1- \frac{K+1}{n})}}{\sqrt{2 \pi (K+1)}}\sqrt{\frac{n}{(n-K-1)}}.
\nonumber  \\
& = c(n;K)
\sum_{r=K+1}^{ \lfloor \frac{n}{2} \rfloor }
\left(\frac{r}{n}\right)^{r(K-1)}
\nonumber \\
& = c(n;K)\hspace{-1mm}
\left(\hspace{-0.5mm}\frac{K+1}{n}\hspace{-0.5mm}\right)^{K^2-1}
\hspace{-1.5mm} +
c(n;K)\hspace{-1.5mm}
\sum_{r=K+2}^{ \lfloor \frac{n}{2} \rfloor }
\left(\frac{r}{n}\right)^{r(K-1)}
\label{eq:koutbounds:con_1_law_last_step} 
\end{align}
with $c(n;K)$ given by (\ref{eq:koutbounds:c(K)}). 
Additionally, we have the bound below derived in \cite{Yagan2013Pairwise}:
for all $n \geq n(K)$ yielding
\[
\sum_{r=K+2}^{ \lfloor \frac{n}{2} \rfloor }
\left(\frac{r}{n}\right)^{r(K-1)}
\leq
\left \lfloor \frac{n}{2} \right \rfloor
\cdot
\left(\frac{K+2}{n}\right)^{(K+2)(K-1)} .
\]

Substituting in (\ref{eq:koutbounds:con_1_law_last_step}) and noting that $P(n;K)=1-\bP{ C_n(K)^c }$, we  obtain (\ref{eq:koutbounds:LowerBoundForConnectivity}).

\subsection{Proof of Lemma~\ref{corr:one_soln}}
First, $\alpha=0$ is a solution to this equation since when $\alpha=0$, both $I_{\alpha}(r,r)$ and  $c\alpha$ are zero. Also, when $c=1$, $\alpha=1/2$ is a solution of the equation since $I_{1/2}(r,r)=1/2$. The derivative of both terms with respect to $\alpha$ is:
\begin{align}
    \frac{\partial (I_{\alpha}(r,r))}{\partial \alpha} &= \frac{\alpha^{r-1}(1-\alpha)^{r-1}}{B(r,r)} , \quad
    \frac{\partial (c \alpha)}{\partial \alpha} = c
\end{align}
It can be seen that the derivative of $c\alpha$ is a constant, and  $\frac{\left(\alpha(1-\alpha)\right)^{r-1}}{B(r,r)}=0$ when $\alpha=0$ and $\frac{\left(\alpha(1-\alpha)\right)^{r-1}}{B(r,r)}$ is monotone increasing in the range $0<\alpha \leq 1/2$. For the case where $c=1$; $\alpha=0$ and $\alpha=1/2$ is a solution to the equation, hence for some $0<\alpha^{**}<1/2$ that satisfies $\frac{\left(\alpha^{**}(1-\alpha^*)\right)^{r-1}}{B(r,r)} = 1$, it must hold that $\frac{\partial (I_{\alpha}(r,r))}{\partial \alpha} < \frac{\partial (\alpha)}{\partial \alpha} =1 $ when $0< \alpha<\alpha^{**}$, and $\frac{\partial (I_{\alpha}(r,r))}{\partial \alpha} > \frac{\partial (\alpha)}{\partial \alpha} =1 $ when $ \alpha^{**} <\alpha \leq 1/2$. This is because if such $\alpha^{**}$ such that $0<\alpha^{**}<1/2$ does not exist, $\alpha=1/2$ can't be a solution to the equation $I_{\alpha}(r,r) = c\alpha$. Now, considering the case for arbitrary $0< c \leq 1$, since $\frac{\alpha^r(1-\alpha)^r}{B(r,r)}=0$ is monotone increasing, there can only be one $0<\alpha^*<1/2$ such that $\frac{(\alpha^*)^r(1-\alpha^*)^r}{B(r,r)} = c$. This means that $c\alpha$ is increasing faster than $I_{\alpha}(r,r)$ in the region $0<\alpha<\alpha^*$, hence there can't be a solution to $I_{\alpha}(r,r)=c\alpha$ in this region. Further, $I_{\alpha}(r,r)$ is increasing faster than $c\alpha$ in the region $\alpha^*<\alpha\leq 1/2$, hence there can be at most one solution to the equation $I_{\alpha}(r,r) = c\alpha$ in the region $0 <\alpha \leq 1/2$. Now, consider the fact that  $I_{\alpha^*}(r,r) < c\alpha^*$, and $I_{1/2}(r,r) = 1/2 \geq c/2$ when $\alpha=1/2$. Combining this with the fact that both functions are continuous, there must be at least one solution to the equation $I_{\alpha}(r,r)=c \alpha$ for $0<c \leq 1$ in the range $\alpha^*<\alpha\leq 1/2$. Combining this with previous statement (that there can be at most one solution in this range), it can be concluded that there is only one solution to the equation $I_{\alpha}(r,r) = c\alpha$ in the range $0<\alpha \leq 1/2$ where $0<c<1$. \qed

\subsection{Proof of Lemma~\ref{corrol:prob1}}
For node $v$, after making one selection, the number of nodes available to choose from decreases so the probability of choosing a node in $S_m^c$ changes at each selection. For example, the probability of choosing a node in $S_m^c$ in the first selection is $\frac{n-m}{n-1}$ and the probability of choosing a node in $S_m^c$ in the second selection is $\frac{n-m-1}{n-2}$ if a node in $S_m^c$ was selected in the first selection and it is $\frac{n-m}{n-2}$ otherwise. Based on this, the probability of selecting a node in $S_m^c$ at the $i^{th}$ selection out of $K_n$ selections can be expressed as $\frac{n-m-j}{n-i}$, $1 \leq i \leq K_n$, $0 \leq j < i$ where $j$ denotes the number of nodes already chosen from the set $S_m^c$ before the $i^{th}$ selection. Since we are considering the case of choosing less than $r$ nodes in $S_m^c$, we have that $j < r$, and with this constraint the lowest possible value of $\frac{n-m-j}{n-i}$ occurs when $j=r-1$ and $i=r$, and hence it is $\frac{n-m-r+1}{n-r}$. This gives a lower bound on the probability of selecting a node in $S_m^c$ in one of the $K_n$ selections and at the same time is an upper bound on the probability of not  selecting a node in $S_m^c$ in one of the $K_n$ selections, and hence it is an upper bound for choosing less than $r$ nodes. \qed

\subsection{Preliminary steps for proving Theorems  \ref{theorem:thmc_1} - \ref{theorem:thmg_5}} \label{sec:proof_start}

 Since the preliminary steps related to the proofs of \ref{theorem:thmc_1} - \ref{theorem:thmg_5} follow a similar framework \cite{IT23sood}, we summarize the main steps here.
 \begin{definition}[Cut]
\cite[Definition 6.3]{MeiPanconesiRadhakrishnan2008}
For a graph $\mathbb{G}$ defined on the node set $\nodes$, a \emph{cut} is a non-empty subset $S \subset \nodes$ of nodes {\em isolated} from the rest of the graph. Namely,  $S \subset \nodes$ is a cut if there is no  edge between $S$ and $S\comp=\nodes \setminus S$. If $S$ is a cut, then so is $S\comp$.
\label{def:cut}
\end{definition}{}
Note that the above definition follows \cite{IT23sood, MeiPanconesiRadhakrishnan2008} and is different from the notion of a cut defined as a subset of nodes that partitions graph into two disjoint subsets.
First, recall that the metrics connectivity  and  the  size  of  the  giant  component  under node removals are defined for the graph $\hhdn$, where the set $\mathcal{N}_D$ of nodes is removed from the graph $\hhn$ and $\mathcal{N}_R:=\mathcal{N} \setminus \mathcal{N}_D$, the remaining nodes in $\hhdn$. Let $\mathcal{E}_n (K_n, \gamma_n; S)$ denote the event that  $S \subset \nodes_R$ is a cut in $\hhdn$ as per Definition~\ref{def:cut}. The event $\mathcal{E}_n (K_n, \gamma_n; S)$ occurs if no nodes in $S$ pick neighbors in $S\comp$, and no nodes in $S\comp$ pick neighbors in $S$. Note that nodes in $S$ or $S\comp$ can still pick neighbors in the set $\nodes_D$. Thus, we have
\begin{align}
\mathcal{E}_n (K_n, \gamma_n; S) =
\bigcap_{i \in S} \bigcap_{j \in S\comp}
\left(
\left \{ i \not \in \Gamma_{\mathcal{N_R},j} \right \}
\cap 
\left \{ j \notin \Gamma_{\mathcal{N_R},i} \right \}
\right). \nonumber
\end{align}

Let $\mathcal{Z}(\lambda_n;K_n, \gamma_n)$ denote the event that $\hhdn$ has no cut $S \subset \nodes_R $ with size  $\lambda_n \leq |S| \leq n-\gamma_n - \lambda_n$ where  $\gamma:\N_0 \rightarrow  \N_0$ is a sequence such that $\lambda_n \leq (n-\gamma_n)/{2} \ \forall n$. 
In other words, $\mathcal{Z}(\lambda_n;K_n, \gamma_n)$ is the event that there are no cuts in $\hhdn$ whose size falls in the range $[\lambda_n, n-\gamma_n-\lambda_n]$. 
\begin{lemma}
\cite[Lemma 8]{IT23sood} For any sequence $x: \mathbb{N}_0 \rightarrow \mathbb{N}_0$ such that $\lambda_n \leq \lfloor (n-\gamma_n)/3 \rfloor$ for all $n$, we have
\begin{align}
    \mathcal{Z}(\lambda_n;K_n, \gamma_n) \Rightarrow |C_{max}(n, K_n, \gamma_n)| > n - \gamma_n - \lambda_n.
\end{align}
\label{lemma:gc}
\end{lemma}
\vspace{-5mm}
Lemma \ref{lemma:gc} states that if the event $\mathcal{Z}(\lambda_n;K_n, \gamma_n)$  holds, then the size of the largest connected component of $\hhdn$ is greater than $n - \gamma_n - \lambda_n$; i.e.,  there are less than $\lambda_n$ nodes outside of the giant component of $\hhdn$. Also note that $\hhdn$ is connected if $\mathcal{Z}(\lambda_n;K_n, \gamma_n)$ takes place with $\lambda_n=1$, since a graph is connected if no node is outside the giant component. In order to establish the Theorems \ref{theorem:thmc_1}-\ref{theorem:thmg_5}., we need to show that $\lim_{n \to \infty} \pr [\mathcal{Z}(\lambda_n;K_n, \gamma_n)\comp] = 0$ with $\lambda_n$, $K_n$ and $\gamma_n$ values as stated in each Theorem. 
From the definition of $\mathcal{Z}(\lambda_n;K_n, \gamma_n)$, we have
\begin{align}
\mathcal{Z}(\lambda_n;K_n, \gamma_n) & = \bigcap_{S \in \mathcal{P}_n: ~\lambda_n\leq  |S| \leq \lfloor \frac{n-\gamma_n}{2} \rfloor}  \left(\mathcal{E}_n({K}_n,{\gamma_n}; S)\right)\comp, \nonumber
\end{align}
where $\mathcal{P}_n$ is the collection of all non-empty  subsets of $\nodes_R$. Complementing both sides and using the union bound, we get
\begin{align}
\pr\left[\left(\mathcal{Z}(\lambda_n;K_n, \gamma_n)\right)\comp\right] &\leq \hspace{-3mm}  \sum_{ S \in \mathcal{P}_n: \lambda_n \leq |S| \leq \lfloor \frac{n-\gamma}{2} \rfloor } \pr[ \mathcal{E}_n ({K}_n,{\gamma_n}; S) ] \nonumber \\
&=\hspace{-1mm} \sum_{r=\lambda_n}^{ \left\lfloor \frac{n-\gamma}{2} \right\rfloor } \hspace{-1mm}
 \sum_{S \in \mathcal{P}_{n,r} } \pr[\mathcal{E}_n ({K}_n,{\gamma_n}; S)] \label{eq:BasicIdea+UnionBound},
\end{align}
where  $\mathcal{P}_{n,r} $ denotes the collection of all subsets of $\nodes_R$ with exactly $r$ elements.
For each $r=1, \ldots , \left\lfloor (n-\gamma_n)/2\right\rfloor$, we can simplify the notation by denoting $\mathcal{E}_{n,r} ({K}_n,{\gamma_n})=\mathcal{E}_n ({K}_n,{\gamma_n} ; \{ 1, \ldots , r \} )$. From the exchangeability of the node labels and associated random variables, we have
\[
\pr[ \mathcal{E}_n({K}_n,{\gamma_n} ; S) ] = \pr[ \mathcal{E}_{n,r}({K}_n,{\gamma_n}) ], \quad S \in
\mathcal{P}_{n,r}.
\]
$|\mathcal{P}_{n,r} | = {n-\gamma_n \choose r}$, since there are ${n-\gamma_n \choose r}$ subsets of $\nodes_R$ with r elements. Thus, we have
\begin{equation*}
\sum_{S \in \mathcal{P}_{n,r} } \pr[\mathcal{E}_n ({K}_n,{\gamma_n} ; S) ] 
= {n-\gamma_n\choose r} ~ \pr[\mathcal{E}_{n,r} ({K}_n,{\gamma_n})]. 
\label{eq:ForEachr}
\end{equation*}
Substituting this into (\ref{eq:BasicIdea+UnionBound}), we obtain 
\begin{align}
\hspace{-.4mm} \pr\left[\left(\mathcal{Z}(\lambda_n;K_n,\gamma_n)\right)\comp\right] \leq \hspace{-1mm} \sum_{r=\lambda_n}^{ \left\lfloor \frac{\hspace{-.5mm} n-\gamma}{2} \right\rfloor } \hspace{-1mm} 
{n-\gamma_n \hspace{-.4mm} \choose r } \hspace{-.4mm}  \pr[ \mathcal{E}_{n,r}({K}_n,{\gamma_n})] \hspace{-1mm} 
\label{eq:Z_bound}
\end{align}

Remember that $\mathcal{E}_{n,r}({K}_n,{\gamma_n})$ is the event that the $n-\gamma_n-r$ nodes in $S$ and $r$ nodes in $S\comp$ do not pick each other, but they can pick nodes from the set $\nodes_D$. Thus, we have:
 \begin{align}
\pr [\mathcal{E}_{n,r}({K}_n,{\gamma_n})] & =   \left( \dfrac{{\gamma_n+r-1 \choose K_n}}{{n-1 \choose K_n}} \right)^{r} \left( \dfrac{{n-r-1 \choose K_n}}{{n-1 \choose K_n}} \right)^{n-\gamma_n-r} \nonumber\\
&\leq \left(\dfrac{\gamma_n+r}{n}\right)^{rK_n} 
\left(\dfrac{n-r}{n}\right)^{K_n(n-\gamma_n-r)} \nonumber
\end{align}
Abbreviating $\pr\left[\mathcal{Z}(1;K_n, \gamma_n)\comp\right]$ as $P_Z$, we get from (\ref{eq:Z_bound}) that
\begin{align}
\hspace{-1mm} P_Z \leq \hspace{-1mm}  \sum_{r=\lambda_n}^{ \left\lfloor \frac{n-\gamma_n}{2} \right\rfloor } \hspace{-1mm} {\hspace{-.5mm}n-\gamma_n\hspace{-.5mm}\choose r} \hspace{-1mm} \left(\hspace{-.5mm}\dfrac{\hspace{-.5mm}\gamma_n+r\hspace{-.5mm}}{n}\right)^{\hspace{-1mm} r K_n} 
\hspace{-1mm} \left(\hspace{-.5mm}\dfrac{n-r}{n}\hspace{-.5mm} \right)^{\hspace{-1mm} K_n(n-\gamma_n-r)\hspace{-.5mm}} 
\label{eq:gc_pz0}
\end{align}
Using the upper bound on binomials \eqref{eq:binomial_property} again, we have
\begin{align}
P_Z &\leq \sum_{r=\lambda_n}^{ \left\lfloor \frac{n-\gamma_n}{2} \right\rfloor } \left( \frac{n-\gamma_n}{r}\right)^r \left( \frac{n-\gamma_n}{n-\gamma_n -r}\right)^{n-\gamma_n-r} \nonumber\\ &\quad \cdot \left(\dfrac{\gamma_n +r}{n}\right)^{rK_n} 
\left(\dfrac{n-r}{n}\right)^{K_n(n-\gamma_n -r)} 
\label{eq:gc_pz}
\end{align}

In order to establish the Theorems, we need to show that (\ref{eq:gc_pz}) goes to zero in the limit of large n for $\lambda_n$, $\gamma_n$ and $K_n$ values as specified in each Theorem. 

Since they will be referred to frequently throughout the proofs, we also include here the following bounds.
\begin{align}
    1\pm x \leq e^{\pm x} \label{eq:expon_upper}
\end{align}
\vspace{-2mm}
\begin{align}
{n \choose m} \leq \left(\frac{n}{m}\right)^m \left(\frac{n}{n-m}\right)^{n-m}, \quad \forall m=1,\ldots,n
\label{eq:binomial_property} 
\end{align}

\subsection{A Proof of Theorem \ref{theorem:thmc_1}}

Recall that in Theorem \ref{theorem:thmc_1}, we have $\gamma_n = \alpha n$ with $0<\alpha<1$ and that we need  $\lambda_n=1$ for connectivity. Using  \eqref{eq:expon_upper}
in  (\ref{eq:gc_pz}), we have
\begin{align}
P_Z \leq \sum_{r=1}^{ \left\lfloor \frac{n-\alpha n}{2} \right\rfloor } \left(\dfrac{n -\alpha n}{r}\right)^r e^r \left(\alpha + \dfrac{r}{n}\right)^{rK_n} e^{\frac{-r K_n (n-\alpha n -r)}{n}} \nonumber
\end{align}
We will show that the right side of the above expression goes to zero as $n$ goes to infinity. Let
 $$A_{n,r,\alpha}: = \left(\dfrac{n -\alpha n}{r}\right)^r e^r \left(\alpha + \dfrac{r}{n}\right)^{rK_n}e^{\frac{-rK_n(n-\alpha n -r)}{n}}.$$
We write
\begin{align*}
P_Z \leq \sum_{r=1}^{ \left\lfloor n/\log n \right\rfloor } A_{n,r,\alpha} + \sum_{r=\left\lfloor n/\log n \right\rfloor}^{ \left\lfloor \frac{n-\alpha n}{2} \right\rfloor } A_{n,r,\alpha} := S_1 + S_2,
\end{align*}
and show that both $S_1$ and $S_2$ go to zero as $n \to \infty$.
We start with the first summation $S_1$.
\begin{align}
S_1  & \! \begin{multlined}[t]
 \leq \sum_{r=1}^{ \left\lfloor n/\log n  \right\rfloor } \left((1 -\alpha )en\cdot e^{  K_n \log(\alpha+ \frac{1}{\log n})  -K_n (1-\alpha -\frac{1}{\log n}) } \right)^r \nonumber
 \end{multlined}
 \end{align}
Next, assume as in the statement of Theorem \ref{theorem:thmc_1} that 
\begin{align}
    K_n = \frac{c_n \log n}{1 - \alpha - \log \alpha}, \quad n=1,2,\ldots
\label{eq:proof1_k}
\end{align}
 for some sequence $c: \mathbb{N}_0 \to \mathbb{R}_+$ such that $\lim_{n \to \infty}c_n = c$ with $c>1$. 
Also define 
\begin{align*}
a_n &:=  (1 -\alpha )en\cdot e^{  K_n\log(\alpha+ \frac{1}{\log n})  -K_n(1-\alpha -\frac{1}{\log n}) }  \\
& = (1 -\alpha )en\cdot e^{  -\frac{c_n \log n}{1 - \alpha - \log \alpha}\left(1-\alpha -\frac{1}{\log n} - \log(\alpha+ \frac{1}{\log n}) \right) } 
\\
& = (1 -\alpha )e n^{1-c_n} \cdot e^{\frac{c_n}{1- \alpha - \log \alpha}\left(1- \log n \cdot \log(1+ \frac{1}{\alpha \log n}) \right)}
\\
& = O(1) n^{1-c_n}
\end{align*}
where we substituted $K_n$ via (\ref{eq:proof1_k}) and used the fact that $ \log n \cdot \log(1+ \frac{1}{\alpha \log n}) =  \frac{1}{\alpha} +o(1)$. Taking the limit as $n \to \infty$ and recalling that 
$\lim _{n \to \infty} c_n = c >1$, we see that $\lim _{n \to \infty} a_n = 0$. Hence, for large $n$, we  have
\begin{align}
S_1 \leq \sum_{r=1}^{ \left\lfloor n/\log n  \right\rfloor } \left( a_n \right)^r \leq \sum_{r=1}^{ \infty} \left( a_n \right)^r = \frac{a_n}{1-a_n}
\end{align}
where the geometric sum converges by virtue of $\lim _{n \to \infty} a_n = 0$. Using this, it is clear that 
$\lim _{n \to \infty}S_1 = 0$.

Now, consider the second summation $S_2$.

\begin{align}
S_2 &\leq\! \sum_{r=\lfloor n/\log n\rfloor}^{\left\lfloor (n-\alpha n)/2 \right\rfloor}
  \left(\dfrac{(n-\alpha n)e}{n/\log n}\right)^r
  \left(\dfrac{\alpha n + \tfrac{n-\alpha n}{2}}{n}\right)^{r K_n}\nonumber\\
  &\quad \cdot e^{\frac{-rK_n}{n}\bigl(n-\alpha n - \tfrac{n-\alpha n}{2}\bigr)} \nonumber\\
&\! \leq \sum_{r=\lfloor n/\log n\rfloor}^{\left\lfloor (n-\alpha n)/2 \right\rfloor}
  \left((1-\alpha)e\log n \cdot e^{K_n\log\!\bigl(\tfrac{1+\alpha}{2}\bigr)\;-\;K_n\tfrac{1-\alpha}{2}}\right)^r \nonumber
\end{align}

Next, we define 
\begin{align}
b_n &:= (1 - \alpha )e\log n \cdot e^{ -K_n \left(  \frac{1-\alpha}{2}  - \log(\frac{1+\alpha}{2}) \right)} \nonumber\\
& = (1 - \alpha )e\log n \cdot e^{ -\frac{c_n \log n}{1 - \alpha - \log \alpha} \left(  \frac{1-\alpha}{2}  - \log(\frac{1+\alpha}{2}) \right)} \nonumber
\end{align}
where we substituted for $K_n$ via (\ref{eq:proof1_k}). Taking the limit as ${n \to \infty}$ we see that
$
\lim _{n \to \infty} b_n = 0
$
upon noting that $ \frac{1-\alpha}{2}  - \log(\frac{1+\alpha}{2}) >0$ and $\lim _{n \to \infty} c_n =c >1$.
 With  arguments similar to those used in the case of $S_1$, we can show that when $n$ is large, $S_2 \leq b_n/(1-b_n)$, leading to $S_2$ converging to zero as $n$ gets large.
With $P_Z \leq S_1 + S_2$, and both $S_1$ and $S_2$ converging to zero when $n$ is large, we establish the fact that $P_Z$ converges to zero as $n$ goes to infinity. This result also yields the desired conclusion $\lim _{n \to \infty} P(n,K_n,\gamma_n) =1$ in Theorem \ref{theorem:thmc_1} since $P_Z = 1-P(n,K_n,\gamma_n)$.  \myendpf
\subsection{A Proof of Theorem \ref{theorem:thmc_2}}
We will first start with part (a) of Theorem \ref{theorem:thmc_2}. \\
\emph{Part a)} Recall that in part (a), $\gamma_n = o(\sqrt{n})$ and  we need $\lambda_n = 1$ for connectivity. Using this and \eqref{eq:expon_upper} in \eqref{eq:gc_pz}, we get
{
\begin{align}
P_Z &\leq\! \sum_{r=1}^{\left\lfloor \frac{n-\gamma_n}{2} \right\rfloor}
  \left(\dfrac{n-\gamma_n}{r}\right)^r
  \left(\frac{n-\gamma_n}{n-\gamma_n-r}\right)^{n-\gamma_n-r} \nonumber\\
  &\quad\left(\dfrac{\gamma_n+r}{n}\right)^{r K_n}
  \left(\frac{n-r}{n}\right)^{K_n(n-\gamma_n-r)} \nonumber\\
&\leq\! \sum_{r=1}^{\left\lfloor \frac{n-\gamma_n}{2} \right\rfloor}
  \left(\dfrac{n-\gamma_n}{r}\right)^r
  \left(1 + \frac{r \gamma_n}{n(n-\gamma_n-r)}\right)^{n-\gamma_n-r} \nonumber\\
  &\quad
  \left(\dfrac{\gamma_n+r}{n}\right)^{r K_n}
  \left(\frac{n-r}{n}\right)^{(K_n-1)(n-\gamma_n-r)} \nonumber\\
&\leq\! \sum_{r=1}^{\left\lfloor \frac{n-\gamma_n}{2} \right\rfloor}
  \left(1+\dfrac{\gamma_n}{r}\right)^{r}
  \left(\dfrac{\gamma_n+r}{n}\right)^{r (K_n-1)}
  e^{\frac{-r(K_n-1)(n-\gamma_n-r)}{n}} \nonumber
\end{align}
}

We will show that the right side of the above expression goes to zero as $n$ goes to infinity. Let
 $$A_{n,r,\gamma_n}: = \left( 1+\dfrac{\gamma_n }{r}\right)^{r}   \left( \dfrac{\gamma_n + r}{n}\right)^{r (K_n-1)} e^{\frac{-r(K_n-1)(n-\gamma_n -r)}{n}} $$
We write
\begin{align*}
P_Z \leq \sum_{r=1}^{ \left\lfloor \sqrt{n} \right\rfloor } A_{n,r,\gamma_n} + \sum_{r=\left\lceil \sqrt{n} \right\rceil}^{ \left\lfloor \frac{n-\gamma_n}{2} \right\rfloor } A_{n,r,\gamma_n} := S_1 + S_2,
\end{align*}
and show that both $S_1$ and $S_2$ go to zero as $n \to \infty$.
We start with the first summation $S_1$.
\begin{align}
S_1 & \leq \sum_{r=1}^{ \left\lfloor \sqrt{n} \right\rfloor }  \left( 1+\frac{\gamma_n }{r}\right)^{r}    \left( \frac{\gamma_n + r}{n}\right)^{r (K_n-1)}  e^{\frac{-r(K_n-1)(n-\gamma_n -r)}{n}} \nonumber \\
& \leq \! \begin{multlined}[t] \sum_{r=1}^{ \left\lfloor \sqrt{n} \right\rfloor }  \left(e^{\log \left( 1+\gamma_n \right)  + (K_n-1) \left[  \log \left( \frac{\gamma_n + \sqrt{n}}{n}\right) - \frac{n-\gamma_n -\sqrt{n}}{n}\right] }\right)^r \nonumber
\end{multlined} \nonumber
\end{align}
Next, assume as in the statement of Theorem \ref{theorem:thmc_2}(a) that \\ $K_n\geq 2, \  \forall n$. Also define
\begin{align*}
a_n &:= e^{\log \left( 1+\gamma_n \right)  + (K_n-1) \left[  \log \left( \frac{\gamma_n + \sqrt{n}}{n}\right) - \frac{n-\gamma_n -\sqrt{n}}{n}\right] }
\\
& \leq e^{\log \left( 1+\gamma_n \right)  + \log \left( 1+\frac{\gamma_n}{\sqrt{n}} \right) -  \log \left( \sqrt{n}\right)} e^{- \frac{n-\gamma_n -\sqrt{n}}{n} }
\\
& = O(1) e^{\log \left( 1+\gamma_n \right) -  \log \left( \sqrt{n}\right)}
\end{align*}
Taking the limit as $n \to \infty$ and recalling that $\gamma_n = o(\sqrt{n})$, we see that
$\lim _{n \to \infty} a_n = 0$. Hence, for large $n$, we  have
\begin{align}
S_1 \leq \sum_{r=1}^{ \left\lfloor \sqrt{n}  \right\rfloor } \left( a_n \right)^r \leq \sum_{r=1}^{ \infty} \left( a_n \right)^r = \frac{a_n}{1-a_n}
\end{align}
where the geometric sum converges by virtue of $\lim _{n \to \infty} a_n = 0$. Using this once again, it is clear from the last expression that 
$\lim _{n \to \infty}S_1 = 0$.

Now, consider the second summation $S_2$.
\begin{align}
S_2 & \leq \! \begin{multlined}[t] \sum_{r=\left\lceil \sqrt{n} \right\rceil}^{ \left\lfloor \frac{n-\gamma_n}{2} \right\rfloor }  \left(e^{\frac{r\gamma_n}{\sqrt{n}}  + (K_n-1) \left[  \log \left( \frac{n + \gamma_n }{2n}\right) - \frac{n-\gamma_n }{2n}\right] }\right)^r \nonumber
\end{multlined} \nonumber
\end{align}

Again assume as in the statement of Theorem \ref{theorem:thmc_2}(a) that $K_n\geq 2$. Next, we define=
\begin{align*}
b_n &:= e^{\frac{r\gamma_n}{\sqrt{n}}  + (K_n-1) \left[  \log \left( \frac{n + \gamma_n }{2n}\right) - \frac{n-\gamma_n }{2n}\right] }
\\
& \leq e^{\frac{r\gamma_n}{\sqrt{n}}  +   \log \left( \frac{1 }{2}\right) +   \log \left( \frac{n }{n+\gamma_n}\right) - \frac{1}{2} + \frac{\gamma_n }{2n} }
\\
& = O(1) e^{-  \log \left( \sqrt{n}\right)}
\end{align*}
Taking the limit as $n \to \infty$ and recalling that $\gamma_n = o(\sqrt{n})$, we see that
$\lim _{n \to \infty} b_n = 0$. Hence, for large $n$, we  have
\begin{align}
S_2 \leq \sum_{r=\left\lceil \sqrt{n} \right\rceil}^{ \left\lfloor \frac{n-\gamma_n}{2} \right\rfloor }  \left( b_n \right)^r \leq \sum_{r=\left\lceil \sqrt{n} \right\rceil}^{ \infty} \left( b_n \right)^r = \frac{(b_n)^{\sqrt{n}}}{1-b_n}
\end{align}
where the geometric sum converges by virtue of $\lim _{n \to \infty} b_n = 0$. Using this once again, it is clear from the last expression that 
$\lim _{n \to \infty}S_2 = 0$. With $P_Z \leq S_1 + S_2$, and both $S_1$ and $S_2$ converging to zero when $n$ is large, we establish the fact that $P_Z$ converges to zero as $n$ goes to infinity. This result also yields the desired conclusion $\lim _{n \to \infty} P(n,K_n,\gamma_n) =1$ in Theorem \ref{theorem:thmc_2}(a) since $P_Z = 1-P(n,K_n,\gamma_n)$. \\ \emph{Part b)} We now continue with the proof of Theorem \ref{theorem:thmc_2}(b). Recall that we had $\gamma_n = \Omega(\sqrt{n})$ and $\gamma_n = o(n)$. Using this and \eqref{eq:expon_upper}
in (\ref{eq:gc_pz}), we get
\begin{align}
P_Z &\leq\! \sum_{r=1}^{\left\lfloor \frac{n-\gamma_n}{2} \right\rfloor}
  \left(\dfrac{n-\gamma_n}{r}\right)^r
  \left(\frac{n-\gamma_n}{n-\gamma_n-r}\right)^{n-\gamma_n-r}\nonumber\\
  &\quad \cdot
  \left(\dfrac{\gamma_n + r}{n}\right)^{rK_n}  e^{\frac{-rK_n(n-\gamma_n-r)}{n}} \nonumber\\
&\leq\! \sum_{r=1}^{\left\lfloor \frac{n-\gamma_n}{2} \right\rfloor}
  e^{(1-\gamma_n/n)r}
  \left(\dfrac{\gamma_n + r}{r}\right)^{r}
  \left(\dfrac{\gamma_n + r}{n}\right)^{r(K_n-1)}\nonumber\\
  &\quad\cdot
  e^{\frac{-rK_n(n-\gamma_n-r)}{n}} \nonumber\\
&\leq\! \sum_{r=1}^{\left\lfloor \frac{n-\gamma_n}{2} \right\rfloor}
  \exp\!\Bigl(r\bigl[(K_n-1)\bigl(\log\bigl(\tfrac{\gamma_n + r}{n}\bigr)
    + \tfrac{r}{n} + \tfrac{\gamma_n}{n} -1\bigr) \nonumber\\
  &\quad
    + \log\bigl(1 + \gamma_n\bigr)
    + \tfrac{n-\gamma_n}{2n}\bigr]\Bigr)
\label{eq:proof3_2}
\end{align}

Next, assume as in the statement of Theorem \ref{theorem:thmc_2}(b) that
\begin{align}
    K_n = \frac{\log(\gamma_n+1)}{\log2 + 1/2} + w(1), \quad n=1,2,\ldots
\label{eq:proof3_2a}
\end{align}
Since $K_n-1 > 0, \quad \forall n=1,2,\ldots$, and noting that $r\leq \left\lfloor \frac{n-\gamma_n}{2} \right\rfloor$ in (\ref{eq:proof3_2}), we have 
\begin{align}
&(K_n-1) \left( \log \left( \frac{\gamma_n + r}{n}\right) + \frac{r}{n} + \frac{\gamma_n}{n} -1  \right) \nonumber\\
&\leq
(K_n-1) \left( \log \left( \frac{\gamma_n + \frac{n-\gamma_n}{2}}{n}\right) + \frac{\frac{n-\gamma_n}{2}}{n} + \frac{\gamma_n}{n} -1  \right)  \nonumber
\end{align}
Using this, we get
\begin{align}
P_Z &\leq\! \sum_{r=1}^{\left\lfloor \frac{n-\gamma_n}{2} \right\rfloor}
  \exp\Bigl(r\bigl[
    (K_n-1)\bigl(\log\bigl(\tfrac{n+\gamma_n}{2n}\bigr)
      - \tfrac{n-\gamma_n}{2n}\bigr) \nonumber\\
   & + \log\bigl(1+\gamma_n\bigr)
    + \tfrac{n-\gamma_n}{2n}
  \bigr]\Bigr)
\end{align}

Next, define 
\begin{align}
    a_n:=e^{(K_n-1)\left(\log \left( \frac{n + \gamma_n}{2n}\right) - \frac{n-\gamma_n}{2n}   \right) + \log \left(1 + \gamma_n \right) 
+ \frac{n-\gamma_n}{2n}}
\end{align}
Recall that $\gamma_n=o(n)$, so we have $\lim _{n \to \infty} \gamma_n/n = 0$. Using this, and substituting $K_n$ via (\ref{eq:proof3_2a}), we get
\begin{align}
    \lim _{n \to \infty} a_n &= \lim _{n \to \infty} \left[ e^{\left(\frac{\log(\gamma_n+1)}{\log 2 +1/2} + w(1)\right)\cdot \left(-log 2 - \frac{1}{2}   \right) + \log \left(1 + \gamma_n \right) 
+ \frac{1}{2}}\right] \nonumber \\
&= \lim _{n \to \infty} \left[ e^{ -w(1)\cdot \left(log 2 + 1/2   \right) - \log \left(1 + \gamma_n \right)+ \log \left(1 + \gamma_n \right) + \frac{1}{2}}\right] \nonumber \\
&= \lim _{n \to \infty} \left[ o(1) e^{ -w(1)\cdot \left(log 2 + 1/2  \right)} \right] = 0
\end{align}
Hence, for large $n$, we  have
\begin{align}
P_Z \leq \sum_{r=1}^{ \left\lfloor \frac{n-\gamma_n}{2} \right\rfloor } \left( a_n \right)^r \leq \sum_{r=1}^{ \infty} \left( a_n \right)^r = \frac{a_n}{1-a_n}
\end{align}
where the geometric sum converges by virtue of $\lim _{n \to \infty} a_n = 0$. Using this, it is clear from the last expression that 
$\lim _{n \to \infty}P_Z = 0$. This result also yields the desired conclusion $\lim _{n \to \infty} P(n,K_n,\gamma_n) =1$ in Theorem \ref{theorem:thmc_2}(b) since $P_Z = 1-P(n,K_n,\gamma_n)$. This result, combined with the proof of part a, concludes the proof of Theorem \ref{theorem:thmc_2}. 
\myendpf

\subsection{A Proof of Theorem \ref{theorem:thmg_3}}

Recall that in Theorem \ref{theorem:thmg_3}, we have $\gamma_n = o(n)$ and $\lambda_n = \Omega(\sqrt{n})$. Using \eqref{eq:expon_upper} in (\ref{eq:gc_pz}),  we have
\begin{align}
P_Z &\leq\! \sum_{r=\lambda_n}^{\left\lfloor \frac{n-\gamma_n}{2} \right\rfloor}
  \left(\dfrac{n-\gamma_n}{r}\right)^r
  \left(\frac{n-\gamma_n}{n-\gamma_n-r}\right)^{n-\gamma_n-r} \nonumber\\ &\quad \cdot
  \left(\dfrac{\gamma_n + r}{n}\right)^{r K_n}
  \left(\frac{n-r}{n}\right)^{K_n(n-\gamma_n-r)}
  \nonumber\\
&\leq\! \sum_{r=\lambda_n}^{\left\lfloor \frac{n-\gamma_n}{2} \right\rfloor}
  \left(\dfrac{n-\gamma_n}{r}\right)^r
  \left(1 + \frac{r \gamma_n}{n(n-\gamma_n-r)}\right)^{n-\gamma_n-r} \nonumber\\ &\quad \cdot
  \left(\dfrac{\gamma_n + r}{n}\right)^{r K_n}
  \left(\frac{n-r}{n}\right)^{(K_n-1)(n-\gamma_n-r)}
  \nonumber\\
&\leq\! \sum_{r=\lambda_n}^{\left\lfloor \frac{n-\gamma_n}{2} \right\rfloor}
  \left(\dfrac{n-\gamma_n}{r}\right)^r
  e^{\frac{r\gamma_n}{n}}
  \left(\dfrac{\gamma_n + r}{n}\right)^{r}
  \left(\dfrac{\gamma_n + r}{n}\right)^{r (K_n-1)}
 \nonumber\\ &\quad \cdot e^{\frac{-r(K_n-1)(n-\gamma_n-r)}{n}}
  \nonumber\\
&\leq\! \sum_{r=\lambda_n}^{\left\lfloor \frac{n-\gamma_n}{2} \right\rfloor}
  \left(1+\dfrac{\gamma_n}{r}\right)^{r}
  \left(\dfrac{\gamma_n + r}{n}\right)^{r (K_n-1)}
 \nonumber\\ &\quad \cdot e^{\frac{-r(K_n-1)(n-\gamma_n-r)}{n}}
  \nonumber
\end{align}
Next, assume as in the statement of Theorem \ref{theorem:thmg_3} that
\begin{align}
    K_n > 1 + \frac{\log(1+\gamma_n / \lambda_n)}{\log 2 + 1/2}
    \label{eq:proof_3_3}
\end{align}
Since $K_n > 1$, we have
\begin{align}
P_Z & \leq \! \begin{multlined}[t] \sum_{r=\lambda_n}^{ \left\lfloor \frac{n-\gamma_n}{2} \right\rfloor }  \left( 1+\dfrac{\gamma_n }{\lambda_n}\right)^{r}    \left( \dfrac{n + \gamma_n }{2n}\right)^{r (K_n-1)}  e^{\frac{-r(K_n-1)(n-\gamma_n )}{2n}}   \nonumber
\end{multlined} \nonumber 
\end{align}
We will show that the right side of the above expression goes to zero as $n$ goes to infinity. Let
 $$a_n: = e^{\log\left(1+\frac{\gamma_n }{\lambda_n}\right) + (K_n-1)\left[ \log\left(\frac{n+\gamma_n}{2n}\right) - \frac{n-\gamma_n}{2n}\right]} $$
Recall that $\gamma_n=o(n)$, so we have $\lim_{n\to \infty}\gamma_n/n=0$. Using this, and substituting for $K_n$ via (\ref{eq:proof_3_3}), we get
\begin{align}
    a_n & < e^{\log\left(1+\frac{\gamma_n }{\lambda_n}\right) + \left( \frac{\log(1+\gamma_n / \lambda_n)}{- \log(1/2) - 1/2}\right)\left[ \log\left(\frac{n+\gamma_n}{2n}\right) - \frac{n-\gamma_n}{2n}\right]}  
\end{align}
Taking the limit $n\to \infty$, we have
\begin{align}
    \lim_{n\to \infty} a_n & <   \lim_{n\to \infty} e^{\log\left(1+\frac{\gamma_n }{\lambda_n}\right) -  \log(1+\gamma_n / \lambda_n)} = e^0 = 1 
\end{align}
Hence, for large $n$, we  have
\begin{align}
P_Z \leq \sum_{r=\lambda_n}^{ \left\lfloor \frac{n-\gamma_n}{2} \right\rfloor } \left( a_n \right)^r \leq \sum_{r=\lambda_n}^{ \infty} \left( a_n \right)^r = \frac{(a_n)^{\lambda_n}}{1-a_n}
\end{align}
where the geometric sum converges by virtue of $\lim _{n \to \infty} a_n < 1$ and $\lim _{n \to \infty} \lambda_n = w(1)$. Using this, it is clear from the last expression that 
$\lim _{n \to \infty}P_Z = 0$. This result also yields the desired conclusion $\lim _{n \to \infty} P_G(n,K_n,\gamma_n,\lambda_n) =1$ in Theorem \ref{theorem:thmg_3} since $P_Z = 1-P_G(n,K_n,\gamma_n,\lambda_n)$. This concludes the proof of Theorem \ref{theorem:thmg_3}. 

\myendpf

\subsection{A Proof of Theorem \ref{theorem:thmg_5}}
Recall that in Theorem \ref{theorem:thmg_5}, we have $\gamma_n = \alpha n$ with $\alpha$ in $(0,1)$, and $\lambda_n < \frac{(1-\alpha)n}{2}$. Using $\gamma_n = \alpha n$  in (\ref{eq:gc_pz}), we get
\begin{align}
P_Z &\leq\! \sum_{r=\lambda_n}^{\left\lfloor \frac{n-\alpha n}{2} \right\rfloor}
  \left(\dfrac{n-\alpha n}{r}\right)^r
  \left(\frac{n-\alpha n}{n-\alpha n-r}\right)^{n-\alpha n-r} \nonumber\\ & \quad \cdot
  \left(\dfrac{\alpha n + r}{n}\right)^{r K_n}
  \left(\frac{n-r}{n}\right)^{K_n(n-\alpha n-r)}
  \nonumber\\
&\leq\! \sum_{r=\lambda_n}^{\left\lfloor \frac{n-\alpha n}{2} \right\rfloor}
  (1-\alpha)^r\,e^{\alpha r}
  \left(1+\dfrac{\alpha n}{r}\right)^{r}
  \left(\dfrac{\alpha n + r}{n}\right)^{r(K_n-1)} \nonumber\\ & \quad \cdot
  e^{-r(K_n-1)\frac{(n-\alpha n-r)}{n}}
  \label{eq:proof3_5_2}
\end{align}

Next, assume as in the statement of Theorem \ref{theorem:thmg_5} that 
\begin{align}
    K_n > 1 + \frac{ \log(1+\frac{\alpha n}{\lambda_n}) + \alpha + \log(1-\alpha) }{\frac{1-\alpha}{2} +  \log2 - \log(1+\alpha)   }, \quad n=1,2,\ldots
\label{eq:proof3_5a}
\end{align}
Since $K_n > 1$, we have
\begin{align}
P_Z
&\leq\! \sum_{r=\lambda_n}^{\left\lfloor \frac{n-\alpha n}{2} \right\rfloor}
  (1-\alpha)^r\,e^{\alpha r}
  \left(1+\dfrac{\alpha n}{\lambda_n}\right)^r
  \left(\tfrac{1+\alpha}{2}\right)^{r (K_n-1)} \nonumber\\ &\cdot
  e^{-r(K_n-1)\left(\tfrac{1-\alpha}{2}\right)}
  \nonumber
\end{align}
Also define 
\vspace{1mm}
\begin{align}
a_n &:=  e^{\alpha +  \log(1-\alpha) + \log(1+\frac{\alpha n}{\lambda_n}) + (K_n-1)\left[  \log \left(\frac{1+\alpha}{2}\right) - \left( \frac{1-\alpha}{2}\right) \right] } \nonumber \\
&<  e^{\alpha +  \log(1-\alpha) + \log(1+\frac{\alpha n}{\lambda_n}) - \left( \alpha +  \log(1-\alpha) + \log(1+\frac{\alpha n}{\lambda_n}) \right) } = 1 \nonumber
\end{align}
where we substituted $K_n$ via (\ref{eq:proof3_5a}). Taking the limit as $n \to \infty$, we see that $\lim_{n \to \infty} a_n < 1$. Hence, for large n, we have
\begin{align}
P_Z \leq \sum_{r=\lambda_n}^{ \left\lfloor \frac{n-\alpha n}{2} \right\rfloor } \left( a_n \right)^r \leq \sum_{r=\lambda_n}^{ \infty} \left( a_n \right)^r = \frac{(a_n)^{\lambda_n}}{1-a_n} \nonumber
\end{align}
where the geometric sum converges by virtue of $\lim _{n \to \infty} a_n < 1$ and $\lim _{n \to \infty} \lambda_n = w(1)$. Using this, it is clear from the last expression that 
$\lim _{n \to \infty}P_Z = 0$. This result also yields the desired conclusion $\lim _{n \to \infty} P_G(n,K_n,\gamma_n,\lambda_n) =1$ in Theorem \ref{theorem:thmg_5} since $P_Z = 1-P_G(n,K_n,\gamma_n,\lambda_n)$. This concludes the proof of Theorem \ref{theorem:thmg_5}.
\myendpf

\subsection{Empirical investigation into the algebraic connectivity of random K-out graphs.}
\begin{figure}[tbhp]
  \centering
  \includegraphics[width=0.44\textwidth]{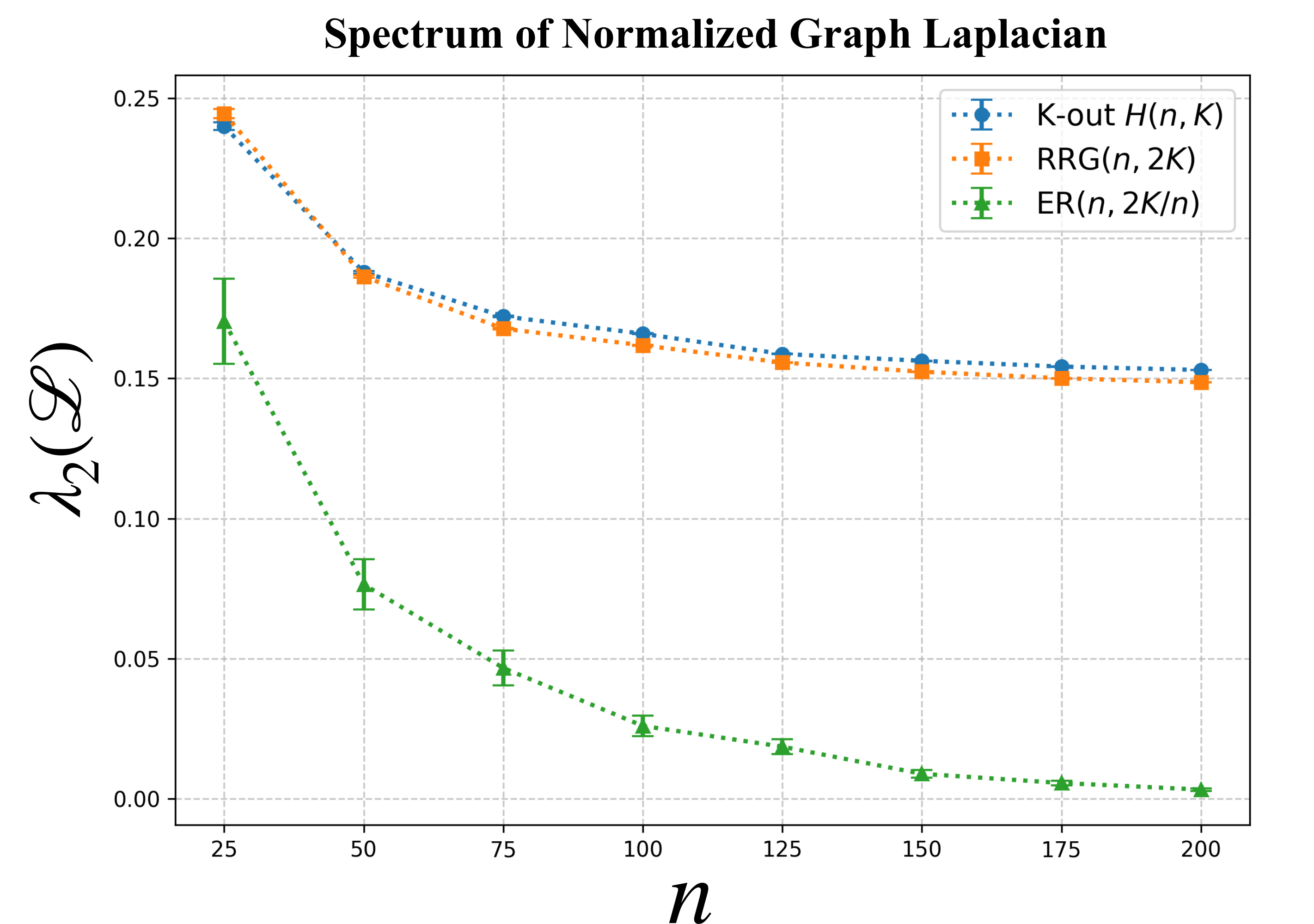}\hfill
  \includegraphics[width=0.44\textwidth]{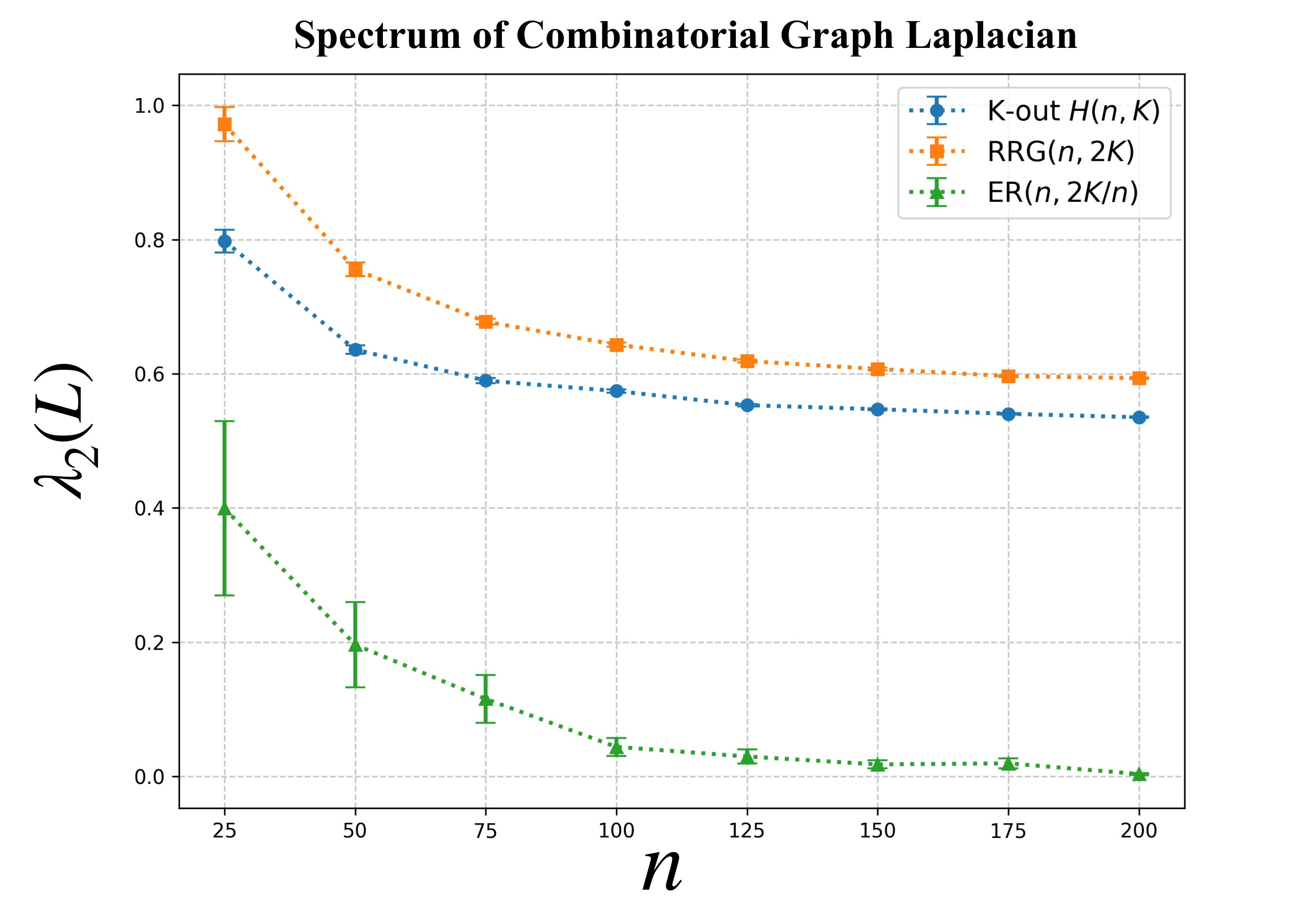}
  \caption{Empirical comparison of the algebraic connectivity for different random graph models: 
    We plot the second smallest eigenvalues, $\lambda_2(\mathcal{L})$ and $\lambda_2(L)$, for the normalized and combinatorial graph Laplacians. The X-axis shows the number of nodes. The Y-axis shows the average of 500 independent realizations for each random graph model: random K-out graphs (with $K=2$), random regular graphs (RRG), and Erdős–Rényi (ER) graphs. Parameters for RRG and ER graphs are chosen to keep the mean node degree fixed. Random K-out graphs closely follow RRGs, while ER graphs show both $\lambda_2(\mathcal{L})$ and $\lambda_2(L)$ rapidly decaying to zero, indicating a disconnected network.}
  \label{fig:alg_connectivity}
\end{figure}

Complementary to the study of graphs from a probabilistic combinatorial lens (as in the field of the random graph), there are many different ways to represent and derive graph properties, including algebraic/spectral approaches based on the graph's adjacency matrix and Laplacian and formulation of optimization problems. Due to correlations in edge generation, computing the spectrum of random matrices corresponding to the K-out graph adjacency matrices is challenging. However, a quantity deduced from the graph spectrum, known as the \emph{algebraic connectivity}, is important in several applications such as \emph{graph conductance}, \emph{mixing time} of Markov chains, and the \emph{convergence} of distributed optimization algorithms\cite{hoory2006expander,shah2009gossip, pirani2023graph}. In Figure~\ref{fig:alg_connectivity}, we perform simulations probing the algebraic connectivity for random K-out graphs. We present an overview of reasoning about the connectivity properties of a graph through spectral approaches below.

Let \( L(\mathcal{G}) := D - A \) denote the \emph{combinatorial} graph Laplacian of an undirected graph \(\mathcal{G}(V, E)\) with adjacency matrix \(A\), where \(D_{ii} = \sum_{j} A_{ij}\); \(D_{ij} = 0\) for \(i \neq j\). 
{For a graph $\mathcal{G}$ with combinatorial Laplacian $L(\mathcal{G})$, the \emph{normalized} Laplacian $\lnormalized(\mathcal{G})$ is related to $L(\mathcal{G})$ through
$$L(\mathcal{G}) = (D)^{1/2} \lnormalized(\mathcal{G})(D)^{1/2}, $$
where $D$ is the diagonal matrix of node degrees for $\mathcal{G}$. The eigenvalues of $L(\mathcal{G})$ are equal to those of $\lnormalized D$. Let $\lambda_1(L(\mathcal{G}))\leq \lambda_2 (L(\mathcal{G})) \cdots \leq \lambda_n (L(\mathcal{G})) \leq \lambda_n (L(\mathcal{G}))$ denote the eigenvalues of $L(\mathcal{G})$. The eigenvalue $\lambda_1(L(\mathcal{G})) = 0$ corresponds to the vector of all ones for all graphs. The second smallest eigenvalue $\lambda_2(L(\mathcal{G})) > 0$ if and only if the graph $\mathcal{G}$ is connected.  For this reason, $\lambda_2(L(\mathcal{G})$ is often referred to as the \emph{algebraic connectivity} of $\mathcal{G}$. The same is true for $\lambda_2(\mathcal{L}(\mathcal{G}))$, i.e., $\lambda_2(\mathcal{L}(\mathcal{G}))= 0$ if and only if $\mathcal{G}$ is disconnected. Furthermore, $0\leq\lambda_2(\mathcal{L}(G))\leq2$}. From the perspective of network design, having $\lambda_2(\mathcal{L}(G))>0$ corresponds to one of the notions of \emph{expansion} through the celebrated Cheeger's Inequality in spectral graph theory; see \cite{hoory2006expander} for equivalent definitions of expanders and a survey on their properties. \cite{pirani2023graph}

We plot the second smallest eigenvalues (algebraic connectivity),  $\lambda_2(\mathcal{L})$ and $\lambda_2(L)$ respectively corresponding to the normalized and combinatorial graph Laplacian. The X-axis represents the number of nodes in the network. For each random graph model: random K-out graphs, random regular graphs (RRG), and Erd\H{o}s-R\'enyi (ER) random graphs, we report the second smallest eigenvalue averaged over 500 independent realizations on the Y Axes. For the random K-out graph $\mathbb{H}(n;K)$, we set $K=2$. We construct RRG \cite{kim2003generating} and ER graphs with parameters chosen to keep the mean node degree fixed. For the setting considered, we observe that Random K-out graphs closely follow the RRGs, whereas for the ER graphs, both $\lambda_2(\mathcal{L})$ and $\lambda_2(L)$ rapidly decay to zero, indicating a disconnected network.


\end{document}